\documentclass[%
 letterpaper,
 reprint,
superscriptaddress,
 amsmath,amssymb, amsfonts,
 aps, prx
]{revtex4-2}

\usepackage{hyperref}
\usepackage{graphicx}
\usepackage{dcolumn}
\usepackage{bm}
\usepackage{lipsum}
\usepackage{txfonts}
\usepackage{xcolor}

\newcommand{\bra}[1]{\langle #1|}
\newcommand{\ket}[1]{| #1 \rangle}
\newcommand{\op}[1]{\hat{#1}}
\newcommand{\mc}[1]{\mathcal{#1}}

\newcommand{\tr}[1]{\mathrm{Tr}\{#1\}}
\newcommand{\proj}[1]{\ket{#1}\bra{#1}}
\newcommand{\erw}[1]{\left\langle #1 \right\rangle}

\newcommand{\bkt}[2]{\bra{#1} #2 \rangle}
\newcommand{\out}[2]{\ket{#1} \bra{#2}}

\newcommand{\ev}[1]{\mathbb{E}[#1]}
\newcommand{\he}{\mathrm{He}_n^{[\sigma]}(D)}
\newcommand{\hen}[1]{\mathrm{He}_{#1}^{[\sigma]}(D)}
\newcommand{\hex}[2]{\mathrm{He}_{#1}^{[\sigma]}(#2)}

\begin{document}

\preprint{}

\title{Quantum thermodynamics of continuous feedback control}

\author{Kacper Prech}
\email{kacper.prech@unibas.ch}
\affiliation{Department of Physics and Swiss Nanoscience Institute, University of Basel, Klingelbergstrasse 82, 4056 Basel, Switzerland}
\author{Joël Aschwanden}
\author{Partick P. Potts}
\affiliation{Department of Physics and Swiss Nanoscience Institute, University of Basel, Klingelbergstrasse 82, 4056 Basel, Switzerland}

\begin{abstract}
The laws of thermodynamics are a cornerstone for describing nanoscale and open quantum systems. However, formulating these laws for systems under continuous feedback control and under experimentally relevant conditions is challenging. In this work, we lay out a formalism for the laws of thermodynamics in an open quantum system under continuous measurement and feedback described by a Quantum Fokker Planck Master Equation. We derive expressions for work, heat, and measurement-induced energy changes, and we investigate entropy production and fluctuation theorems. We illustrate our results with a continuous version of a measurement-driven Szilard engine, as well as a work extraction scheme in a two-level system under bang-bang control. Our results provide insights into the energetics as well as the irreversibility of classical and quantum systems under continuous feedback control.

\end{abstract}

\maketitle

\section{Introduction}

Macroscale devices such as as engines and refrigerators are governed by the laws of thermodynamics~\cite{Groot1969NonequilibriumT}. The first law of thermodynamics states that the total energy change can be separated into work and heat. Work is an energy change mediated by degrees of freedom that we have control over, meaning that it can be harvested from or supplied to the device externally, whereas heat originates from interactions with an environment, which we cannot control. The second law states that the entropy production, which depends on the heat, cannot be negative, restricting which processes the system can undergo.

Our ability to measure and control systems in nanoscale and quantum regimes, where fluctuations play an essential role, has increased significantly in the last decades~\cite{Ciliberto2017, Minev2018, Guerlin2007, Delglise2008, Kurzmann2019, Hofmann2016, Gillett2010, Armen2002, Zhou2012, Riste2012, Champagne2013}. In order to understand the thermodynamics of such microscopic systems, definitions of work and heat have been extended to the classical stochastic 
regime~\cite{sekimoto1997, Seikimoto1998,seikimotoBook, PostModern, 
Seifert_2012} and to open quantum systems~\cite{Pusz, Alicki, Potts_2021}. 
The second law of thermodynamics has likewise been generalized to stochastic classical and quantum systems~\cite{Landi2021, Esposito_2011, Esposito_corr, spohn_ent, lindblad_ent, Potts_2021}. This formulation of the laws of thermodynamics finds applications, for instance, in investigating and realizing microscopic heat engines and refrigerators~\cite{MitchisonPotts, Ghosh2018, Levy2018, Rossnagel}. Moreover, going beyond average values, work, heat and entropy production have also been defined as stochastic quantities along individual trajectories both in the classical regime~\cite{Seifert_2005, Jarzynski_1997a, Jarzynski_1997b, Crooks_1999, Crooks_2000} and for open quantum systems~\cite{Horowitz_2012, Horowitz_2013, Manzano_2015, Manzano_2018, Manzano_2022, Breuer2005, Hekking2013}. A powerful result following from this formalism is a Fluctuation Theorem (FT)~\cite{Seifert_2005, Jarzynski_1997a, Jarzynski_1997b, Crooks_1999, Crooks_2000, Horowitz_2013, Manzano_2015, Manzano_2018, Manzano_2022}, which can be considered a generalization of the second law of thermodynamics.

Measurement and feedback is of fundamental interest in the field of open quantum systems as it enables enhanced control. A particularly useful type of feedback is continuous feedback control~\cite{Wiseman1993, Wiseman1994, Doherty1999, Korotkov2001, Zhang2017, BelavkinBook, Wiseman_Milburn_book, Jacobs_book, Annby2022}, where a controller performs a continuous measurement on the system and uses the obtained outcomes to influence the system. Among applications of continuous feedback protocols are cooling and trapping~\cite{UrsoCooling, UshevCooling, desousa2024cooling, kumasaki2025thermodynamicapproachquantumcooling, Debiossac}, quantum state stabilization~\cite{Vijay2012Stab, SmithStab, SayrinStab, Feng2011, Liu2010}, squeezing~\cite{WisemannSqu}, error correction~\cite{Sarovar2004}, and charging a battery~\citep{Mitchison2021chargingquantum}.

Feedback control has significant consequences for the laws of thermodynamics since the concept of entropy is closely related to information. This has been realized in early thought experiments including Maxwell's demon and Szilard's engine~\cite{maxwell1872theory, Szilard1929berDE, Rex}, where information on microscopic degrees of freedom seemingly allow to overcome the second law of thermodynamics 
Various models of implementing Maxwell's demon have been investigated in classical and quantum systems~\cite{Lloyd, annbyandersson2024maxwells, Elouard2017b, Annby2020, Sanchez2019, Schmitt2023, Esposito2012, Strasberg2013, Schaller2011, Averin2011, DeffnerDemon, Strasberg2017, Quan2006, Ito2014} as well as experimentally in superconducting circuits~\cite{Cottet2017, Masayuma, Naghiloo2017}, metallic islands~\cite{Koski_2014, Koski_2014b, Koski2015, Barker2022}, photonic systems~\cite{Vidrighin2016}, mechanical oscillators~\cite{archambault2024firstpassageinformationengine, archambault2024inertialeffectsdiscretesampling}, NMR systems~\cite{Camati2017}, silicon-vacancy centers~\cite{yada2024experimentallyprobingentropyreduction}, and Brownian particles~\cite{Tobaye_2010, Roldn2013, Paneru2018, Saha2022, Admon2018}.
In the modern understanding of the second law of thermodynamics for feedback controlled systems, the entropy production is constrained by an information term arising from correlations between the controller and the system's state~\cite{Sagawa2012, Potts2018, prech2023, Yada2022, Sagawa2008, Sagawa2010}. Similarly, in the FT, the stochastic entropy production is corrected by an additional stochastic information term, which was derived for both classical~\cite{Potts2018, Sagawa2012, Horowitz2010} and quantum systems~\cite{prech2023, Yada2022} under continuous feedback control.

Measurement and feedback also has a profound impact on the first law of thermodynamics. Any quantum measurement causes a backaction on the density matrix of the system~\cite{Nielsen_Chuang_2010}, which may change its energy~\cite{Elouard2017, Jacobs2009, Brandner_2015, Yi2017, abdelkhalek2018}. This has been dubbed \textit{quantum heat}~\cite{Elouard2017} or measurement energy~\cite{Jacobs2009, Brandner_2015} (hereafter we shall use the latter). 
Previous research on the energetics of quantum systems under measurement and feedback mostly focused either on discrete measurement and feedback protocols~\cite{Gong2016, Jacobs2009, Brandner_2015, Elouard2017b} or solely on the effect of measurement~\cite{Elouard2017, Rossi2020, Belenchia2020, Landi2022, Belenchia2022, Kewming2022, Elouard2025}, including an experimental investigation~\cite{Rossi2020}. However, there is only limited research concerning the energetics of continuous measurement and feedback protocols~\cite{Alonso2016, Bhandari2022, Yanik, Efficiently}.

Recently, a powerful approach to describing continuous feedback in experimentally realistic conditions has been derived: the Quantum Fokker-Planck Master Equation (QFPME)~\cite{Annby2022}. It describes the joint time-evolution of a quantum state and the measurement outcome used for feedback. This framework includes a finite bandwidth for the detector and applies to both linear and nonlinear feedback. A particularly important instance of the latter is bang-bang control, where the feedback is changed when the measurement outcome passes a threshold. This makes the QFPME a versatile tool for various feedback control schemes including optimal-control methods~\cite{Cavina2018}. 
The QFPME has already been applied to model quantum-to-classical transition in a double quantum dot implementation of Maxwell's demon~\cite{annbyandersson2024maxwells}, trapping a harmonic oscillator~\cite{desousa2024cooling}, and generation of entanglement in an open quantum system~\cite{Diotallevi2024}.

In this work, we investigate the laws of thermodynamics and FTs for an open quantum system under measurement and feedback described by the QFPME. First, we obtain expressions for the average power $P$, heat current $ J $, and measurement-energy rate $ \dot{E}_{\rm M} $. To illustrate the importance of our results, we apply these definitions to investigate a continuous version of the measurement-driven heat engine of Ref.~\cite{Elouard2017b}.
Next, we find a novel FT and second law of thermodynamics for classical systems, where a new stochastic term, which we name the measurement entropy, enters on equal footing with entropy production.
The measurement entropy originates from the detector's delay and thus has a different character than information terms in Refs.~\cite{Sagawa2012, Potts2018, prech2023, Yada2022}. We illustrate the FT and the second law on a two-level system subjected to a threshold-like (bang-bang) feedback protocol. We then extend our findings obtained for classical systems to the quantum regime with the aid of the Keldysh quasi-probability distribution~\cite{Hofer2017, Nazarov2003, Levitov1996, Hofer2016, Clerk2011}.

This paper is structured as follows. We begin by introducing the general setting of feedback control and the QFPME in Sec.~\ref{sec:QFPME}. In Sec.~\ref{sec:1law}, we present our results concerning the first law of thermodynamics. Our results about the fluctuation theorems and the second law of thermodynamics are provided in Sec.~\ref{sec:2law}. Lastly, discussion and conclusions feature in Sec.~\ref{sec:discussion}.

\section{Quantum Fokker Planck Master Equation}
\label{sec:QFPME}

\begin{figure} 
\includegraphics[width=0.9\linewidth]{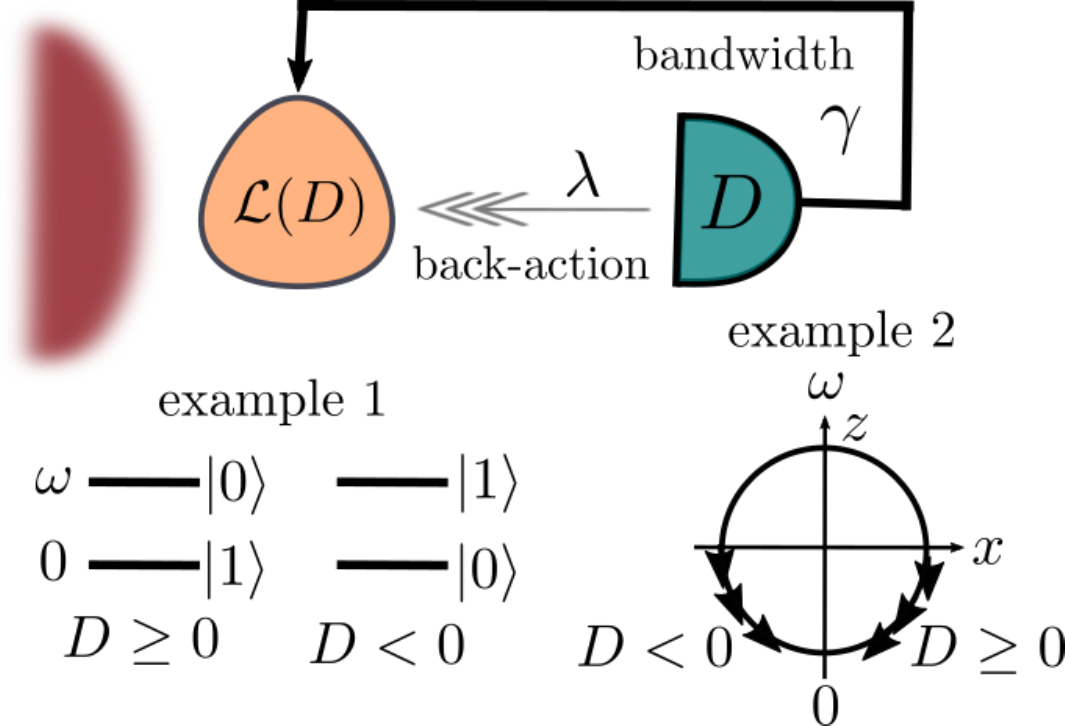}
	\caption{The general setting of the QFPME. A quantum system interacting with its environment is continuously measured by a detector with strength $\lambda$, which causes back-action on the system with the strength. The instantaneous outcome of the detector $D$ is affected by the bandwidth $\gamma$ and fed back into the superoperator $\mc{L}(D)$ that controls the dynamics of the system.
    Sketches "example 1" and "example 2" illustrate the two-level system under bang-bang control introduced in  Sec.~\ref{sec:2level} and the continuous-measurement-driven engine introduced in Sec.~\ref{sec:conteng}.
    \label{fig:feedbackpicture}}
\end{figure}

The general setup of the QFPME is illustrated in Fig.~\ref{fig:feedbackpicture}. We consider a continuous measurement of the Hermitian observable $\op{A}$~\cite{Jacobs2006rev, Bednorz2012} in an open quantum system. Feedback is controlled by an instantaneous measurement outcome $D$ of a detector with a bandwidth $\gamma$, which filters out high frequency noise from the outcomes of the continuous measurement~\cite{Whitley2010, Sarovar2004, Sarovar2005, Feng2011, Liu2010, WarszawskiI, WarszawskiII}, thus introducing a delay.

The time evolution of the joint state $\hat{\rho}_t(D)$ of the system and the measurement outcome follows the QFPME~\cite{Annby2022}:
\begin{equation}
\label{eq:QFPME}
\begin{split}
    \partial_t\hat{\rho}_t(D) &= \mathcal{L}(D)\hat{\rho}_t(D)+\lambda\mathcal{D}[\hat{A}]\hat{\rho}_t(D) \\
    & -\gamma\partial_D\mathcal{A}(D)\hat{\rho}_t(D)+\frac{\gamma^2}{8\lambda}\partial^2_D\hat{\rho}_t(D),
\end{split}
\end{equation}
where $\mathcal{A}(D)\hat{\rho}=\frac{1}{2}\{\hat{A}-D,\hat{\rho}\}$ and $\mc{D}[\op{O}]\op{\rho} = \op{O} \op{\rho} \op{O}^\dagger - \{\op{O}^\dagger \op{O}, \op{\rho} \}/2$.
The first term on the right hand side of Eq.~\eqref{eq:QFPME} describes feedback with a $D$-dependent Liouvillian $\mc{L}(D)$. The second term is a dephasing in the eigenbasis of the observable $\hat{A}$ with a rate $\lambda$, which is the effect of the measurement backaction. The measurement rate $\lambda$ quantifies its strength and invasiveness, with $\lambda \to \infty$ corresponding to a continuous projective measurement. The third and fourth terms, called drift and diffusion, describe the time-evolution of the detector's measurement outcome. 

In many situations of practical interest, where the system is weakly coupled to the environment, the superoperator $\mathcal{L}(D)$ corresponds to a Lindblad master equation~\cite{Pusz, Alicki, Potts_2021, Albash2012}
\begin{equation}
\label{eq:ME}
    \mathcal{L}(D) \hat{\rho}_t = -i[\hat{H}_t(D), \hat{\rho}_t ] + \mathcal{L}_{\rm B} \hat{\rho}_t,
\end{equation}
where $\mathcal{L}_{\rm B}$ describes interaction with the environment and $\hat{H}_t(D)$ is a $D$-dependent feedback Hamiltonian, which may also optionally include an explicit time dependence not resulting from feedback. Crucially, the Hamiltonian may be an arbitrary function of the measurement outcome. In this work, we will consider linear feedback, with $\hat{H}(D) \propto D$, as well as threshold-like feedback, where, e.g., $\op{H}_+$ and $\op{H}_-$ are applied when $D \geq D_{\rm th}$ and $D < D_{\rm th}$ respectively. Threshold-like feedback is also referred to as bang-bang control and is useful, for instance, for optimizing protocols for work extraction in driven open quantum systems~\cite{Cavina2018} and quantum error correcting~\cite{Sarovar2004}. 
In the QFPME-formalism~\eqref{eq:QFPME}, the distribution of the measurement outcome at time $t$ is given by $P_t(D) = \text{Tr} \{ \hat{\rho}_t(D) \}$. The quantum state of the system without any knowledge of the measurement outcome is given by $\hat{\rho}_t = \int dD \hat{\rho}_t(D) $, whereas the quantum state conditioned on the outcome $D$ is obtained by $\hat{\rho}_t(D) / P_t(D)$. Expectation values are computed by averaging over the measurement outcomes, as well as the quantum mechanical ensemble:
\begin{equation}
\label{eq:avg}
    \erw{\op{O}(D)} := \int dD \tr{\hat{O}(D) \hat{\rho}_t(D) }.
\end{equation}
This definition will be essential to evaluate energy terms in the QFPME.


We emphasize that the QFPME can also be applied to classical (incoherent) systems that are measured with a finite bandwidth. In this case $[\hat{A}, \hat{\rho}_t(D)] = 0$, i.e., there is no measurement backaction.

\section{The first law of thermodynamics}
\label{sec:1law}

In this section we present our investigation of the first law of thermodynamics for systems described by the QFPME. We begin with a brief summary of how heat and work are defined in Lindblad master equations and then proceed with our results for feedback controlled systems and their derivation. Our findings are then illustrated with examples.

\subsection{Power and heat current in the Lindblad master equation}

Here we consider open quantum systems without any feedback control that evolve in time under the Lindblad master equation $\partial_t \op{\rho}_t = \mc{L}(D) \op{\rho}_t$, i.e., according to Eq.~\eqref{eq:ME} but treating $D$ as a fixed parameter instead of a dynamical variable. The definitions of heat and work in Lindblad master equations have been debated in the literature~\cite{Novotny2002,Levy2014,Trushechkin2016,Hofer2017njp,Gonzalez2017,DeChiara2018, Hegwill2011}. Due to the approximations made in deriving master equations, the laws of thermodynamics may generally not be guaranteed. It is however always possible to obtain a thermodynamically consistent framework \cite{Potts_2021}. In this work, we focus on two particularly simple thermodynamic frameworks that cover many applications. The first framework~\cite{Pusz, Alicki} is obtained by defining the internal energy as
\begin{equation}
\label{eq:PowerHeatME}
   U \equiv  \erw{\op{H}_t(D)} = \tr{ \op{\rho}_t  \op{H}_t(D)}, 
\end{equation}
the power as the change in energy due to the change in the Hamiltonian,
\begin{equation}
\label{eq:PowerME}
    P = \tr{ \op{\rho}_t \partial_t \op{H}_t(D)} = \erw{\partial_t \op{H}_t(D)},
\end{equation}
and the heat current as the change in energy due to the change in the density matrix,
\begin{equation}
\label{eq:HeatME}
    J = \tr{\op{H}_t(D) \partial_t \op{\rho}_t} = \tr{\op{H}_t(D) \mc{L}_{\rm B} \op{\rho}_t} = \erw{\mc{L}_{\rm B}^\dagger \op{H}_t(D)},
\end{equation}
where $\mc{L}_{\rm B}^\dagger$ is the adjoint of $\mc{L}_{\rm B}$.
We note that the first law of thermodynamics, $\partial_t U = P+J$, is valid by construction.

The second framework relies on the thermodynamic Hamiltonian $\hat{H}_{\rm TD}$ introduced in Ref.~\cite{Potts_2021}. In particular, there are scenarios where a thermodynamically consistent framework can be obtained with a time-independent thermodynamic Hamiltonian obeying
\begin{equation} \label{eq:HTDtime}
    \partial_t\hat{H}_t(D) = i[\hat{H}_t(D),\hat{H}_{\rm TD}(D)],
\end{equation}
an example is presented in Sec.~\ref{sec:conteng} below.
In this framework, the internal energy and heat current are defined with the help of $\hat{H}_{\rm TD}(D)$:
\begin{equation}
\label{eq:HTD}
   U \equiv  \erw{\op{H}_{\rm TD}(D)},\hspace{.5cm} J =  \erw{\mc{L}_{\rm B}^\dagger \op{H}_{\rm TD}(D)},
\end{equation}
while the power is still given by Eq.~\eqref{eq:PowerME}.
Again, the first law of thermodynamics is obeyed, due to Eq.~\eqref{eq:HTDtime}. 




\subsection{The first law for the QFPME}

For quantum systems under continuous measurement and feedback described by the QFPME~\eqref{eq:QFPME}, the rate of change of the average energy can be separated into power $P$, heat current $J$, and the energy flow due to the measurement $\dot{E}_{\rm M}$:
\begin{equation}
\label{eq:PowerHeatQFPME}
    \partial_t U = \partial_t \int{dD} \tr{\op{U}(D)\op{\rho}_t(D) }  = P + J + \dot{E}_{\rm M},
\end{equation}
where the operator $\hat{U}(D)$ depends on the thermodynamic framework (in this work, it is equal to $\hat{H}_t(D)$ or $\hat{H}_{\rm TD}(D)$). The power is given by 
\begin{equation}
\label{eq:PowerQFPME}
    P  = \gamma \erw{ \mathcal{A}(D)\partial_D\hat{U}(D) }+\frac{\gamma^2}{8\lambda} \erw{\partial^2_D\hat{U}(D) } +\erw{ \partial_t\hat{H}_t(D) } ,
\end{equation}
and the heat current reads
\begin{equation}\label{eq:HeatQFPME}
    J = \erw{\mc{L}_{\rm B}^\dagger \op{U}(D)}.
\end{equation}
Finally, the energy flow due to the measurement reads
\begin{equation} \label{eq:EmQFPME}
    \dot{E}_{\rm M} = \lambda \erw{\mc{D}[\hat{A}] \op{U}(D)}.
\end{equation}
The average values above are defined according to Eq.~\eqref{eq:avg}. Equation~\eqref{eq:PowerHeatQFPME} constitutes the first law of thermodynamics for open quantum systems under continuous feedback control in analogy to Eq.~\eqref{eq:PowerHeatME}. A derivation and justification is presented in the subsection below.

The power $P$ consists of the three terms on the right-hand side of Eq.~\eqref{eq:PowerQFPME}. The first and the second one are a direct result of the drift and diffusion terms describing how the (thermodynamic) Hamiltonian changes in time due to the feedback. Importantly, the second term involving the second derivative with respect to $D$ plays a role for nonlinear feedback. The third term in power as well as the heat current  are equivalent to the expressions in the absence of feedback. The measurement energy rate $\dot{E}_{\rm M}$~\eqref{eq:EmQFPME} quantifies the change in energy due to the continuous measurement of $\op{A}$ as a result of the measured operator not commuting with the Hamiltonian. This term is also known as \textit{quantum heat}~\citep{Elouard2017}.

\subsection{Stochastic energetics in the QFPME}
While the first law in Eq.~\eqref{eq:PowerHeatQFPME} can be derived from the QFPME directly, the interpretation of heat and work is particularly illuminating in the stochastic formulation of the QFPME~\cite{Annby2022}. In this formulation, the central quantity is the density matrix conditioned on all previous measurement outcomes $\hat{\rho}_{\rm c}$. Its time-evolution is determined by the Belavkin equation, a stochastic differential equation
\begin{equation}
\label{eq:Belavkin}
    d\hat{\rho}_{\rm c}=\mathcal{L}(D)\hat{\rho}_{\rm c} dt+\lambda\mathcal{D}[A]\hat{\rho}_{\rm c} dt+\sqrt{\lambda}\{\hat{A}-\langle\hat{A}\rangle_{\rm c},\hat{\rho}_{\rm c} \} dW,
\end{equation}
where $\langle\hat{A}\rangle_{\rm c} = \tr{\op{A} \hat{\rho}_{\rm c}}$ and $dW$ is a Wiener increment with the expectation value $\ev{dW} = 0$ and $dW^2 = dt$~\cite{Jacobs_random}.

The measurement outcome is determined by an Ornstein-Uhlenbeck process, described by the stochastic differential equation~\cite{Annby2022}
\begin{equation}
\label{eq:dDito}
    dD_{\rm c} = \gamma \left( \langle\hat{A}\rangle_{\rm c} -D_{\rm c} \right)dt+\frac{\gamma}{2\sqrt{\lambda}}dW,
\end{equation}
where we use the subscript 'c' to denote that this is the measurement outcome conditioned on all previous measurement outcomes and not the argument in a distribution. In the Ornstein-Uhlenbeck process~\eqref{eq:dDito}, the terms on the right-hand site are called drift and diffusion, respectively. The drift describes how $D_{\rm c}$ approaches the conditional expectation value of $\op{A}$, wheres the diffusive part corresponds to the noise in the measurement outcome.

The left-hand side of Eq.~\eqref{eq:PowerHeatQFPME}, can then be equivalently expressed as
\begin{equation}
\label{eq:EnergyQFPME}
\begin{split}
    \partial_t \erw{ \hat{U} } &= \mathbb{E} \left[ \text{Tr} \left\{  \frac{d(\op{U}\op{\rho}_{\rm c})}{dt} \right\} \right]  \\
    &=\mathbb{E} \left[ \text{Tr} \left\{  \frac{d\op{U}\op{\rho}_{\rm c} + \op{U}d\op{\rho}_{\rm c} + d\op{U}d\op{\rho}_{\rm c}}{dt} \right\} \right],
\end{split}
\end{equation}
where $\ev{\cdot}$ denotes the average over all trajectories the measurement outcome can take.  For brevity of notation we have dropped the $D$-dependence in $\op{U} \equiv \op{U}(D_{\rm c})$, and we write $d\hat{U} \op{\rho}_{\rm c} = (d\hat{U}) \op{\rho}_{\rm c}$. We note that since $\hat{U}$ depends on $D_{\rm c}$, it is also a stochastic quantity. However, we refrain from giving it a subscript since it only depends on the current measurement outcome, not its full history in contrast to $\hat{\rho}_{\rm c}$ and $D_{\rm c}$.

To calculate the terms appearing in Eq.~\eqref{eq:EnergyQFPME}, we find the differential of the internal energy operator $\hat{U}$, which describes another Ornstein-Uhlenbeck process
\begin{equation}
\label{eq:dHito}
    d\op{U} = \left[ \partial_t \op{U} + \gamma \left( \langle \op{A}\rangle_{\rm c} -D_{\rm c} \right)\partial_D \op{U} + \frac{\gamma^2}{8 \lambda} \partial^2_D \op{U} \right]dt + \frac{\gamma}{2\sqrt{\lambda}} \partial_D \op{U} dW,
\end{equation}
where again $\hat{U}$ and its derivatives are evaluated at $D_{\rm c}$. A derivation of this equation can be found in App.~\ref{app:stochastic1}.
Since both $d\hat{U}$ and $d \op{\rho}_{\rm c}$ include the Wiener increment $dW$, and since $dW^2=dt$, the term $d\op{U} d \op{\rho}_{\rm c}$ is linear in $dt$ and thus contributes to Eq.~\eqref{eq:EnergyQFPME}.
Combining the Belavkin equation~\eqref{eq:Belavkin} with the differential in Eq.~\eqref{eq:dHito} allows us to obtain the expectation values of all three differential terms in Eq.~\eqref{eq:EnergyQFPME}, which are given by Eqs.~\eqref{eq:HdrhoRate},~\eqref{eq:rhodHRate}, and~\eqref{eq:dHdrhoRate}, respectively; see App.~\ref{app:stochastic2} for details.


When $\op{U}(D)=\op{H}_t(D)$, the differential terms in Eq.~\eqref{eq:EnergyQFPME} have a particularly transparent interpretation. 
Just like in the absence of feedback, we may use Eq.~\eqref{eq:dHito} to identify energy changes due to the change in Hamiltonian as work and energy changes due to changes in the density matrix as heat. In this approach, the term including $d\hat{U}\hat{\rho}_{\rm c}$ contributes to work, while the term $\hat{U}d\hat{\rho}_{\rm c}$ contributes to heat. As we discuss in more detail below, the remaining term including $d\hat{U}d\hat{\rho}_{\rm c}$ also contributes to work.

We first consider the heat-like contributions to the energy change
\begin{equation}
    \frac{\ev{\tr{\op{H}_t d\op{\rho}_{\rm c}}}}{dt} = J + \dot{E}_{\rm M},
    \label{eq:heatlike}
\end{equation}
which includes the heat current $J$~\eqref{eq:HeatQFPME} and the measurement energy rate $\dot{E}_{\rm M}$~\eqref{eq:EmQFPME}. This identification motivates the term \textit{quantum heat}~\cite{Elouard2017}. However, it has been shown that the energy contribution from the measurement cannot always be interpreted as heat but sometimes rather acts as a source of work~\cite{Jacobs2009,Elouard2017,Elouard2025}. We further note that in the thermodynamic framework where $\hat{U}(D) = \hat{H}_{\rm TD}(D)$, the power due to the explicit time-dependence of the Hamiltonian, $\erw{\partial_t \op{H}_t(D)}$, is included in $\ev{\tr{\op{H}_{\rm TD} d\op{\rho}_{\rm c}}}/dt$. 
This further illustrates that care has to be taken in identifying energy flows as heat or work.

The power reads
\begin{equation}
    \label{eq:powerlike}
    P = \frac{\ev{\tr{d\op{H}_t(\op{\rho}_{\rm c}+d\op{\rho}_{\rm c}}}}{dt}.
\end{equation}
This can be motivated by considering the feedback control in infinitesimal time increments. In each increment, a measurement is performed which updates the density matrix and the measurement outcome. Then, the Hamiltonian is updated by feeding back the new measurement outcome. These two steps can be illustrated as:
\begin{equation}
    \tr{\op{H}_t \op{\rho}_{\rm c} } \rightarrow \tr{\op{H}_t (\op{\rho}_{\rm c} + d \op{\rho}_{\rm c})} \rightarrow \tr{(\op{H}_t + d\op{H}_t) (\op{\rho}_{\rm c} + d \op{\rho}_{\rm c})}.
\end{equation}
Work is performed in the second step, where the Hamiltonian is changed, resulting in the power given in Eq.~\eqref{eq:powerlike}, see also App.~\ref{app:stochastic2} for details.

\subsection{Applicability to classical systems} \label{sec: clas energ}
Our results are equally applicable to classical systems under continuous feedback control. 
Suppose a classical system can occupy a set of microstates (configurations) $\{a\}$ with corresponding $D$-dependent energies $\omega_a(D)$. We can define the joint probability of the system $a$ and the measurement outcome $D$ at a given time as $p_a(D)$.

Formally, $\op{\rho}_t(D) = \sum_a p_a(D) \proj{a}$ and $\op{H}_t(D) = \sum_a \omega_a(D) \proj{a}$, where $\bkt{a}{a^{'}} = \delta_{a, a^{'}}$, and we can write the eigendecomposition of the observable as $\op{A} = \sum_a \xi_a \proj{a}$.

Then, the measurement energy $\dot{E}_{\rm M}$ vanishes. This is expected because there is no measurement backaction. The three terms that appear in the power $P$ in Eq.~\eqref{eq:PowerQFPME} [with $\op{U}(D) = \op{H}_t(D)$] can simply be found using the following expressions:
\begin{equation}
    \begin{split}
        & \erw{\partial^2_D\hat{H}_t(D) } = \sum_a \int dD p_a(D) \partial_D^2 \omega_a(D), \\
        &\erw{ \mathcal{A}(D)\partial_D\hat{H}_t(D) } = \sum_a \int dD p_a(D) (a-D) \partial_D \omega_a(D) ,\\
        & \text{and} \quad \erw{\partial_t\hat{H}_t(D) } = \sum_a \int dD p_a(D) \partial_t \omega_a(D). \\
    \end{split}
\end{equation}
Notably, the nonlinearity of feedback also has consequences for the energetics of classical systems.

\subsection{Example 1: two-level system under bang-bang control} \label{sec:2level}

\begin{figure}
\includegraphics[width=0.9\linewidth]{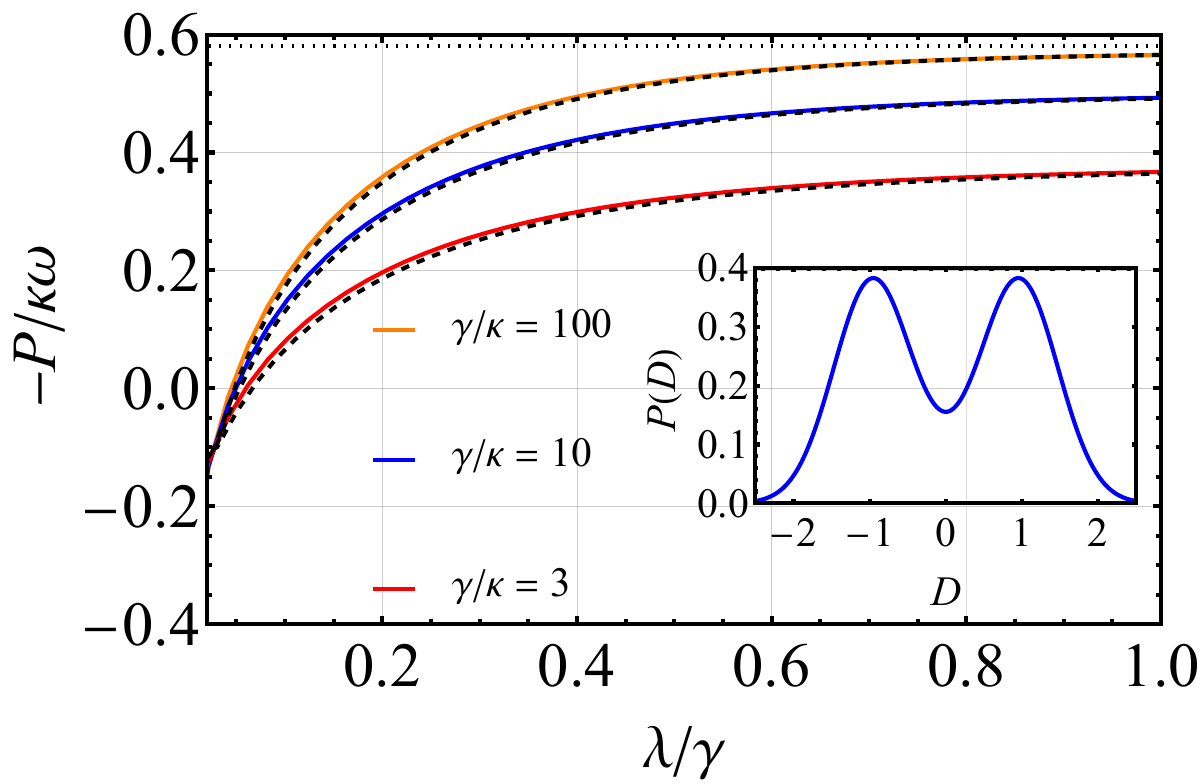}
	\caption{Extracted power for the two-level system under bang-bang control. The power is obtained from Eq.~\eqref{eq:Power2lvl} and shown for $\gamma/\kappa = 100$ (yellow), $\gamma/\kappa = 10$ (blue), and $\gamma/\kappa = 3$ (red). Black dashed lines represent corresponding heat currents, $J/\kappa \omega$, obtained from Eq.~\eqref{eq:Heat2lvl}. The horizontal black dotted line indicates the maximal possible power, $-P/(\kappa \omega) = n_{\rm B} \simeq 0.58$. The inset shows the probability $P(D)$ as a function of $D$ for $\gamma/\kappa = 10$ and $\lambda/\gamma = 0.5$. In all plots, $\omega= k_{\rm B}T$. 
	\label{fig:FirstLawClas}}
\end{figure}

As a first example to illustrate the first law of thermodynamics, we investigate a classical two-level system under bang-bang control (threshold-like feedback), introduced in Ref.~\cite{Annby2022}. The system, which is illustrated in Fig.~\ref{fig:feedbackpicture} ("example 1"), has two discrete energy levels, $\ket{0}$ and $\ket{1}$, with energy gap $\omega$ (hereafter we set $\hbar = 1$). The $D$-dependent Hamiltonian is given by [and we will use $\hat{U}(D)=\hat{H}(D)]$
\begin{equation} \label{eq:H2}
    \op{H}(D) = \theta(D)  \omega \proj{0} + (1-\theta(D)) \omega \proj{1},
\end{equation}
where $\theta(D)$ is the Heavyside step function. We continuously measure the observable $\op{A} = \op{\sigma}_z = \proj{1} - \proj{0}$, which means that the sign of $D$ indicates which state is occupied. Since $[\op{A}, \op{H}] = 0$, the system remains diagonal in the energy eigenbasis. The system is weakly coupled to a bosonic reservoir with the Bose-Einstein distribution $n_{\rm B} = \left[\exp{(\frac{\omega}{k_{\rm B}  T}) -1} \right]^{-1}$, where $k_{\rm B}$ and $T$ are the Boltzmann constant and temperature of the reservoir, respectively. The Liouvillian in the QFPME is given by (note that $[\hat{H}(D),\hat{\rho}(D)]=0$)
\begin{equation} \label{eq:LiouClas}
    \mc{L}(D) =  \theta(D) \mc{L}_+ + (1-\theta(D)) \mc{L}_-,
\end{equation}
where
\begin{equation} \label{eq:LiouPlus}
    \mc{L}_+ \op{\rho} = \kappa n_{\rm B} \mc{D}[\op{\sigma}] \op{\rho} + \kappa (n_{\rm B} +1) \mc{D}[\op{\sigma}^\dagger] \op{\rho}
\end{equation}
and
\begin{equation} \label{eq:LiouMinus}
    \mc{L}_- \op{\rho} = \kappa n_{\rm B} \mc{D}[\op{\sigma}^\dagger] \op{\rho} + \kappa (n_{\rm B} +1) \mc{D}[\op{\sigma}] \op{\rho}.
\end{equation}
Here, $\op{\sigma} = \out{0}{1}$ and $\kappa$ is characterizes the coupling strength to the environment.

In this setup, work is extracted in the following way: starting in the ground state, the system can be excited with rate $\kappa n_{\rm B}$. This state of change is ideally detected through the measurement outcome $D$ changing sign. When this happens, the ground and excited energy levels are swapped through feedback. The system is then again in its ground state, ready absorb the next quantum of energy $\omega$. 
For a perfect measurement, the reservoir is thus cooled one quantum at a time.
However, measurement inaccuracy and the detector's delay may result in heating instead.

For the power, we find from Eq.~\eqref{eq:PowerQFPME}
\begin{equation} \label{eq:Power2lvl}
    P = -\gamma \omega P(D=0) + \frac{\gamma^2}{8 \lambda} \omega \partial_D \left[  p_1(D)  -  p_0(D)   \right]|_{D = 0},
\end{equation}
where $p_0(D) = \bra{0} \op{\rho}_t(D) \ket{0}$, $p_1(D) = \bra{1} \op{\rho}_t(D) \ket{1}$, and $P(D) = p_0(D) + p_1(D)$.
For the heat current, we find from Eq.~\eqref{eq:HeatQFPME}
\begin{equation} \label{eq:Heat2lvl}
    J = \omega \kappa n_{\rm B} (1 - \eta) - \omega \kappa (1+n_{\rm B}) \eta,
\end{equation}
where
\begin{equation} \label{eq:eta}
    \eta = \int_0^\infty  p_0(D) dD + \int_{-\infty}^0  p_1(D) dD ;
\end{equation}
denotes the error probability, i.e., the probability that the sign of $D$ does not correspond to the state the system is in.
The smaller the error probability, the more heat is absorbed by the system and extracted as power via the feedback mechanism.
Since $[\op{A}, \op{\rho}_t(D)] = 0$, the measurement energy vanishes as expected, $\dot{E}_{\rm M} = 0$. Details can be found in App.~\eqref{sec:energeticsClas}.

The steady-state $P(D)$ that the measurement outcome takes the value $D$ is illustrated in the inset of Fig.~\ref{fig:FirstLawClas}. Two peaks located at $D=1$ and $D=-1$, which are eigenvalues of the measurement observable $\op{A} = \proj{1} - \proj{0}$, indicate that the measurement outcome is likely to accurately represent the actual state of the system, and smearing of the probability is caused by measurement imprecision and the finite bandwidth.
The extracted power, $-P$, computed using Eq.~\eqref{eq:Power2lvl}, is illustrated in Fig.~\ref{fig:FirstLawClas}. We note that in the steady state, $P+J=0$. As anticipated, increasing the measurement strength $\lambda$ results in a higher rate of work extraction as errors are suppressed. Furthermore, the higher the ratio $\gamma/\kappa$ is, the faster the detector and the more work is extracted. In the limit $\gamma \gg \kappa$, the measurement outcome $D$ can adapt to each transition before the subsequent transition occurs. The extracted power is saturated at $-P=\omega\kappa n_B$ for $\lambda\gg \gamma\gg \kappa$. This limit corresponds to error free work extraction, $\eta\rightarrow 0$. 
In App.~\ref{sec:ssClas}, we show how the probability and power in the steady state are computed.

\subsection{Example 2: Continuous-measurement-driven engine} \label{sec:conteng}

To illustrate our results in the quantum regime, we investigate an engine that is driven by a continuous measurement, i.e., the measurement energy flow $\dot{E}_{\rm M}$ is converted into work. The considered engine is a continuous version of the engine introduced in  Ref.~\cite{Elouard2017b}, where a projective measurement increases the energy of a qubit, which is then extracted as work by applying a unitary rotation. 

We consider a qubit with energy splitting $\omega$ that is coherently driven with a strength proportional to the measurement outcome
\begin{equation} \label{eq:Hlab}
    \hat{H}_{\rm lab}(D)=\frac{\omega}{2}\hat{\sigma}_z+ i g D \left(e^{i\omega t}\hat{\sigma}-e^{-i\omega t}\hat{\sigma}^{\dagger}\right).
\end{equation}
where $\op{\sigma}_{x, y, z}$ are standard Pauli matrices, and $\hat{\sigma}=(\hat{\sigma}_x+i\hat{\sigma}_y)/2$. The subscript $"\mathrm{lab}"$ signifies a laboratory frame.
The setup is shown in Fig.~\ref{fig:feedbackpicture} ("example 2"). Here we consider weak drives, i.e., $g\ll\omega$, which allows for a thermodynamically consistent description with the thermodynamic Hamiltonian
\begin{equation} \label{eq: H TD example}
    \hat{U} = \op{H}_{\rm TD} = \frac{\omega}{2}\hat{\sigma}_z.
\end{equation}
The internal energy of the system is thus fully determined by the projection of its Bloch vector on the $z$-axis.

In a frame rotating with $\frac{\omega}{2}\hat{\sigma}_z$, the Hamiltonian is given by (see App.~\ref{app:rotating} for details)
\begin{equation} \label{eq:H}
    \hat{H}(D) = gD \op{\sigma}_y,
\end{equation}
which corresponds to a rotation around the $y$-axis on the Bloch sphere with a direction set by the sign of $D$. We continuously measure the observable $\op{A} = \op{\sigma}_x$, which means the measurement outcome $D$ indicates a position of the quantum state on the $x$-axis of the Bloch sphere. If the sign of $D$ and the actual position on the $x$-axis are in agreement, the Hamiltonian in Eq.~\eqref{eq:H} drives the system towards the ground state of $\hat{\sigma}_z$, extracting work. However, if they are in disagreement, we supply work by driving towards the excited eigenstate of $\hat{\sigma}_z$. The time evolution of $\op{\rho}_t(D)$ is described by the QFPME~\eqref{eq:QFPME}, where the system is additionally weakly coupled to a bosonic reservoir with Bose-Einstein distribution $n_{\rm B}$. Its effect is described by the superoperator
\begin{equation} \label{eq:bath}
    \mc{L}_{\rm B} \op{\rho} = \kappa n_{\rm B} \mc{D}[\op{\sigma}^\dagger] \op{\rho} + \kappa (1+n_{\rm B}) \mc{D}[\op{\sigma}] \op{\rho}.
\end{equation}

The power extracted in this engine can be computed from Eq.~\eqref{eq:PowerQFPME} and reads (note that $\hat{H}_{\rm TD}$ is independent of $D$)
\begin{equation} \label{eq:PowerEng}
    P =  \erw{\partial_t \op{H}_{\rm lab}(D)}_{\rm lab} = -\omega g \erw{ D \op{\sigma}_x},
\end{equation}
where $\langle \bullet\rangle_{\rm lab}$ denotes an average in the lab frame, while the average without a subscript corresponds to the rotating frame. The measurement energy flow, which provides the energy source for this engine, is given by
\begin{equation} \label{eq:MEEng}
    \dot{E}_{\rm M} = - \lambda \omega \erw{ \op{\sigma}_z } ,
\end{equation}
and the heat current reads
\begin{equation} \label{eq:HeatEng}
    J =  - \frac{\omega}{2}\kappa\left[1+ (2n_B+1) \langle \op{\sigma}_z \rangle\right], 
\end{equation}
see App.~\ref{app:energetics} for details.

\begin{figure}
    \includegraphics[width=0.9\linewidth]{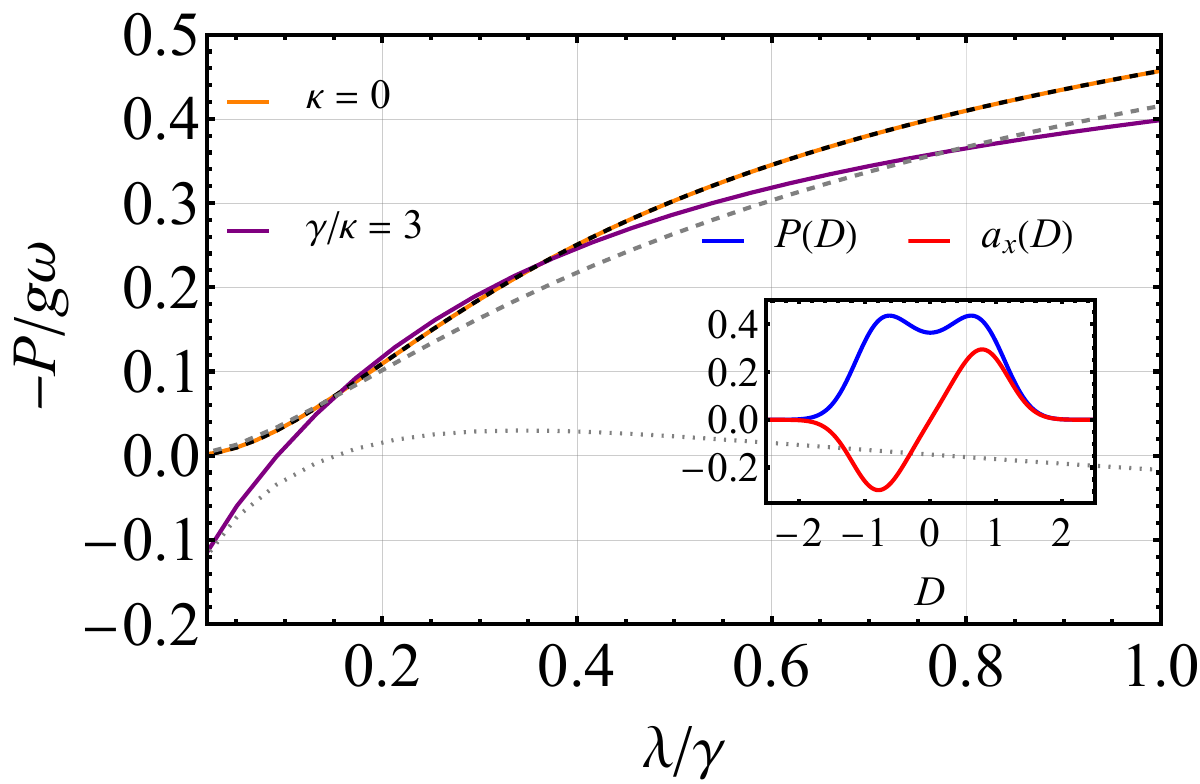}
	\caption{The first law of thermodynamics for the continuous-measurement-driven engine under the QFPME. The extracted power, $-P/(g \omega)$, obtained with Eq.~\eqref{eq:PowerEng} as a function of $\lambda/\gamma$ for no reservoir (orange) and for $\gamma/\kappa = 3$ (purple), where $\gamma/g = 1$. The dashed lines represent the measurement energy rates, $\dot{E}_{\rm M}/(g \omega)$, obtained with Eq.~\eqref{eq:MEEng}, for the system without (black dashed) and with the reservoir (gray dashed). The dotted gray line is the heat current $J/(g \omega)$ in Eq.~\eqref{eq:HeatEng} for $\kappa=\gamma/3$.  The inset shows the probability $P(D)$ and $a_x(D) = \tr{\op{\sigma}_x \op{\rho}(D)}$ as a function of $D$ for $\kappa=0$ with $\lambda = \gamma = g$. In all plots, $\omega = k_{\rm B}T$.
		\label{fig:FirstLawQuant}}
\end{figure}

The performance of this engine is illustrated in Fig.~\ref{fig:FirstLawQuant}.
Generally, the extracted power, $-P$, increases as the measurement strength $\lambda$ increases. This is a consequence of the fact that the measurement energy flow, the fuel of this engine, increases for stronger measurements. Indeed, in the absence of a reservoir ($\kappa=0$), the first law implies $-P =\dot{E}_{\rm M}$ in the steady state. The presence of an environment generally has a detrimental effect on the extracted power. For very weak measurements ($\lambda\ll\gamma$), no energy is provided by the measurement and power is dissipated as heat into the environment. For very strong measurements, the system is mostly close to the eigenstates of $\hat{\sigma}_x$, and thus $\langle \hat{\sigma}_z\rangle\simeq 0$ which results in a heat current of $J=-\omega\kappa/2$. We note however that $\dot{E}_\text{M}$ remains finite as it is proportional to $\lambda$. In this limit, the energy provided by the measurement fuels the engine, but a part of it is also dissipated for $\kappa\neq 0$. Finally, we note that there is a range in $\lambda/\gamma$ where heat is extracted from the environment. This is because the drive moves the system toward the ground state, making energy absorption from the environment more likely. This extracted heat partly results in enhanced work extraction, but it also reduces the energy provided by the measurement.

\section{Fluctuation theorems and the second law of thermodynamics}
\label{sec:2law}
In the second part of our work, we derive a novel FT and a second law of thermodynamics using the feedback-control formalism of the QFPME~\eqref{eq:QFPME}. After a brief introduction to FTs for feedback-controlled systems based on the existing literature, we introduce the FT derived from the QFPME for classical systems and contrast it to previous results. We then demonstrate how our results can be extended to the quantum regime.

Beginning with systems without feedback control, the FT, which was first introduced for classical systems~\cite{Seifert_2005, Jarzynski_1997a, Jarzynski_1997b, Crooks_1999, Crooks_2000} and later generalized to open quantum systems~\cite{Horowitz_2013, Manzano_2015, Manzano_2018, Manzano_2022, Campisi_2009}, is given by
\begin{equation} \label{eq:FT}
   \frac{P_{\rm B}[\bar{\mu}]}{P[\mu]} = e^{-\sigma[\mu]} \implies \erw{e^{-\sigma}} = 1,
\end{equation}
where the equality on the left is called the detailed FT, and the equality on the right, implied by the detailed FT, is the integral FT. The key element is the stochastic entropy production $\sigma[\mu]$ along a single trajectory $\mu$. In classical systems, we consider phase-space trajectories~\cite{Seifert_2005, Jarzynski_1997a, Jarzynski_1997b, Crooks_1999, Crooks_2000} and in open quantum systems quantum-jump trajectories~\cite{Horowitz_2012, Horowitz_2013, Manzano_2015, Manzano_2018, Manzano_2022, Breuer2005, Hekking2013} associated with the unraveling of a Lindblad master equation. Here, $\bar{\mu}$ is obtained by time-reversing $\mu$, and the subscript $"\textrm{B}"$ signifies a backward version of the experiment, where any time-dependent parameter is time reversed. The FT implies that the probability of the time-reversed trajectory $\bar{\mu}$ occurring is exponentially suppressed by the stochastic entropy production $\sigma[\mu]$. Importantly, the second law of thermodynamics, 
\begin{equation} \label{eq:SL}
    \langle \sigma \rangle = \sum_\mu P[\mu] \sigma[\mu] \geq 0,
\end{equation}
which states that the average entropy production cannot be negative, follows from the FT, which can, therefore, be seen as a generalization of the second law.

Under feedback control, Eqs.~\eqref{eq:FT} and~\eqref{eq:SL} no longer apply, and the average entropy production in the system can become negative. 
In the modern formulation of information thermodynamics~\cite{Potts2018, Sagawa2010, Sagawa2012, prech2023, Yada2022, Horowitz2010, Gong2016}, the stochastic entropy production $\sigma[\mu]$ is viewed on an equal footing to an information term $I[\mu]$, giving rise to modified versions of the FTs and the second law,
\begin{equation} \label{eq:FTSLmf}
        \frac{P_{\rm B}[\bar{\mu}]}{P[\mu]} = e^{-\sigma[\mu] - I[\mu]} \implies \erw{e^{-\sigma - I}} = 1 \implies \langle \sigma \rangle  \geq -\langle I \rangle,
\end{equation}
where now $\mu$ is a joint trajectory of the system and the measurement outcomes used for feedback, and the information term $I[\mu]$ arises from correlations between the trajectory of the system and the measurement outcomes.
These relations have been derived for both classical~\cite{Potts2018, Sagawa2010, Sagawa2012, Horowitz2010} and quantum systems~\cite{prech2023, Yada2022}, and different information terms are known: transfer entropy~\cite{Sagawa2012} or coarse-grained entropy~\cite{Potts2018} for classical systems and quantum-classical transfer entropy~\cite{Yada2022} or coarse-grained entropy~\cite{prech2023} for quantum systems. Below we present relations in the spirit of Eq.~\eqref{eq:FTSLmf} that we obtained for systems described by the QFPME.



\subsection{Classical systems}

\subsubsection{Fluctuation theorems and the second law} \label{sec:ftclas}
We start with presenting our results for a classical system, which may occupy a set of microstates (configurations) $\{a\}$.
Let us denote as $a_n$ the microstate of the system at the time $t_n = n \delta t$ for $n = 0, 1, ..., N$, where $N = \tau/\delta t$ and $\tau$ is the duration of the experiment. We continuously monitor the state of the system. In the absence of measurement and feedback control, the trajectory $\mu$ is given by the tuple $\mathbf{a} := (a_0, a_1, ... , a_N)$. It becomes continuous in the limit $\delta t \to 0$ for fixed $\tau$.
In the same way, we can define a trajectory of the detector's measurement outcomes $\mathbf{D}$, such that the trajectory $\mu = (\mathbf{a}, \mathbf{D})$ includes both the system states and the measurement outcomes. We assume that the state of the system $a$ is monitored, i.e. $\op{A} = \sum_a a \proj{a}$.

From the QFPME, we derived the novel detailed and integral FTs, 
\begin{equation} \label{eq: FT detailed}
    \frac{P[\bar{\mathbf{a}}, \bar{\mathbf{D}}]}{P[\mathbf{a}, \mathbf{D}]} = e^{-\sigma[\mathbf{a}, \mathbf{D}] - \sigma_{\rm m}[\mathbf{a}, \mathbf{D}]} \implies \erw{e^{-\sigma - \sigma_{\rm m}}} = 1,
\end{equation}
which form a counterpart to Eq.~\eqref{eq:FTSLmf}. Here, $\bar{\mathbf{a}}$ and $ \bar{\mathbf{D}}$ are time-revered versions of $\mathbf{a}$ and $\mathbf{D}$. The left-hand side of the detailed FT in Eq.~\eqref{eq: FT detailed} is the ratio of probabilities that the system and detector follow specific trajectories and their time-reversed versions.
The term $\sigma_{\rm m}$, which appears instead of the information term $I$ and originates from feedback control, is named by us ``measurement entropy" and is implicitly defined as
\begin{equation} \label{eq: meas ent def}
    e^{ - \sigma_{\rm m}[\mathbf{a}, \mathbf{D}]} := \frac{P_{\rm m}[\bar{\mathbf{D}}|\bar{\mathbf{a}}]}{P_{\rm m}[\mathbf{D}|\mathbf{a}]} ,
\end{equation}
where $P_{\rm m}[\mathbf{D}|\mathbf{a}]$ is the probability of obtaining the trajectory of measurement outcomes $\mathbf{D}$ when the system's trajectory is fixed to be $\mathbf{a}$. 
It is completely determined by an Ornstein-Uhlenbeck process with a time-dependent drift term given by $\mathbf{a}$, see App.~\ref{app:generalsetting}.
We find that the measurement entropy can be cast into a stochastic It\^o integral
\begin{equation} \label{eq: meas ent for}
        \sigma_{\rm m} = \frac{8 \lambda}{\gamma} \left( \int_0^\tau (a_t -D_t) dD_t \right) - \gamma \tau -  \ln \frac{P_{\rm ini}[D_\tau|a_\tau]}{P_{\rm ini}[D_0|a_0]}
\end{equation}
where $a_t$ is the state of the system at the time $t$, and $dD_t$ is given by the stochastic It\^o differential equation~\eqref{eq:dDito} but with $a_t$ replacing the expectation value of $\op{A}$:
\begin{equation} \label{eq: dD a}
    dD_t = \gamma(a_t - D_t)dt + \frac{\gamma}{2\sqrt{\lambda}} dW.
\end{equation}
The time-discrete representation of Eq.~\eqref{eq: meas ent for} is presented in Eq.~\eqref{eq: meas ent dis}. The last term in Eq.~\eqref{eq: meas ent for} is determined by the initial $P_{\rm ini}$ probability distribution of the measurement outcomes. 
The asymmetry between $P_{\rm m}[\mathbf{D}|\mathbf{a}]$ and $P_{\rm m}[\bar{\mathbf{D}}|\bar{\mathbf{a}}]$, which results in a non-vanishing measurement entropy $\sigma_{\rm m}$, arises from the detector's delay due to its finite bandwidth and is an essential feature of feedback control described by the QFPME formalism. The measurement entropy captures the fact that it is more likely to observe measurement outcomes that lag behind than measurement outcomes that anticipate the behavior of the system.  
Details about the general setting of feedback control based on the QFPME are presented in App.~\ref{app:generalsetting}, and Eqs.~\eqref{eq: FT detailed} and~\eqref{eq: meas ent for} are derived in App.~\eqref{app:derFTclas}.

Similarly to the second law in Eq.~\eqref{eq:FTSLmf}, from the FT in Eq.~\eqref{eq: FT detailed}, we obtain the second law of thermodynamics for QFPME-based feedback by applying Jensen's inequality:
\begin{equation} \label{eq: SL meas ent}
    \erw{\sigma} \geq - \erw{\sigma_{\rm m}}
\end{equation}
We find that the rate of change of the average of the measurement entropy can be expressed as
\begin{equation} \label{eq: meas ent avg}
    \partial_t \langle \sigma_{\rm m} \rangle =  8 \lambda  \left( \left\langle (D_t - a_t)^2 \right\rangle - \frac{\gamma}{8\lambda} \right). 
\end{equation}
The first term denotes the mean squared error of the measurement outcome. The second term is the mean squared error of the measurement outcome if the system remains in a fixed state $a$. In this case, the distribution of the measurement outcome tends to a Gaussian with mean $a$ and variance $\gamma/(8\lambda)$. The measurement entropy thus quantifies the excess mean squared error, which arises due to the dynamics of the system, which is only captured by the measurement outcome with a delay. Equation.~\eqref{eq: meas ent avg} is derived in App.~\ref{app: meas ent avg}.

\subsubsection{Comparison with other approaches to feedback control} \label{sec: differences clas}

We wish to highlight conceptual differences between the QFPME-based feedback and the formalism of Refs.~\cite{Potts2018, Sagawa2012, Horowitz2010} which result in the FTs discussed in the beginning of this section. In deriving these FTs, the main challenge is that measurement and feedback cannot be time-reversed, since measurement occurs before feedback. This requires the choice of a backward experiment that is not equal to the time-reversed forward experiment. As shown in Ref.~\cite{Potts2018}, different choices leading to different information quantifiers are possible.

For the QFPME, the feedback is instantaneous and therefore time-reversing the measurement outcome results in the time-reversed feedback protocol. Causality in the QFPME is not enforced by employing feedback after the measurement outcome is obtained, but rather by the finite reaction time of the detector, which ensures that the measurement outcome reacts to changes in the system with a finite delay. For this reason, an experiment described by the QFPME can be time-reversed and no auxiliary backward experiment needs to be introduced. 

We further note that previous works typically assume that the measurement is time-reversal symmetric \cite{Potts2018, Sagawa2012, Horowitz2010}, i.e., that $P_{\rm m}[\bar{\mathbf{D}}|\bar{\mathbf{a}}]=P_{\rm m}[\mathbf{D}|\mathbf{a}]$ such that the measurement entropy vanishes. For an example of how the measurement entropy may enter previous FTs in the presence of measurement and feedback, see the supplemental material of \cite{Potts2018}.


\subsubsection{Coarse graining}

Since the idea behind measurement and feedback protocols is to gather and use information about the system in the form of the trajectory of measurement outcomes, $\mathbf{D}$, and not the systems's trajectory, we may only have access to $\mathbf{D}$ without any knowledge of $\mathbf{a}$. Motivated by this scenario, we provide a FT for just the detector's trajectory using coarse graining.
Let us define the effective entropy production coarse-grained over $\mathbf{a}$, the inaccessible part of the trajectory: 
\begin{equation} \label{eq: ent cg}
    e^{-\Sigma[\mathbf{D}]} := \int d\mathbf{a} e^{-\sigma[\mathbf{a}, \mathbf{D}]} e^{-\sigma_{\rm m}[\mathbf{a}, \mathbf{D}]} P[\mathbf{a}|\mathbf{D}],
\end{equation}
where $P[\mathbf{a}|\mathbf{D}] = P[\mathbf{a}, \mathbf{D}] /P[\mathbf{D}] $ is the conditional probability, with $P[\mathbf{D}] = \int d\mathbf{a} P[\mathbf{a}, \mathbf{D}]$. Then, the coarse-grained counterpart to Eq.~\eqref{eq: FT detailed} holds,
\begin{equation} \label{eq: FT cg}
    P[\mathbf{D}] e^{-\Sigma[\mathbf{D}]} = P[\bar{ \mathbf{D} }] \implies  \langle e^{-\Sigma[\mathbf{D}]} \rangle =1,
\end{equation}
which depends only on $\mathbf{D}$.
For details, see App.~\ref{app:coarse-graining}.





\subsection{Example: two-level system under bang-bang control}
\label{sec:QuantumEngine}

\begin{figure}
    \includegraphics[width=0.9\linewidth]{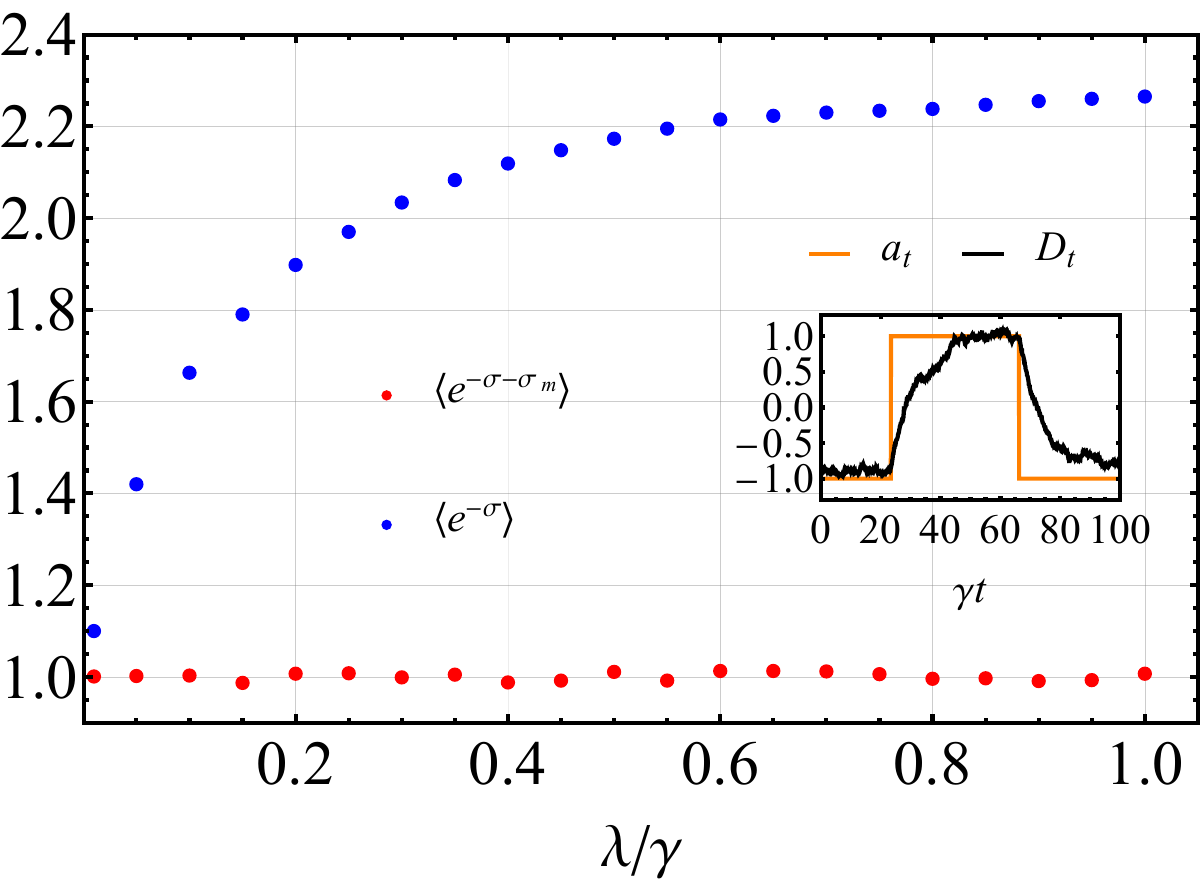}
	\caption{Integral FT for the two-level system under bang-bang control. Our FT from Eq.~\eqref{eq: FT detailed} (red dots) is compared to the standard integral FT (blue dots). The ensemble averages are obtained with Monte-Carlo simulations of trajectories of the system and detector, where the total duration $\tau = 10/\gamma$ is split into $10^3$ steps, and each point is obtained using $10^8$ trajectories. The stochastic measurement entropy $\sigma_{\rm m}$ is calculated with Eq.~\eqref{eq: meas ent for}, where the initial detector's distributions $P[D|a]$ are Gaussian distributions centered around $a$ and with variance $\gamma/(8 \lambda)$ [see Eq.~\eqref{eq: fast det lim}]. The stochastic entropy production reads $\sigma = -m \omega /(k_{\rm B} T)$, where $m$ is the net number of extracted energy quanta. The inset depicts a pair of sample trajectories of the system (orange) and detector (black) as a function of time with $\lambda/\gamma = 10$. In all plots, $\omega = k_{\rm B}T$ and $\gamma/\kappa = 10$.
		\label{fig:ftClas}}
\end{figure}

To illustrate our results, we investigate the classical two-level system under bang-bang control, which we have already used to illustrate the first law of thermodynamics (see Sec.~\ref{sec:2level} for details). 

\subsubsection{The FT and the second law}
We first show a sample trajectory of the system's state $a_t$ and the measurement outcome $D_t$ as a function of the time $t$, which is illustrated in the inset of Fig.~\ref{fig:ftClas}. The state $a_t$ jumps at random times, and $D_t$ drifts towards $a_t$ simultaneously exhibiting random Gaussian fluctuations. Deviations of $D_t$ from $a_t$ that go beyond the Gaussian fluctuations are responsible for a buildup of the measurement entropy $\erw{\sigma_{\rm m}}$ according to Eq.~\eqref{eq: meas ent avg}.

In Fig.~\ref{fig:ftClas}, we illustrate our integral FT in Eq.~\eqref{eq: FT detailed}, showing good agreement between the analytical results and numerical Monte-Carlo simulations. To obtain each data point, we numerically simulated a large number of trajectories and computed their corresponding stochastic measurement entropy $\sigma_{\rm m}$ [using Eq.~\eqref{eq: meas ent for}] and stochastic entropy production $\sigma$.
The entropy production depends entirely on the heat exchanged with the environment (equivalently extracted work), $\sigma = -m \omega /(k_{\rm B} T)$, where $m$ is the net number of energy quanta $\omega$ flowing from the environment into the system in one trajectory (equivalently, the net number of quanta extracted through feedback control). The boundary term that typically contributes to the entropy production vanishes because at all times the system is found in each state with probability $1/2$. The number $m$ is fully specified by $\mathbf{a}$ and $\mathbf{D}$ via the feedback protocol described in Sec.~\ref{sec:2level}.
In addition, Fig.~\ref{fig:ftClas} features the standard FT for systems without feedback control, $\erw{e^{-\sigma}}$, which now deviates from $1$.

\begin{figure}
\includegraphics[width=0.9\linewidth]{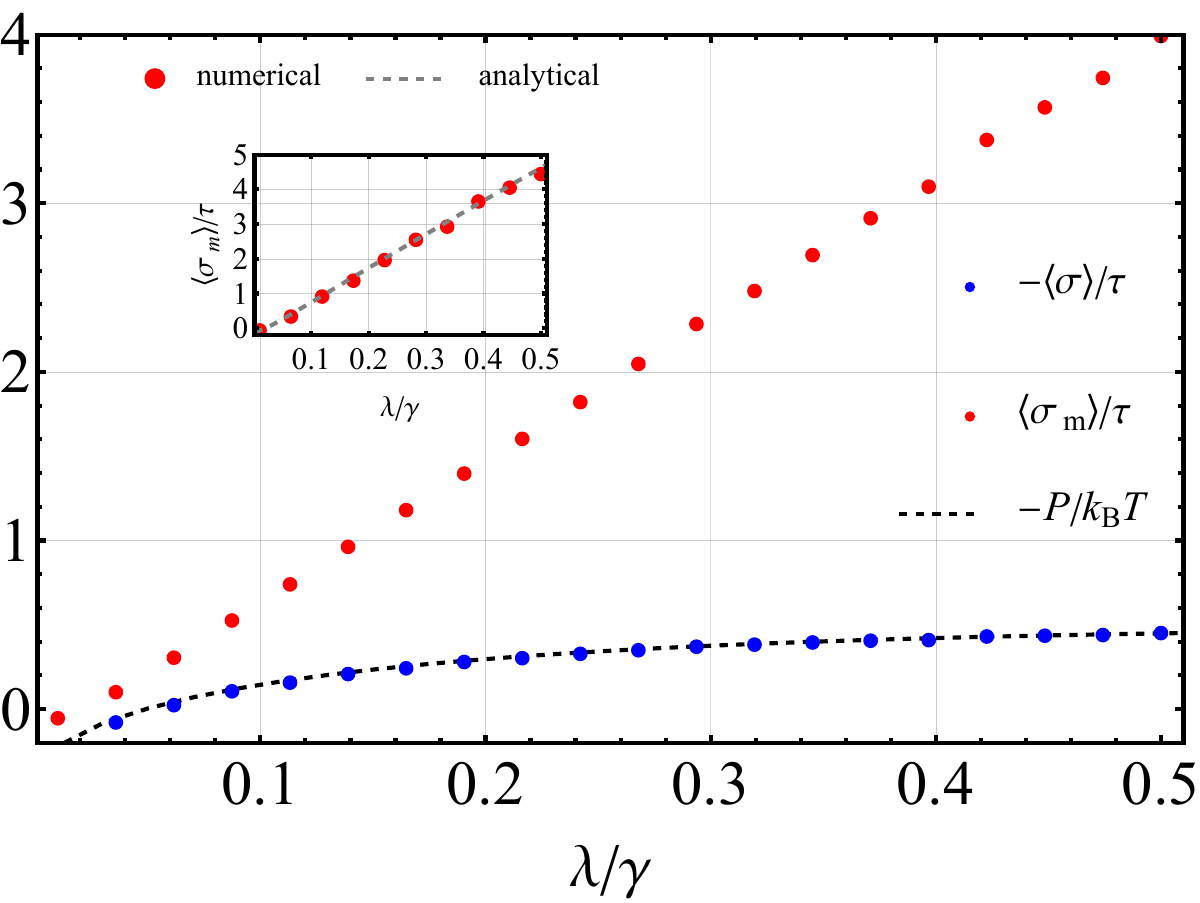}
	\caption{The second law of thermodynamics for the two-level system under bang-bang control. Both $\erw{\sigma_{\rm m}}/\tau$ (red dots) and $-\erw{\sigma}/\tau$ (blue dots) are computed with Monte-Carlo simulations using Eq.~\eqref{eq: meas ent for} and $\sigma = -m \omega/(k_{\rm B} T)$ for $\gamma/\kappa = 10$, where $\tau$ is the time duration and $m$ is the total number of extracted energy quanta. The black dashed line corresponds to the average entropy production obtained using the power $P$ in Eq.~\eqref{eq:Power2lvl}. The inset illustrates the agreement between $\erw{\sigma_{\rm m}}$ obtained numerically (red dots) with Eq.~\eqref{eq: meas ent for} and analytically with Eq.~\eqref{eq: meas ent fast} (gray dashed line) in the fast-detector limit ($ \gamma/\kappa = 100$). In all plots, $\omega = k_{\rm B} T $.
		\label{fig:slClas}}
\end{figure}

The corresponding second law of thermodynamics, Eq.~\eqref{eq: SL meas ent}, is illustrated in Fig.~\ref{fig:slClas}, where $\erw{\sigma}/\tau$ (blue dots) and $\erw{\sigma_{\rm m}}/\tau$ (red dots) are computed using the same method as for the FT above, with $\tau$ being the duration of the experiment. The average entropy production, determined by the power extracted from the system as a result of feedback control, is indeed bounded from above by the average measurement entropy. For large measurement strength, the second law becomes a lose bound. This is expected since in this case the measurement outcome shows a clear delay, which is not masked by noise. Observing the time-reversed becomes highly unlikely as this would correspond to the measurement outcome anticipating the behavior of the system.
In addition, in the plot we include the average entropy production rate obtained by computing $P/(k_{\rm B}T)$ from Eq.~\eqref{eq:Power2lvl}, which we have calculated in the first part of the paper dedicated to the first law of thermodynamics. There is a good agreement between these two methods, as evidenced by blue dots laying on top of the black dashed line.

\subsubsection{Fast-detector limit}

When the detector is much faster than the dynamics of the system ($\gamma \gg \kappa$),
the steady-state rate of change of the average measurement entropy $\erw{\sigma_{\rm m}}$ can be obtained analytically:
\begin{equation} \label{eq: meas ent fast}
    \partial_t \erw{\sigma_{\rm m}} =  \frac{8  \lambda \kappa }{\gamma} \left[ 1 + 2n_{\rm B} - \text{Erf}(2 \sqrt{\lambda/\gamma}) - \frac{e^{-4\lambda/\gamma}}{2 \sqrt{\pi \lambda/\gamma} }  \right],
\end{equation}
where $\text{Erf}(\bullet)$ is the error function; see App.~\ref{app:fast_det} for a derivation.
In order to demonstrate this approximation, in the inset of Fig.~\ref{fig:slClas}
we show $\erw{\sigma_{\rm m}}$ obtained with Eq.~\eqref{eq: meas ent fast} and with a numerical simulation using Eq.~\eqref{eq: meas ent for} for a system with $\gamma/\kappa = 100$.



Under the fast-detector approximation, the detailed fluctuation theorem for the net number of extracted quanta in the long-time limit, $m$, reads
\begin{equation} \label{eq:FTm}
    \frac{P[-m]}{P[m]} = e^{m\left[\omega/(k_{\rm B}T) - \log{\left(\frac{1-\eta}{\eta} \right)} \right]},
\end{equation}
where $P[m]$ is the probability of $m$ net extracted quanta in a trajectory, $\log{\left(\frac{1-\eta}{\eta} \right)}$ is an information term due to feedback control, and $\eta = \left[1 - \text{Erf} \left(2 \sqrt{\lambda/\gamma} \right) \right]/2$ is the probability of erroneously applying feedback [see Eq.~\eqref{eq:eta}]. This result was obtained with the QFPME in Ref.~\cite{Annby2022} following the approach of Ref.~\cite{Esposito2012}. We show that Eq.~\eqref{eq:FTm} can be likewise derived using our FT in Eq.~\eqref{eq: FT detailed} with the aid of coarse-graining, which is demonstrated in App.~\ref{app:FTm}.


\subsection{Quantum systems}

\subsubsection{Fluctuation theorems} \label{sec:ftquant}

Having presented the FTs in Eq.~\eqref{eq: FT detailed} and the second law in Eq.~\eqref{eq: SL meas ent} for classical systems, we now show their counterparts for quantum dynamics. For open quantum systems described the Lindblad master equation
\begin{equation}
    \partial_t \op{\rho}_t = -i[\op{H}_t, \op{\rho}_t] + \sum_k \mc{D}[\op{L}_k] \op{\rho}_t,
\end{equation}
where $\op{L}_k$ are Lindblad jump operators, a trajectory of the system is typically defined via the so-called quantum-jump unraveling of the master equation~\cite{Manzano_2015, Manzano_2018, Manzano_2022, Horowitz_2013}, which enters the probabilities in the FT without feedback control [see Eq.~\eqref{eq:FT}]. It includes outcomes of the projective measurements onto the eigenstates of the initial and final states of the quantum systems as well as a list of jumps, $ ( k_1, k_2, ..., k_M )$, together with their corresponding times, $ (s_1, s_2, ..., s_M )$, which we collectively denote by $\Gamma$.

The quantum system is continuously measured with the observable $\op{A} = \sum_a a \proj{a}$~\citep{Jacobs2006rev, Bednorz2012}, and feedback at the time $t$ is implemented with the measurement outcome $D_t$, just as we have considered for the classical system. However, the difference in the quantum regime is that the system may be in a superposition of eigenstates of the observable $\op{A}$. To define a trajectory for the quantum system, we combine the above-introduced unraveling with trajectories along a closed time contour~\cite{Hofer2016}. See App.~\ref{app: quantum FT and SL} for details. In this method, identity operators $\op{I} = \sum_a \proj{a}$ are inserted at every step $\delta t$ in the forward-in-time and backwards-in-time branches of the closed time contour. Upon performing the Keldysh rotation, the trajectory can be phrased in terms of a classical trajectory, $\mathbf{a}_{\rm c} : = ( a_0^{\rm c}, a_1^{\rm c}, ..., a_N^{\rm c} ) $,  related to diagonal elements of the density matrix, and a quantum trajectory $\mathbf{a}_{\rm q} : = ( a_0^{\rm q}, a_1^{\rm q}, ..., a_N^{\rm q} ) $, associated to superposition states. The quasi-probability distribution of the trajectory, is then expressed as $P[\Gamma, \mathbf{a}_{\rm c}, \mathbf{a}_{\rm q}, \bf{D} ]$. Importantly, it does not have to be positive and can be even complex. However, it is still normalized.

With this formulation, we find the novel FTs for open quantum systems under QFPME-base feedback control:
\begin{equation} \label{eq:FT quant}
    \frac{ P[\bar{\Gamma}, \bar{\bf{a}}_{\rm c}, \bar{\bf{D}} ] }{P[\Gamma, \bf{a}_{\rm c}, \bf{D}]} = e^{-\sigma[\Gamma, \bf{D}] - \sigma_{\rm m}[\mathbf{a}_{\rm c}, \bf{D}] } \implies \left\langle e^{-\sigma - \sigma_{\rm m}} \right\rangle = 1,
\end{equation}
Here, crucially, the stochastic entropy production $\sigma[\Gamma, \bf{D}]$ depends on the unraveling trajectory $\Gamma$, which encompasses all quantum jumps, and the measurement entropy $\sigma_{\rm m}[\bf{a}_{\rm c}, \bf{D}]$ depends on the classical trajectory $\bf{a}_{\rm c}$ but not the quantum trajectory. This allows us to integrate out the quantum trajectory and obtain the quasi-probability distribution $P[\Gamma, \mathbf{a}_{\rm c}, \mathbf{D} ] = \int d \mathbf{a}_{\rm q} P[\Gamma, \mathbf{a}_{\rm c}, \mathbf{a}_{\rm q}, \mathbf{D} ]$, which is real valued but may take on negative values. In the FT, the meaning of $\bar{\Gamma}, \bar{\bf{a}}_{\rm c}$, and $ \bar{\bf{D}} $ is the time-reversed versions of $\bar{\Gamma}, \bar{\bf{a}}_{\rm c}$, and $ \bar{\bf{D}} $, respectively. In the quantum case, we find the expression for the stochastic measurement entropy, 
\begin{equation} \label{eq: meas ent for quant}
        \sigma_{\rm m} = \frac{8 \lambda}{\gamma} \left( \int_0^\tau (a_t^{\rm c} -D_t) dD_t \right) - \gamma \tau -  \ln \frac{P_{\rm ini}[D_\tau]}{P_{\rm ini}[D_0]},
\end{equation}
where the classical Keldysh path $\bf{a}_{\rm c}$ replaces the system's trajectory $\bf{a}$ [cf. Eq.~\eqref{eq: meas ent for}]. See App.~\ref{app: quantum FT and SL} for derivations.

In Sec.~\ref{sec: differences clas}, we discussed conceptual differences between our FT for QFPME-based feedback control and the approaches of Refs.~\cite{Potts2018, Sagawa2012, Horowitz2010} for classical systems. We note that the latter have been generalized to open quantum systems in Refs.~\cite{prech2023, Yada2022}. The same differences are present between our results in the quantum regime and Refs.~\cite{prech2023, Yada2022}.

\subsection{Coarse-graining and the second law}
Since the quasi-probability distribution $P[\Gamma, \bf{a}_{\rm c}, \bf{D}]$ does not need to be positive, the second law does not follow from the integral FT in Eq.~\eqref{eq:FT quant} anymore.
However, $P[\Gamma, \mathbf{D}] = \int d \mathbf{a}_{\rm c}P[\Gamma, \mathbf{a}_{\rm c}, \mathbf{D}]$ represents a valid probability distribution. Therefore, by defining the measurement entropy coarse-grained over the inaccessible classical trajectory $\mathbf{a}_{\rm c}$, 
\begin{equation}
    e^{-\sigma_{\rm m, cg}[\mathbf{D}]} := \int d\mathbf{a}_{\rm c}~ e^{ - \sigma_{\rm m}[\mathbf{a}_{\rm c}, \bf{D}] } P[\mathbf{a}_{\rm c}|\Gamma, \mathbf{D}],
\end{equation}
where $ P[\mathbf{a}_{\rm c}|\Gamma, \mathbf{D}] =  P[\Gamma, \mathbf{a}_{\rm c}, \mathbf{D}]/ P[\Gamma, \mathbf{D}]$,
we obtain the FT
\begin{equation} \label{eq:FT quant cg}
    \frac{ P[\bar{\Gamma}, \bar{\bf{D}} ] }{P[\Gamma, \bf{D}]} = e^{-\sigma[\Gamma, \bf{D}] - \sigma_{\rm m, cg}[\mathbf{D}] } \implies \left\langle e^{-\sigma - \sigma_{\rm m, cg}} \right\rangle = 1.
\end{equation}
Both $\Gamma$ and $\mathbf{D}$ are experimentally accessible, and the probabilities on the left-hand side are nonnegative and normalized, which makes this relation the quantum analogue for the classical result [see Eq.~\eqref{eq: FT detailed}]. Importantly, the coarse-grained measurement entropy $\sigma_{\rm m, cg}[\mathbf{D}]$ depends solely on the trajectory of the detector. The second law follows from the right-hand side of Eq.~\eqref{eq:FT quant cg}:
\begin{equation} \label{eq:SL quant cg}
    \erw{\sigma} \geq - \erw{ \sigma_{\rm m, cg} } .
\end{equation}

\subsection{Example 2: continuous-measurement-driven engine}

\begin{figure}
    \centering
\includegraphics[width=0.99\linewidth]{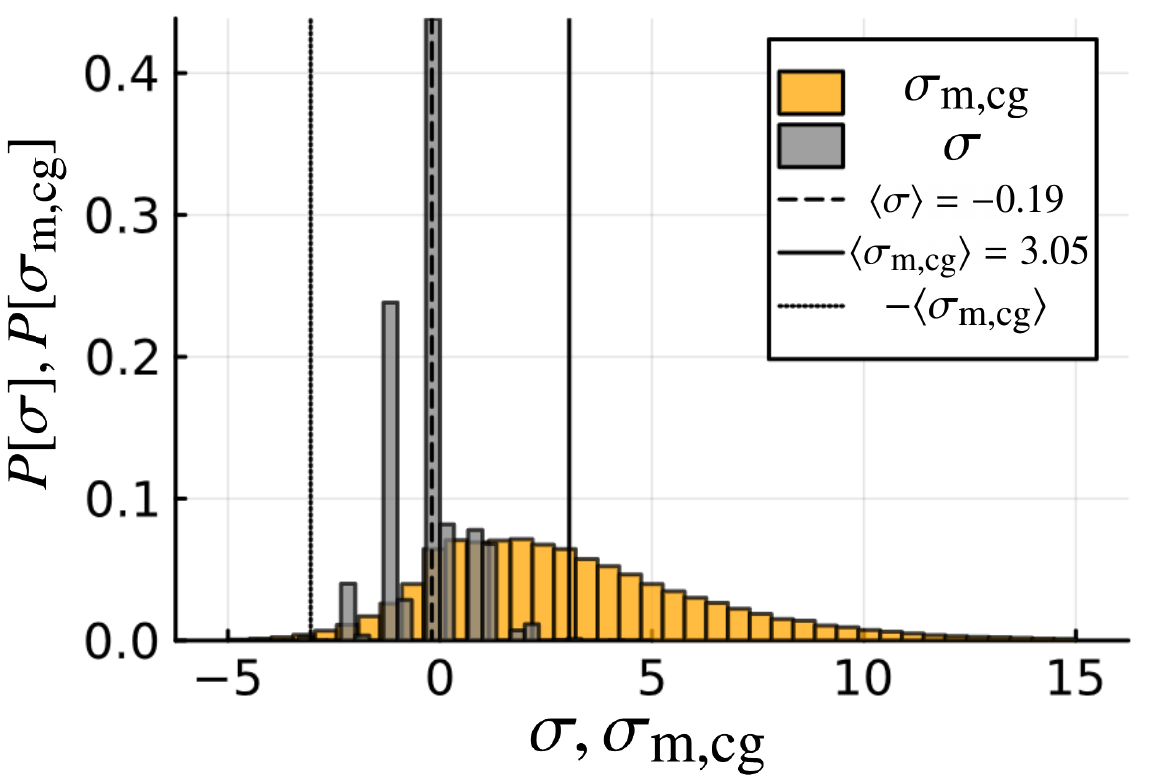}
    \caption{The FT and the second law for the continuous-measurement-driven engine. Gray bars represent a normalized histogram of $\sigma$, which are obtained with Eq.~\eqref{eq:quantent} by numerically sampling trajectories with Monte-Carlo simulations. The black dashed vertical line corresponds to the average $\langle \sigma \rangle$. For each sampled trajectory, we compute the corresponding $\sigma_{\rm m, cg}$ using the detailed FT in Eq.~\eqref{eq:FT quant cg}. Yellows bars show a normalized histogram of $\sigma_{\rm m, cg}$, and the corresponding $\langle \sigma_{\rm m, cg} \rangle$ is marked with the black vertical line. The average $-\langle \sigma_{\rm m, cg} \rangle$ (black dotted vertical line) is positioned on the left from $\langle \sigma \rangle$, which is consistent with the second law in Eq.~\eqref{eq:SL quant cg}. Parameters: $\gamma = 10 \kappa = 5 g = 5 \lambda = \omega$ and $\omega = k_{\rm B}T$. The total duration $\tau = 10/\gamma$ is split into $10^3$ steps, and $10^6$ trajectories were sampled.
    }
    \label{fig:histoq}
\end{figure}

To illustrate the quantity $\sigma_{\rm m, cg}[\mathbf{D}]$ for quantum systems, we resort to the continuous-measurement-driven engine, which was introduced in Sec.~\ref{sec:conteng}, and we apply the detailed FT in Eq.~\eqref{eq:FT quant cg}. Using Monte-Carlo simulations, we numerically sample values of the stochastic entropy production $\sigma$, whose distribution is illustrated in Fig.~\ref{fig:histoq} as a normalized histogram (grey bars). The initial state of the system is given by the thermal state with respect to the thermodynamic Hamiltonian $\frac{\omega}{2} \op{\sigma}_z$. The corresponding expectation value $\langle \sigma \rangle$, which is negative, is marked with a vertical black dashed line.
The stochastic quantity $\sigma$ is computed with Eq.~\eqref{eq:quantent}, where the stochastic entropy production corresponding the lowering ($\op{\sigma}$) and raising ($\op{\sigma}^\dagger$) operators are given by $\omega/(k_{\rm B}T)$ and $-\omega/(k_{\rm B}T)$, respectively. For each numerically sampled trajectory $(\Gamma, \mathbf{D})$, we compute the corresponding probabilities $P[\Gamma, \mathbf{D}]$ and $P[\bar{\Gamma}, \bar{\mathbf{D}}]$, which we use to obtain the associated $\sigma_{\rm m, cg}[\mathbf{D}]$ using the detailed FT in Eq.~\eqref{eq:FT quant cg}. Yellow bars in Fig.~\ref{fig:histoq} show a normalized histogram of $\sigma_{\rm m, cg}$, and the expectation value $\langle \sigma_{\rm m, cg} \rangle$ is marked with a vertical black line. The corresponding $-\langle \sigma_{\rm m, cg} \rangle$ (vertical black dotted line) lies on the left from $\langle \sigma \rangle$ (vertical black dashed line), which illustrates the second law in Eq.~\eqref{eq:SL quant cg}.

\section{Discussion and conclusions}
\label{sec:discussion}


Definitions of work and heat have been well studied in classical and quantum systems without feedback control, but such a formulation has been lacking for continuously measured systems under arbitrary (linear but also nonlinear) feedback protocols. Relying on the QFPME, which describes the time evolution of the joint state of a system and a detector with finite bandwidth, we have derived expressions for the power, heat current, and measurement energy rate with the aid of methods from stochastic calculus. We have applied these definitions to investigate the performance of two different work-extraction toy models: a two-level system under bang-bang control, which is an example of a class of threshold-like feedback protocols, and a continuous-measurement-driven engine, where the quantum measurement provides a source of energy by backaction on the quantum state of the qubit. Since measurement and feedback constitutes a powerful tool to control nanoscale systems, we expect that the first law of thermodynamics presented here will be useful in future investigations of such setups.

The second part of this manuscript is concerned with the second law and FTs for systems that follow the QFPME. While in the existing literature, FTs have been extended to feedback controlled systems in classical and quantum regimes by including information terms that quantify correlations between measurement outcomes and trajectories followed by the system, a common assumption in these studies is a symmetry between forward and backward experiments in the conditional probability of measurement outcomes ($P_{\rm m}[\mathbf{D}|\mathbf{a}]$). As evidenced, this assumption is naturally broken in systems described with the QFPME-formalism, which is a consequence of the finite bandwidth of the detector, a relevant characteristic of experiments. Using the QFPME-framework (as well as Keldysh quasiprobabilities in the quantum regime), we derive the corresponding FTs, where the information term is replaced with the measurement entropy that quantifies the asymmetry in distributions of measurement outcomes. 
Our results are illustrated on both previously used toy models and they shed insight into the role of the measurement in the irreversibility of feedback experiments.

We anticipate the presented framework to be applied to a variety of classical and quantum devices to assess their thermodynamic performance and obtain insights into their limitations. In particular, we unlock this possibility for models where nonlinear protocols may be leveraged to optimize cooling or any other goal. Furthermore, extending the presented framework to different types of detector filter (e.g., a high-pass frequency filter) constitutes an intriguing avenue to be explored.



\appendix

\begin{widetext}

\section{The first law of thermodynamics~\eqref{eq:PowerHeatQFPME}}
\label{app:stochastic}

\subsection{Derivation of $d\op{U}$ in Eq.~\eqref{eq:dHito}} \label{app:stochastic1}
We first show how to obtain $d\op{U}$ in Eq.~\eqref{eq:dHito} using $dD_{\rm c}$ in Eq.~\eqref{eq:dDito} and It\^o's lemma in stochastic calculus. The differential of the Hamiltonian $\op{U} = \op{U}_t(D)$ is given by
\begin{equation}
\label{eq:dH}
    d\op{U} = \partial_t \op{U} dt +  \partial_D \op{U} dD + \frac{1}{2} \partial_t^2 \op{U} (dt)^2 + \frac{1}{2} \partial_D^2 \op{U} (dD)^2 + ... \quad ,
\end{equation}
where $\op{U} = \op{U}_t(D)$, $\partial_D \op{U} = \partial_D \op{U}_t(D)$, and $\partial_D^2 \op{U} = \partial_D^2 \op{U}_t(D)$ ought to be viewed as operators. In particular, they take the random process $D_{\rm c}$ as a value of $D$. Inserting $dD_{\rm c}$ from Eq.~\eqref{eq:dDito} into Eq.~\eqref{eq:dH} for $dD$, results in an expression for $d\op{U} = d\op{U}_t(D_{\rm c})$ involving expressions of the form $dt^n dW^m$ for integers $n \geq 0$ and $m \geq 0$. Following It\^o's rules, in the limit $dt \to 0$, the only remaining terms are proportional to $dt$, $dW$, and $dW^2=dt$.  
This yields the expression
\begin{equation}
    d\op{U} = \left[ \partial_t \op{U} + \gamma\left( \langle\op{A}\rangle_{\rm c} -D \right)\partial_D \op{U} + \frac{\gamma^2}{8 \lambda} \partial^2_D \op{U} \right]dt + \frac{\gamma}{2\sqrt{\lambda}} \partial_D \op{U} dW,
\end{equation}
reproducing Eq.~\eqref{eq:dHito}.

\subsection{Calculation of $\ev{\tr{d\op{U} \op{\rho}_{\rm c}}}$, $\ev{\tr{\op{U} d\op{\rho}_{\rm c}}}$, and $\ev{\tr{d\op{U} d\op{\rho}_{\rm c}}}$}
\label{app:stochastic2}

Next, we derive the expectation values $\ev{\tr{d\op{U} \op{\rho}_{\rm c}}}$, $\ev{\tr{\op{U} d\op{\rho}_{\rm c}}}$, and $\ev{\tr{d\op{U} d\op{\rho}_{\rm c}}}$, which appear in Eq.~\eqref{eq:EnergyQFPME}.
To find $\ev{\tr{\op{U} d\op{\rho}_{\rm c}}}$, we substitute the Belavkin equation~\eqref{eq:Belavkin} for $d\op{\rho}_{\rm c}$:
\begin{equation}
\begin{split} \label{eq:Hdrho}
    \ev{\tr{\hat{U}d\hat{\rho}_{\rm c}}} & = \ev{ \tr{\hat{U}\mathcal{L}(D_{\rm c})\hat{\rho}_{\rm c} dt+\hat{U}\lambda\mathcal{D}[\hat{A}]\hat{\rho}_{\rm c} dt+\hat{U}\sqrt{\lambda}\{\hat{A}-\langle \hat{A}\rangle_{\rm c},\hat{\rho}_{\rm c}\}dW } } \\
    &=  \ev{ \tr{-i\op{U}[\hat{H}, \hat{\rho}_{\rm c}] + \hat{U}\mathcal{L}_{\rm B}\hat{\rho}_{\rm c}  +\hat{U}\lambda\mathcal{D}[\hat{A}]\hat{\rho}_{\rm c}  }}dt\\
    & =  \ev{ \tr{ i \op{\rho}_{\rm c} [\op{H}, \op{U}] }}dt + \ev{ \tr{\hat{\rho}_{\rm c} \mathcal{L}_{\rm B}^\dagger\hat{U}  +\lambda \hat{\rho}_{\rm c} \mc{D}[\op{A}]\hat{U}   }}dt. \\
\end{split}
\end{equation}
To obtain the second equality, we used $\ev{dW} = 0$ and inserted the Liouvillian $\mc{L}(D_{\rm c})$ in Eq.~\eqref{eq:ME} with $\op{H} = \op{H}_t(D_{\rm c})$. In the third equality, we used the cyclic property of the trace.
From Eq.~\eqref{eq:Hdrho}, 
using $\ev{\op{O}(D_{\rm c}) \op{\rho}_{\rm c}} = \int dD \op{O}(D) \op{\rho}_t(D)$, we find
\begin{equation} \label{eq:HdrhoRate}
\begin{split}
    \frac{\ev{\tr{\hat{U} d\hat{\rho}_{\rm c}}}}{dt} &  = \int dD   \tr{\hat{\rho}_t(D) \left( i [\op{H}, \op{U}] + \mathcal{L}_{\rm B}^\dagger\hat{H}  +\lambda  \mc{D}[\op{A}]\hat{H} \right)  }  = \erw{i  [\op{H}, \op{U}]} + \erw{\mathcal{L}_{\rm B}^\dagger\hat{U}} + \lambda \erw{\mc{D}[\op{A}]\hat{U}}. \\
\end{split}
\end{equation}
The first term on the right-hand side vanishes when $\op{U} = \op{H}_t(D)$ and is equal to $\erw{\partial_t \op{H}}$ when $\op{U} = \op{H}_{\rm TD}(D)$ [see Eq.~\eqref{eq:HTDtime}]. The second and third term terms are the heat current $J$ in Eq.~\eqref{eq:HeatQFPME} and the measurement energy $\dot{E}_{\rm M}$ in Eq.~\eqref{eq:EmQFPME}, respectively.
To find $\ev{\tr{d\op{U} \op{\rho}_{\rm c}}}$, we substitute Eq.~\eqref{eq:dHito} for $d\op{U}$:
\begin{equation}
\begin{split}
    \ev{\tr{ d\op{U} \op{\rho}_{\rm c} }} & = \ev{ \tr{ \op{\rho}_{\rm c} ( \partial_t \op{U} + \gamma(\langle\op{A}\rangle_{\rm c} -D )\partial_D \op{U} + \frac{\gamma^2}{8 \lambda} \partial^2_D \op{U} )dt + \frac{\gamma}{2\sqrt{\lambda}} \op{\rho}_{\rm c} \partial_D \op{U} dW }} \\
    &= \ev{ \tr{ \op{\rho}_{\rm c} ( \partial_t \op{U} - \gamma D \partial_D \op{U} + \frac{\gamma^2}{8 \lambda} \partial^2_D \op{U} )}}dt + \gamma \ev{\tr{ \langle\op{A}\rangle_{\rm c} \op{\rho}_{\rm c}  \partial_D \op{U} }}dt,\\
\end{split}
\end{equation}
where we used $\ev{dW} =0$ to show the second equality. Therefore, we have
\begin{equation} \label{eq:rhodHRate}
    \frac{\ev{\tr{ d\op{U} \op{\rho}_{\rm c} }}}{dt} = \erw{\partial_t \op{U}} + \frac{\gamma^2}{8\lambda} \erw{\partial_D^2 \op{U}} - \gamma \erw{D \partial_D \op{U}}  + \gamma \ev{\tr{\langle\op{A}\rangle_{\rm c} \op{\rho}_{\rm c}  \partial_D \op{U} }}.
\end{equation}
To compute $\ev{\tr{d\op{U} d\op{\rho}_{\rm c}}}$, we again use Eqs.~\eqref{eq:dHito} and~\eqref{eq:Belavkin} for $d\op{U}$ and $d\op{\rho}_{\rm c}$, respectively. Multiplication of these two terms results in
\begin{equation}
\begin{split}
    d\op{U} d\op{\rho}_{\rm c} &= \left[ \partial_t \op{U} + \gamma \left( \langle\op{A}\rangle_{\rm c} -D \right)\partial_D \op{U} + \frac{\gamma^2}{8 \lambda} \partial^2_D \op{U} \right] \left[ \mathcal{L}(D_{\rm c})\hat{\rho}_{\rm c} +\lambda\mathcal{D}[A]\hat{\rho}_{\rm c} \right]dt^2 + \left( \frac{\gamma}{2\sqrt{\lambda}} \partial_D \op{U} \right) \left[ \mathcal{L}(D_{\rm c})\hat{\rho}_{\rm c} +\lambda\mathcal{D}[A]\hat{\rho}_{\rm c} \right] dtdW \\
    & + \left[ \partial_t \op{U} + \gamma \left( \langle\op{A}\rangle_{\rm c} -D \right)\partial_D \op{U} + \frac{\gamma^2}{8 \lambda} \partial^2_D \op{U} \right] \left( \sqrt{\lambda}\{\hat{A}-\langle\hat{A}\rangle_{\rm c},\hat{\rho}_{\rm c} \}  \right) dt dW +  \left( \frac{\gamma}{2\sqrt{\lambda}} \partial_D \op{U} \right)  \left( \sqrt{\lambda}\{\hat{A}-\langle\hat{A}\rangle_{\rm c},\hat{\rho}_{\rm c} \}  \right) dW^2.
\end{split}
\end{equation}
We are concerned with the expectation value in the limit  $dt\to 0$. Following the It\^o's rules, $dt^2$ and $dtdW$ approach $0$, and $dW^2$ approaches $dt$, which results in
\begin{equation} \label{eq:dHdrhoRate}
    \frac{\ev{\tr{d\op{U} d\op{\rho}_{\rm c}}}}{dt} = \frac{\gamma \sqrt{\lambda}}{2 \sqrt{\lambda}} \ev{ \tr{ \{\hat{A}-\langle\hat{A}\rangle_{\rm c},\hat{\rho}_{\rm c}\} \partial_D\hat{H} }} = \frac{\gamma}{2} \ev{ \tr{ \op{\rho}_{\rm c} \{\op{A}, \partial_D \op{U} \} - 2\langle\op{A}\rangle_{\rm c} \op{\rho}_{\rm c}  \partial_D \op{U} }} = \frac{\gamma}{2} \erw{\{\op{A}, \partial_D \op{U} \}} - \gamma \ev{ \tr{ \langle\op{A}\rangle_{\rm c} \op{\rho}_{\rm c}  \partial_D \op{U} }},
\end{equation}
where we used the cyclic property of the trace to show the second equality.
By combining Eq.~\eqref{eq:rhodHRate} with Eq.~\eqref{eq:dHdrhoRate} and noticing that the last terms on the right-hand sides of these equations cancel each other, we obtain
\begin{equation} \label{eq:Power App}
\begin{split}
    \frac{\ev{\tr{d\op{U} \op{\rho}_{\rm c}}}}{dt} + \frac{\ev{\tr{d\op{U} d\op{\rho}_{\rm c}}}}{dt} &=   \frac{\gamma^2}{8\lambda} \erw{\partial_D^2 \op{U}} - \gamma \erw{D \partial_D \op{U}}  + \gamma \ev{\tr{\langle\op{A}\rangle_{\rm c} \op{\rho}_{\rm c}  \partial_D \op{U} }} + \frac{\gamma}{2} \erw{\{\op{A}, \partial_D \op{U} \}} - \gamma \ev{ \tr{ \langle\op{A}\rangle_{\rm c} \op{\rho}_{\rm c}  \partial_D \op{U} }} + \erw{\partial_t \op{U}}\\
    & =  \gamma \erw{\mc{A}(D)\partial_D \op{U}} + \frac{\gamma^2}{8\lambda} \erw{\partial_D^2 \op{U}} + \erw{\partial_t \op{U}},
\end{split}
\end{equation}
where we used $\{\op{A}, \partial_D \op{H} \}/2 - D \partial_D \op{H} = \{\op{A} - D, \partial_D \op{H} \}/2 = \mc{A}(D) \partial_D \op{H}$. The first two terms are exactly the drift and diffusion terms contributing to power $P$ in Eq.~\eqref{eq:PowerQFPME}. When $\op{U} = \op{H}_t(D)$, the last term describes changes in the Hamiltonian due to direct time-dependent driving, whereas when $\op{U} = \op{H}_{\rm TD}(D)$, which is not explicitly time dependent, the last term vanishes.

\section{Classical two-level system under bang-bang control}
In this section, we give details about the classical two-level system under bang-bang control in Sec.~\ref{sec:2level}.

\subsection{Power and heat} \label{sec:energeticsClas}
We first show the expression for power in Eq.~\eqref{eq:Power2lvl}. The first derivative of the Hamiltonian~\eqref{eq:H2} with respect to $D$ is given by
\begin{equation}
    \partial_D \op{H} = \partial_D \left( \theta(D) \omega \proj{0} + (1-\theta(D)) \omega \proj{1} \right) = \omega \delta(D) \left( \proj{0} - \proj{1} \right) 
\end{equation}
where we have used $\partial_D \theta(D) = \delta(D)$. Here, $\delta(D)$ is the Dirac-delta distribution, and recall that $\theta(D)$ is the Heavyside step function.
For the drift term in the expression for power~\eqref{eq:PowerQFPME} in the QFPME, we find
\begin{equation}
    \erw{\mc{A}(D) \partial_D \op{H}} = \erw{ \left( \proj{1} - \proj{0} -D \right) \omega \delta(D) \left( \proj{0} - \proj{1} \right) } = \omega \erw{ \delta(D)  \op{I} } = -\omega \int dD \delta(D)\tr{\rho_t(D)} = -\omega \tr{\rho_t(0)} = -\omega P_t(0), 
\end{equation}
where we inserted $\op{A} = \proj{1} - \proj{0}$ and used the identity $D \delta(D) = 0$ to obtain the second equality.
The diffusion term in Eq.~\eqref{eq:PowerQFPME} yields
\begin{equation}
    \erw{\partial_D^2 \op{H}} = \omega \erw{ \partial_D \delta(D) \left( \proj{0} - \proj{1} \right) } = - \omega \left( \partial_D \bra{0} \op{
    \rho}_t(D)
    \ket{0}|_{D = 0} - \partial_D \bra{1} \op{
    \rho}_t(D)
    \ket{1}|_{D = 0}  \right) =  - \omega \left[ \partial_D p_0(D) |_{D = 0} - \partial_D p_1(D) |_{D = 0}  \right] ,
\end{equation}
where we have used the relation $\int_{-\infty}^\infty dD f(D) \partial_D \delta(D) = - \int_{-\infty}^\infty dD \delta(D) \partial_D f(D) $ for a distribution $f(\cdot)$. Here, we have also introduced $p_0(D) = \bra{0} \op{\rho}_t(D) \ket{0}$ and $p_1(D) = \bra{1} \op{\rho}_t(D) \ket{1}$ for brevity of notation. Inserting the above drift and diffusion terms into Eq.~\eqref{eq:PowerQFPME} results in the power in Eq.~\eqref{eq:Power2lvl}.

As far as the heat current is concerned, we start from Eq.~\eqref{eq:HeatQFPME}. The system absorbs and ejects energy quanta $\omega$ with rates $\kappa n_{\rm B}$ and $\kappa (1+n_{\rm B})$. When $D \geq 0$, $\ket{0}$ and $\ket{1}$ are the excited and the ground state, respectively, and the contribution to the heat current is given by $\omega\int_{0}^\infty dD \left(\kappa n_{\rm B} p_1(D) -\kappa (1+n_{\rm B}) p_0(D)\right)$. When $D<0$, the expression is analogous, but the states are switched. Therefore, the expression for heat current reads
\begin{equation}
\begin{split}
    J &= \omega\int_{0}^\infty dD \left(\kappa n_{\rm B} p_1(D) -\kappa (1+n_{\rm B}) p_0(D)\right) + \omega\int_{-\infty}^0 dD \left(\kappa n_{\rm B} p_0(D) -\kappa (1+n_{\rm B}) p_1(D)\right) \\
    & = \omega \kappa n_{\rm B} \left( \int_{0}^\infty dD p_1(D) + \int_{-\infty}^0 dD p_0(D) \right) - \omega \kappa (1+n_{\rm B}) \left(\int_{0}^\infty dD p_0(D) + \omega\int_{-\infty}^0 dD p_1(D)\right),
\end{split}
\end{equation} 
Two integrals in the bracket after $\kappa n_{\rm B}$ is the total probability that $D$ shows the correct state of the system, whereas the two integrals in the bracket after $\kappa (1+n_{\rm B})$ is the total probability that they do not. The latter is the error probability $\eta$ in Eq.~\eqref{eq:eta} in the main text, resulting in Eq.~\eqref{eq:Heat2lvl}.


\subsection{Steady-state solution} \label{sec:ssClas}
From the QFPME~\eqref{eq:QFPME} with the Liouvillian $\mc{L}(D)$~\eqref{eq:LiouClas}, we obtain a pair of coupled differential equations,
\begin{equation} \label{eq:setClas}
    \begin{cases}
        &\partial_t p_0(D) = \mc{J} p_0(D) + \gamma \partial_D p_0(D) + \theta(D) \left( \kappa n_{\rm B} p_1(D) - \kappa (1+n_{\rm B})p_0(D) \right) + (1-\theta(D)) \left( \kappa (1+n_{\rm B}) p_1(D) - \kappa n_{\rm B} p_0(D) \right),  \\
        &\partial_t p_1(D) = \mc{J} p_1(D) - \gamma \partial_D p_1(D)
        + \theta(D) \left( \kappa (1+n_{\rm B}) p_0(D) - \kappa n_{\rm B} p_1(D) \right) + (1-\theta(D)) \left( \kappa n_{\rm B} p_0(D) - \kappa (1+n_{\rm B}) p_1(D) \right),\\
    \end{cases}
\end{equation}
where we have introduced the differential operator $\mathcal{J}\cdot = \gamma\partial_D(D\,\cdot)+\frac{\gamma^2}{8\lambda}\partial_D(\cdot)$. It is convenient to define $q_+(D) = p_0(D)+ p_-(D)$ and $q_-(D) = p_0(D) - p_-(D)$, which allows us to rewrite Eq.~\eqref{eq:setClas} as
\begin{equation} \label{eq:setClasPM}
    \begin{cases}
        &\partial_t q_+(D) = \mc{J} q_+(D) + \gamma \partial_D q_-(D) ,  \\
        &\partial_t q_-(D) = \mc{J} q_-(D) + \gamma \partial_D q_+(D) - (\kappa n_{\rm B} - \kappa (1+n_{\rm B})) q_+ - (\kappa n_{\rm B} + \kappa (1+n_{\rm B}))q_-
        + 2 \theta(D) (\kappa n_{\rm B} - \kappa (1+n_{\rm B})) q_+(D) .\\
    \end{cases}
\end{equation}
To make progress, we introduce the functions
\begin{equation} \label{eq:G}
    G_n = G_n(D) =\frac{e^{-\frac{D^2}{2\sigma}}}{\sqrt{2\pi\sigma}} \he 
\end{equation}
for $n\geq 0$, where
\begin{equation} \label{eq:He}
    \he = \left(\frac{\sigma}{2} \right)^{\frac{n}{2}} \mathrm{H}_n\left( \frac{D}{\sqrt{2 \sigma}} \right)
\end{equation}
are generalized Hermite polynomials with  $\sigma=\frac{\gamma}{8\lambda}$ and $ \mathrm{H}_n(x) = (-1)^n e^{x^2} \partial^n_x e^{-x^2} $ are standard physicist's Hermite polynomials.
The functions $G_n$ fulfill two useful properties:
\begin{equation} \label{eq:Grelations}
    \mathcal{J} G_n =-\gamma n G_{n} \quad \textrm{and} \quad  \partial_D G_n = -\frac{1}{\sigma} G_{n+1}, 
\end{equation}
which follow from two identities obeyed by the generalized Hermite polynomials, $\partial_D \he = n \hen{n-1}$ and $\hen{n+1} = 2D \he - \partial_D \hen{n-1}$. 
Next, we expand $q_{\pm}(D)$ as
\begin{equation} \label{eq:qGexpansion}
    q_\pm(D) = \sum_{n = 0}^\infty q_n^{\pm} G_n(D),
\end{equation}
where the coefficients $q_n^{\pm}$ do not explicitly depend on $D$. We insert this expansion into Eq.~\eqref{eq:setClasPM} and consider the steady-state condition by setting the time-derivatives on the left-hand sides to zero.
Next, we multiply both equations~\eqref{eq:setClasPM} by $\hen{m}$ and integrate over $D$ from $-\infty$ to $\infty$.
Using the orthonormality of the generalized Hermite polynomials,
\begin{equation} \label{eq:ortho}
    \int_{-\infty}^{\infty}He_{n}^{[\sigma]}(D)He_{m}^{[\sigma]}(D)\frac{e^{-\frac{D^2}{2\sigma}}}{\sqrt{2\pi\sigma}}dD =\delta_{n,m}n!\sigma^n,
\end{equation}
and the relation
\begin{equation}
    \int_0^\infty He_{n}^{[\sigma]}(D)He_{m}^{[\sigma]}(D)\frac{e^{-\frac{D^2}{2\sigma}}}{\sqrt{2\pi\sigma}}dD = 
    \begin{cases}
    & \frac{1}{2} \sigma^m m! \quad \text{for $n=m$},\\
    & 0 \quad \text{for $n+m$ even},\\
    &  \frac{(-1)^{(m+n-1)/2}\sqrt{\sigma^{m+n}} m!! (n-1)!!}{\sqrt{2 \pi }(m-n)} \quad \text{for $n$ even and $m$ odd}, \\
    &  \frac{(-1)^{(m+n-1)/2}\sqrt{\sigma^{m+n}} n!! (m-1)!!}{\sqrt{2 \pi }(n-m)} \quad \text{for $n$ odd and $m$ even}, \\
    \end{cases}
\end{equation}
with $(-1)!! = 1$,
we find a set of linear equations:
\begin{equation} \label{eq:linearwithC}
    \begin{cases}
         &0 =  -\gamma m q_m^{+} - \frac{\gamma}{\sigma}  q_{m-1}^{-}   \\
        &0 =  -\gamma m q_m^{-} - \frac{\gamma}{\sigma}  q_{m-1}^{+}  -\kappa (1+2n_{\rm B}) q_m^-
        - 2  \kappa   \mc{C}_m . \\
    \end{cases}
\end{equation}
Here, 
\begin{equation}
    \mc{C}_m = 
    \begin{cases}
        \sum_{n = 0, 2, ...} q_n^+ \frac{(-1)^{(m+n-1)/2}\sqrt{\sigma^{m+n}} m!! (n-1)!!}{\sqrt{2 \pi }(m-n) m! \sigma^m} \quad \text{for $m$ odd},\\
        \sum_{n = 1, 3, ...} q_n^+ \frac{(-1)^{(m+n-1)/2}\sqrt{\sigma^{m+n}} n!! (m-1)!!}{\sqrt{2 \pi }(n-m)m! \sigma^m} \quad \text{for $m$ even},\\
    \end{cases}
\end{equation}
where $q^\pm_n =0$ for $n<0$. We can notice that in Eq.~\eqref{eq:linearwithC}, elements $q_{m}^-$ with even index $m$ are coupled only to elements $q_{m'}^+$ with odd index $m'$ and vice-versa. For even $m$'s and odd $m'$'s, $q_m^- = q_{m'}^+ = 0$ is a solution. The same does not hold for odd $m$'s and even $m'$'s, because from the second line in Eq.~\eqref{eq:linearwithC} with $m = 1$, we would have $q_0^+ = 0$. However, we can show that $q_0^+ = 1$, which follows from normalization of probability:
\begin{equation} \label{eq:Pcondition}
    1 = \int_{-\infty}^\infty q_+(D) dD = \sum_{n = 0}^\infty q_n^+ \int_{-\infty}^\infty G_n(D) dD = \sum_{n = 0}^\infty q_n^+ \int_{-\infty}^\infty G_n(D) \hen{0} dD = \sum_{n=0}^\infty q_n^+ \delta_{n, 0}n! \sigma^n = q_0^+,
\end{equation}
where we inserted Eq.~\eqref{eq:qGexpansion} to obtain the second equality and used $\hen{0} = 1$ in the third equality and Eq.~\eqref{eq:ortho} in the fourth equality.
Therefore, we obtain a set of linear equations for $q^-_m$ and $q^-_{m'}$ terms with only odd $m$'s and even ${m'}$:
\begin{equation} \label{eq:linearwithCl}
    \begin{cases}
         &0 =  -\gamma (2\ell) q_{2\ell}^{+} - \frac{\gamma}{\sigma}  q_{2\ell-1}^{-}   \\
        &0 =  -\gamma ((2\ell+1) q_{2 \ell+1}^{-} - \frac{\gamma}{\sigma}  q_{2 \ell}^{+}  -  \kappa (1+2n_{\rm B}) q_{2 \ell +1}^-
        - 2 \kappa \sum_{k=0}^\infty q^+_{2 k} \frac{(-1)^{\ell +k}\sqrt{\sigma^{-2\ell + 2k -1}} (2 \ell +1)!! (2k-1)!!}{\sqrt{2 \pi }(2\ell -2k + 1 ) (2\ell +1)! }  . \\
    \end{cases}
\end{equation}

To solve them, we make a truncation by considering the first $L$ equations with $\ell = 0, 2, ..., L$. This means we set $q^+_{2 \ell} = q^-_{2\ell +1} = 0 $ for $\ell > L$. The truncation is justified, because Eq.~\eqref{eq:qGexpansion} converges for a sufficiently large $L$. This way we obtain $2L+1$ equations (the first line in Eq.~\eqref{eq:linearwithCl} with $\ell = 0$ yields $0=0$) with $2L+1$ variables, which is a solvable system of equations. The probability $P(D)$ is then found using Eq.~\eqref{eq:qGexpansion}.
The functions $G_n$ are either even or odd depending on the index, $G_n(-D) = (-1)^n G_n(D)$, which results in $q_+(-D) = q_+(D)$ and  $q_-(-D) = q_-(D)$. The occupation probabilities are given by $p_0(D) = (q_+(D) + q_-(D))/2$ and $p_1(D) = (q_+(D) - q_-(D))/2$. They obey the relations $p_0(-D) = p_1(D)$ and $\int_{-\infty}^\infty dD p_0(D) = \int_{-\infty}^\infty dD p_1(D) = 1/2$.
The power $P$ in Eq.~\eqref{eq:Power2lvl} can be obtained with the expression
\begin{equation}
    P = -\omega \gamma \sum_{\ell = 0}^\infty q^+_{2 \ell} G_{2\ell}(0) - \omega \frac{\gamma^2}{8 \lambda} \sum_{\ell = 0}^\infty q^-_{2 \ell + 1} \partial_D G_{2 \ell +1}(D)|_{D=0} = -\omega \gamma \sum_{\ell = 0}^\infty q^+_{2 \ell} G_{2\ell}(0) +\omega \gamma \sum_{\ell = 0}^\infty q_{2 \ell +1}^- G_{2 
    \ell +2}(0),
\end{equation}
where we have applied $\partial_D G_{n}(D) = -(1/\sigma) G_{n+1}(D)$.

\section{Continuous-measurement-driven heat engine}

In this section, we provide details about the continuous-measurement-driven quantum heat engine in Sec.~\ref{sec:QuantumEngine}.

\subsection{Rotating frame} \label{app:rotating}
Let us denote the density matrix in the laboratory frame and in the rotating frame as $\op{\varrho}_t(D)$ and $ \op{\rho}_t(D) = \op{V} \op{\varrho}_t(D) \op{V}^\dagger$, respectively. Here, $\op{V} = e^{it \frac{\omega}{2} \op{\sigma}_z}$. The laboratory-frame density matrix $\op{\varrho}_t(D)$ follows the QFPME~\eqref{eq:QFPME} with the laboratory-frame Hamiltonian $\op{H}_{\rm lab}$~\eqref{eq:Hlab}, and the observable $\op{A}_{\rm lab} = \op{V}^\dagger \op{\sigma}_x \op{V}$ is continuously measured. In the rotating frame, the equation of motion reads
\begin{equation} \label{eq:dynrot}
    \partial_t  \op{\rho}_t(D) = \partial_t \left( \op{V} \op{\varrho}_t(D) \op{V}^\dagger \right) = \partial_t \op{V} \left( \op{\varrho}_t(D) \op{V}^\dagger \right)   +  
 \left( \op{V}  \op{\varrho}_t(D)  \right) \partial_t \op{V}^\dagger +  \op{V} \left(\partial_t \op{\varrho}_t(D) \right) \op{V}^\dagger = i \frac{\omega}{2} [ \op{\sigma}_z, \op{\rho}_t(D)] + \op{V} \left(\partial_t \op{\varrho}_t(D) \right) \op{V}^\dagger,
\end{equation}
where we used $\partial_t \op{V} = i \frac{\omega}{2} \op{\sigma}_z \op{V}$ and  $\partial_t \op{V}^\dagger = -i \frac{\omega}{2} \op{V}^\dagger \op{\sigma}_z$. For $\partial_t \op{\varrho}_t(D)$, we can substitute the QFPME~\eqref{eq:QFPME}, which we write here for convenience:
\begin{equation} \label{eq:dyn}
    \partial_t\hat{\varrho}_t(D) = -i[\op{H}_{\rm lab }, \hat{\varrho}_t(D)] + \mc{L}_{\rm B} \hat{\varrho}_t(D)+\lambda\mathcal{D}[\hat{A}_{\rm lab}]\hat{\varrho}_t(D) 
     -\gamma\partial_D\mathcal{A}_{\rm lab}(D)\hat{\varrho}_t(D)+\frac{\gamma^2}{8\lambda}\partial^2_D\hat{\varrho}_t(D).
\end{equation}
We now compute $\op{V} \cdot \op{V}^\dagger$ applied to each term on the right-hand side. The first term results in
\begin{equation} \label{eq:term1}
    \op{V} \left(-i[\op{H}_{\rm lab }, \hat{\varrho}_t(D)]\right) \op{V}^\dagger = -i [ \op{V} \op{H}_{\rm lab } \op{V}^\dagger, \op{V} \hat{\varrho}_t(D) \op{V}^\dagger ] = - i\frac{\omega}{2}[\op{\sigma}_z , \hat{\rho}_t(D) ] -i[\op{H}, \hat{\rho}_t(D)] , 
\end{equation}
where we used $\op{V} \op{H}_{\rm lab } \op{V}^\dagger = \frac{\omega}{2}\op{\sigma}_z + gD \op{\sigma}_y = \frac{\omega}{2}\op{\sigma}_z + \op{H}$, with $\op{H}$ being the Hamiltonian~\eqref{eq:H} in the rotating frame. Notice that the term $- i\frac{\omega}{2}[\op{\sigma}_z , \hat{\rho}_t(D) ]$ in Eq.~\eqref{eq:term1} cancels with the first term in the right-hand side of Eq.~\eqref{eq:dynrot}. From the second term in Eq.~\eqref{eq:dyn}, where $\mc{L}_{\rm B}$ is given in Eq.~\eqref{eq:bath}, we obtain
\begin{equation} \label{eq:term2}
    \op{V} \left( \mc{L}_{\rm B}  \op{
    \varrho}_t(D) \right) \op{V}^\dagger =  \mc{L}_{\rm B} \op{V} \op{
    \varrho}_t(D) \op{V}^\dagger =  \mc{L}_{\rm B}  \op{
    \rho}_t(D).
\end{equation}
Here, we used the relations $\op{V} \op{\sigma}^\dagger \op{\sigma} \op{V} = \op{\sigma}^\dagger \op{\sigma} $, $\op{V} \op{\sigma} \op{\sigma}^\dagger \op{V} = \op{\sigma} \op{\sigma}^\dagger $, $\op{V}  \op{\sigma} \op{V} = e^{-i\omega t} \op{\sigma} $, and $\op{V}  \op{\sigma}^\dagger \op{V} = e^{i\omega t} \op{\sigma}^\dagger $. From the third term in Eq.~\eqref{eq:dyn} we have
\begin{equation} \label{eq:term3}
    \op{V} \left( \lambda\mathcal{D}[\hat{A}_{\rm lab}]\hat{\varrho}_t(D) \right) \op{V}^\dagger = \lambda \mathcal{D}[\op{V} \hat{A}_{\rm lab} \op{V}^\dagger] \op{V} \hat{\varrho}_t(D) \op{V}^\dagger = \lambda\mathcal{D}[\hat{A}]\hat{\rho}_t(D).
\end{equation}
From the fourth term in Eq.~\eqref{eq:dyn} we find
\begin{equation} \label{eq:term4}
    \op{V} \left( -\gamma\partial_D\mathcal{A}_{\rm lab}(D)\hat{\varrho}_t(D) \right)  \op{V}^\dagger = -\frac{\gamma}{2} \partial_D \{ \op{V} \hat{A}_{\rm lab} \op{V}^\dagger -D, \op{V} \hat{\varrho}_t(D) \op{V}^\dagger  \} = -\frac{\gamma}{2} \partial_D \{  \hat{A} -D, \hat{\rho}_t(D)  \} = -\gamma \partial_D \mathcal{A}(D) \op{\rho}_t(D),
\end{equation}
where we used $\op{V} \hat{A}_{\rm lab} \op{V}^\dagger = \op{A}$.
Lastly, the fifth term in Eq.~\eqref{eq:dyn} yields
\begin{equation} \label{eq:term5}
    \op{V} \left( \partial_D^2 \op{\varrho}_t(D) \right)  \op{V}^\dagger = \partial_D^2 \left( \op{V}  \op{\varrho}_t(D)   \op{V}^\dagger \right) = \partial_D^2   \op{\rho}_t(D)  .
\end{equation}
Collecting together Eqs.~\eqref{eq:term1}-\eqref{eq:term5} leads to the QFPME~\eqref{eq:QFPME} describing the time evolution of the density matrix  $\op{\rho}_t(D)$ in the rotating frame, where the measured observable is $\op{A} = \op{\sigma}_x$, the Hamiltonian $\op{H}$ is given in Eq.~\eqref{eq:H}, and the superoperator $\mc{L}_{\rm B}$ describing the effect of the reservoir is given in Eq.~\eqref{eq:bath}.

\subsection{Power, heat, and measurement energy} \label{app:energetics}
All expressions that appear in the first law of thermodynamics in Eq.~\eqref{eq:PowerHeatQFPME} are in the laboratory frame, which means they are evaluated with respect to $\op{\varrho}_t(D)$. However, we would like to compute them in the rotating frame, with respect to $\op{\rho}_t(D) = \op{V} \op{\varrho}_t(D) \op{V}^\dagger$ with $\op{V} = e^{it \frac{\omega}{2} \op{\sigma}_z}$, which follows the QFPME with the time-independent Hamiltonian $\op{H}$~\eqref{eq:H}. To this end, for the expectation value of $\op{O}$ in the laboratory frame, we write the corresponding expression in the rotating frame:
\begin{equation} \label{eq:labrot}
    \langle\op{O}\rangle_{\rm lab} := \int dD \tr{ \op{O} \op{\varrho}_t(D) } = \int dD \tr{ \op{V} \op{O} \op{V}^\dagger \op{V} \op{\varrho}_t(D) \op{V}^\dagger} =  \int dD \tr{ \op{V} \op{O} \op{V}^\dagger  \op{\rho}_t(D) } =: \erw{\op{V} \op{O} \op{V}^\dagger},
\end{equation} 
which can be employed to find the power~\eqref{eq:PowerQFPME}, heat current~\eqref{eq:HeatQFPME}, and measurement energy~\eqref{eq:EmQFPME}. Below, we compute each element that appears in these quantities.
The thermodynamic Hamiltonian $\op{H}_{\rm TD} = \frac{\omega}{2}\hat{\sigma}_z$ [see Eq.~\eqref{eq: H TD example}] is independent of $D$ and therefore the only term contributing to the power is 
$P =  \erw{\partial_t \op{H}_{\rm lab}}_{\rm lab}$, where $\op{H}_{\rm lab} = \frac{\omega}{2}\hat{\sigma}_z+ i g D \left(e^{i\omega t}\hat{\sigma}-e^{-i\omega t}\hat{\sigma}^{\dagger}\right)$ [see Eq.~\eqref{eq:Hlab}]. Using Eq.~\eqref{eq:labrot}, we find
\begin{equation}
    P = \erw{ \op{V} \left( \partial_t \op{H}_{\rm lab} \right) \op{V}^\dagger } =  -g\omega \erw{ D \op{V} \left(e^{i\omega t} \op{\sigma} + e^{-i\omega t} \op{\sigma}^\dagger  \right) \op{V}^\dagger }  = -\omega g \erw{D ( \op{\sigma} + \op{\sigma}^\dagger )} = -\omega g \erw{D \op{\sigma}_x},
\end{equation}
The above equation is the power in Eq.~\eqref{eq:PowerEng}.
For the measurement energy~\eqref{eq:EmQFPME} we obtain
\begin{equation}
    \dot{E}_{\rm M} = \erw{ \op{V} \left( \lambda \mc{D}[\op{A}_{\rm lab} ] \op{H}_{\rm TD} \right) \op{V}^\dagger } = \lambda \erw{ \mc{D}[\op{V}\op{A}_{\rm lab} \op{V}^\dagger ] \op{V} \op{H}_{\rm TD} \op{V}^\dagger } = \lambda \erw{ \mc{D}[\op{\sigma}_x ] \left( \frac{\omega}{2}\op{\sigma}_z  \right) } = - \lambda \omega \erw{ \op{\sigma}_z }, 
\end{equation}
which recovers Eq.~\eqref{eq:MEEng}.
For the heat current~\eqref{eq:HeatQFPME}, we find
\begin{equation}
\begin{split}
    J =& \erw{ \op{V} \left( \mc{L}^\dagger_{\rm B} \op{H}_{\rm TD}  \right) \op{V}^\dagger } = \erw{ \op{V} \left( \kappa (1+n_{\rm B}) \mc{D}^\dagger [\op{\sigma}] \op{H}_{\rm TD}  \right) \op{V}^\dagger } + \erw{ \op{V} \left( \kappa n_{\rm B} \mc{D}^\dagger [\op{\sigma}^\dagger] \op{H}_{\rm TD}  \right) \op{V}^\dagger }  =
    \kappa (1+n_{\rm B}) \erw{  \mc{D}^\dagger [ \op{\sigma} ] \op{V} \op{H}_{\rm TD} \op{V}^\dagger   } + \kappa n_{\rm B} \erw{   \mc{D}^\dagger [ \op{\sigma}^\dagger ] \op{V} \op{H}_{\rm TD} \op{V}^\dagger   } \\
    = & \kappa (1+n_{\rm B}) \erw{  \mc{D}^\dagger [ \op{\sigma} ] \left( \frac{\omega}{2}\op{\sigma}_z  \right)   } + \kappa n_{\rm B} \erw{   \mc{D}^\dagger [ \op{\sigma}^\dagger ] \left( \frac{\omega}{2}\op{\sigma}_z  \right)  } =  -\frac{\omega}{2} \kappa  - \frac{\omega}{2} \kappa (1+2n_{\rm B}) \langle \op{\sigma}_z \rangle, 
    \end{split}
\end{equation}
which is Eq.~\eqref{eq:HeatEng}.

\subsection{Steady-state solution} \label{app:ss}

The joint state of the quantum system and the measurement outcome $\op{\rho}_t(D)$ can be expressed as
\begin{equation}
    \hat{\rho}_t(D)=\frac{1}{2} \left(P_t(D) \op{I} +a_{x, t}(D)\hat{\sigma}_x+a_{y, t}(D)\hat{\sigma}_y+a_{z, t}(D)\hat{\sigma}_z \right),
\end{equation}
where $\op{I}$ is the identity matrix and $(a_x, a_y, a_z)$ is the Bloch vector for a given $D$ at the time $t$, where $a_j = a_{j, t}(D)$ and $P_t(D) = P$ for brevity (not to be confused with power).
Using the relation $a_j = \tr{\op{\sigma}_j \op{\rho}_t(D)} $ for $j \in \{x, y, z \}$, the QFPME for the continuous-measurement-driven heat engine leads to a set of coupled differential equations,
\begin{equation} \label{eq:coupled diff}
    \begin{cases}
        &\partial_t P =\gamma\partial_D (DP-a_x) +\frac{\gamma^2}{8\lambda}\partial_D^2 P, \\
        &\partial_t a_x=2gD a_z+\gamma\partial_D (Da_x-P)+\frac{\gamma^2}{8\lambda}\partial_D^2a_x  - \frac{\kappa (1+2n_{\rm B})}{2} a_x,  \\
        &\partial_t a_y=-2\lambda a_y+\gamma\partial_D (D a_y)+\frac{\gamma^2}{8\lambda}\partial_D^2 a_y - \frac{ \kappa (1+2n_{\rm B})}{2} a_y,  \\
        & \partial_t a_z =-2gD a_x -2\lambda a_z +\gamma\partial_D (D a_z+\frac{\gamma^2}{8\lambda}\partial_D^2 a_z - \kappa (1+2n_{\rm B}) b_z   - \kappa P. \\
    \end{cases}
\end{equation}
Since we are concerned with the steady state, we may set the left-hand sides to zero. Using the differential operator $\mathcal{J}\cdot = \gamma\partial_D(D\,\cdot)+\frac{\gamma^2}{8\lambda}\partial_D(\cdot)$, which has already been introduced, we can simplify the four equations above to
\begin{equation} \label{eq:coupled diff simpl}
    \begin{cases}
        &0 = \mathcal{J} P -\gamma\partial_D a_x, \\
        &0=2gDa_z +\mathcal{J} a_x-\gamma\partial_D P - \frac{ \kappa (1+2n_{\rm B})}{2} a_x, \\
        & 0=-2\lambda a_y+\mathcal{J} a_y -  \frac{\kappa (1+2n_{\rm B})}{2} a_y, \\
        & 0 =-2gD a_x -2\lambda a_z +\mathcal{J} a_z -  \kappa (1+2n_{\rm B}) b_z   - \kappa P . \\
    \end{cases}
\end{equation}
Next, we expand the Bloch coefficients using the functions $G_n$ defined in Eq.~\eqref{eq:G}:
\begin{equation} \label{eq:Gexpansion}
    P = \sum_{n = 0}^\infty p_n G_n \quad \mathrm{and} \quad a_j = \sum_{n = 0}^\infty a^{(j)}_n G_n,
\end{equation}
    which we insert into Eq.~\eqref{eq:coupled diff simpl}. We note that $p_n$ and $a_n^{(j)}$ do not depend on $D$. We apply the relations in Eq.~\eqref{eq:Grelations} and multiply each equation by $\hen{m}$.
    Integrating over $D$ from $-\infty$ to $\infty$ and utilizing the identities 
    $\partial_D \he = n \hen{n-1}$ and $\hen{n+1} = 2D \he - \partial_D \hen{n-1}$ as well as the orthonormality~\eqref{eq:ortho} of Hermite polynomials leads to a recursive scheme
\begin{equation} \label{eq:recursive Bloch}
    \begin{cases}
        & 0 =mp_m-\frac{1}{\sigma}a_{m-1}^{(x)}, \\
        & 0=- \left(2\lambda + \frac{\kappa (1+2n_{\rm B})}{2} \right) a_m^{(y)}-\gamma m a_m^{(y)}, \\
        &0=ga_{m-1}^{(z)}+g(m+1)\sigma a_{m+1}^{(z)}-\frac{\gamma}{2}m a_m^{(x)}+\frac{\gamma}{2\sigma}p_{m-1} - \frac{ \kappa (1+2n_{\rm B})}{4} b_m^{(x)} , \\
        & 0=-ga_{m-1}^{(x)}-g(m+1)\sigma a_{m+1}^{(x)}-\lambda a_m^{(z)}-\frac{\gamma}{2}ma_m^{(z)} -a_m^{(z)} \frac{ \kappa (1+2n_{\rm B})}{2} - \frac{\kappa }{2} p_m,\\
    \end{cases}
\end{equation}
where we set $a^{(j)}_m = 0$ and $p_m = 0$ when $m<0$.
From the second equation we immediately see that $a_m^{(y)} = 0$. In the remaining three equations, "even" terms $a^{x}_{m = 2\ell}$ are coupled to "odd" terms $a^{z}_{m = 2\ell \pm 1}$ and $p_{m = 2\ell \pm 1}$, and vice-versa. Therefore, Eq.~\eqref{eq:recursive Bloch} actually forms two recursive sets of linear equations, one for "even" $a^{x}_{m = 2\ell}$ terms,
\begin{equation} \label{eq:rec even}
    \begin{cases}
        & 0  =(2\ell+1)p_{2\ell+1}-\frac{1}{\sigma}a_{2\ell}^{(x)}, \\
        & 0 = ga_{2\ell-1}^{(z)}+g(2\ell+1)\sigma a_{2l+1}^{(z)} - \left( \frac{\gamma}{2}(2\ell) + \frac{ \kappa (1+2n_{\rm B})}{4} \right) a_{2\ell}^{(x)}+\frac{\gamma}{2\sigma}p_{2\ell-1}, \\
        &0=-ga_{2\ell}^{(x)}-g(2\ell+2)\sigma a_{2\ell+2}^{(x)}- \left( \lambda + \frac{ \kappa (1+2n_{\rm B})}{2} \right) a_{2\ell+1}^{(z)}-\frac{\gamma}{2}(2\ell+1)a_{2\ell+1}^{(z)} - \frac{\kappa }{2} p_{2\ell + 1}.
    \end{cases}
\end{equation}
and one for "odd" $a^{x}_{m = 2\ell+1}$ terms,
\begin{equation} \label{eq:rec odd}
    \begin{cases}
        & 0  =(2\ell)p_{2\ell}-\frac{1}{\sigma}a_{2\ell-1}^{(x)}, \\
        & 0 = ga_{2\ell}^{(z)}+g(2\ell+2)\sigma a_{2\ell+2}^{(z)} - \left( \frac{\gamma}{2}(2\ell+1) + \frac{\kappa (1+2n_{\rm B})}{4} \right) a_{2\ell+1}^{(x)}+\frac{\gamma}{2\sigma}p_{2\ell}, \\
        &0=-ga_{2\ell-1}^{(x)}-g(2\ell+1)\sigma a_{2\ell+1}^{(x)}- \left( \lambda + \frac{ \kappa (1+2n_{\rm B})}{2} \right) a_{2\ell}^{(z)}-\frac{\gamma}{2}(2\ell)a_{2\ell}^{(z)} - \frac{\kappa }{2} p_{2\ell }.
    \end{cases}
\end{equation}
In Eq.~\eqref{eq:rec even}, $a^{(x)}_{2\ell} = a^{(z)}_{2\ell+1} = p_{2\ell+1} = 0$ is a solution. 
An all-zero solution does not hold in Eq.~\eqref{eq:rec odd} because we would have $p_0 = 0$ from the second line with $\ell=0$, but $p_0 = 1$ as a result of normalization, similarly to Eq.~\eqref{eq:Pcondition}. 
Since the function $G_n(D) = (-1)^n G_n(-D)$ is either even or odd depending on the index, $P = \sum_{\ell = 0}^\infty p_{2\ell} G_{2\ell}$ and $a_z = \sum_{\ell = 0}^\infty a^{(z)}_{2\ell} G_{2\ell}$ will be even functions of $D$, whereas $a_x = \sum_{\ell = 0}^\infty a^{(x)}_{2\ell+1} G_{2\ell+1}$ will be odd.

The recursive scheme in Eq.~\eqref{eq:rec odd} can be solved with a cutoff as follows. From $\ell = 0, 1, ..., L$, we obtain $3L +2$ equations (the first line for $\ell = 0$ yields a trivial equation). These equation involve $L+1$ terms $p_{0}, p_2, ..., p_{2L}$, $L+2$ terms $a_0^{(z)}, a_2^{(z)}, ..., a_{2L+2}^{(z)}$, and $L+1$ terms $a_1^{(x)}, a_3^{(x)}, ..., a_{2L+1}^{(x)}$. We impose a cutoff after $L+1$ nonzero terms in the expansion of $P$, $a_z$, and $a_x$ in Eq.~\eqref{eq:Gexpansion}, meaning we set $a_{2L+2}^{(z)} = 0$. This is justified for a large cutoff $L$, because the expansion~\eqref{eq:Gexpansion} should converge for all values of $D$. In addition, we know that $p_0 = 1$. Therefore, we have $3L+2$ coupled linear equations consisting of $3L+2$ variables, which can be solved.

The method to compute the steady state described above can be used to obtain the thermodynamic quantities appearing in the first law in Eq.~\eqref{eq:PowerHeatQFPME}. For the power~\eqref{eq:PowerEng}, we find
\begin{equation}
\begin{split}
    P &= - \omega g \erw{D\op{\sigma}_x} = -g \omega \int_{-\infty}^\infty dD D a_x = -g \omega \sum_{\ell = 0}^\infty a_{2\ell+1} \int_{-\infty}^\infty  G_{2\ell+1} D dD = -g \omega \sum_{\ell = 0}^\infty a_{2\ell+1} \int_{-\infty}^\infty  G_{2\ell+1} \hen{1} dD \\
    &=  -g \sigma \omega a_1^{(x)} = -\frac{\omega g \gamma }{8\lambda} a_1^{(x)},
\end{split}
\end{equation}
where we used $\tr{\op{\sigma}_x \op{\rho}_t(D)} = a_x$ in the second equality, $a_x = \sum_{\ell = 0}^\infty a^{(x)}_{2\ell+1} G_{2\ell+1}$ in the third equality, $\hen{1} = D$ in the fourth equality, the orthonormality~\eqref{eq:ortho} in the fifth equality, and $\sigma=\frac{\gamma}{8\lambda}$ in the sixth equality. Similarly, the measurement energy~\eqref{eq:MEEng} and heat current~\eqref{eq:HeatEng} reduce to
\begin{equation}
    \dot{E}_{\rm M} = - \lambda \omega \erw{ \op{\sigma}_z }  = - \lambda \omega \int_{-\infty}^\infty dD D a_z = - \lambda \omega a_0^{(z)}
\end{equation}
 and
\begin{equation}
    J = -\frac{\omega}{2} \kappa - \frac{\omega}{2}  \kappa (1+2n_{\rm B}) \langle \op{\sigma}_z \rangle  = -\frac{\omega}{2}  \kappa  - \frac{\omega}{2} \kappa (1+2n_{\rm B}) a_0^{(z)},
\end{equation}
respectively.
Finally, from the third line in Eq.~\eqref{eq:rec odd} with $\ell = 0$, where $p_0 = 1$, it follows that $0= -g \sigma a_{1}^{(x)}- \left( \lambda + \frac{ \kappa }{2}(1+2n_{\rm B}) \right) a_{0}^{(z)} - \frac{\kappa }{2} $. This means that $P + J + \dot{E}_{\rm M} = 0$, which is consistent with the steady-state solution.
\end{widetext}

\clearpage
\newpage
\mbox{~}
\clearpage
\newpage

\section{Fluctuation theorems for classical systems} \label{app:derivationsclassical}

\subsection{General setting }
\label{app:generalsetting}
When introducing the general setting of the QFPME-based feedback control, we largely follow the notation of Refs.~\cite{Sagawa2012, Potts2018}.
At the start of the forward experiment, the system and detector are prepared in the states $a_0$ and $D_0$ according the initial joint distribution $P_{\rm ini}[a_0, D_0]$. The system is driven with a protocol $\mathbf{\Lambda} := (\Lambda_1, ... , \Lambda_N)$ such that the probability $P[a_n|a_{n-1}, \Lambda_{n}]$ of transition $a_{n-1} \to a_n$ depends on $\Lambda_{n}$, where $\mathbf{\Lambda}$ encompasses all settings of experimental "knobs". Feedback control is implemented by making the protocol $\mathbf{\Lambda}$ dependent on the trajectory $\mathbf{D} := (D_0, D_1, ..., D_N, D_{N+1})$ of the detector such that $\Lambda_n(\mathbf{D}) = \Lambda(D_n)$ depends only on $D_n$, which is required to describe the dynamics of the system with the QFPME formalism. We thereby ignore an explicit time-dependence in the protocol for simplicity. Here, each $D_{n \geq 1}$ is contingent upon $(a_0, ..., a_{n-1})$ and $(D_0, ..., D_{n-1})$, which is characterized by the probability $P_{\rm m}[D_n|(a_0, ..., a_{n-1}), (D_0, ..., D_{n-1})]$. A general way to write the joint probability of the system's and measurement outcome's trajectories is given by [cf. Eq.~(47) in Ref.~\cite{Sagawa2012} and Eq.~(5) in Ref.~\cite{Potts2018}]
\begin{equation} \label{eq:prob ad}
\begin{split}
     P[\mathbf{a}, \mathbf{D}] =& P_{\rm ini}[a_0, D_0] P[D_{1}|(a_0), (D_0)] P[a_1| a_0, \Lambda(D_1)] \\
     &\times P[D_{2}|(a_0, a_1), (D_0, D_1)] P[a_2| a_1, \Lambda(D_2)] \times ... \\ & \times P[a_N| a_{N-1}, \Lambda(D_{N})] P[D_{N+1}| (a_0, ..., a_N), (D_0, ..., D_N)] \\
     =:& P_{\rm m}[\mathbf{D}|\mathbf{a}] P[\mathbf{a}|\mathbf{\Lambda}(\mathbf{D})],
\end{split}  
\end{equation}
where we grouped together
\begin{equation} \label{eq:prob a}
     P[\mathbf{a}|\mathbf{\Lambda}(\mathbf{D})] = P_{\rm ini}[a_0]  P[a_1| a_0, \Lambda(D_1)] ...   P[a_N| a_{N-1}, \Lambda(D_{N})]
\end{equation}
and
\begin{equation} \label{eq:prob d}
    P_{\rm m}[\mathbf{D}|\mathbf{a}] =  P_{\rm ini}[D_0|a_0] P[D_{1}|a_0, D_0] ... P[D_{N+1}| (a_0, a_1, ..., a_N)],
\end{equation}
with $P_{\rm ini}[a_0, D_0] = P_{\rm ini}[D_0|a_0] P_{\rm ini}[a_0]$ and $P_{\rm ini}[a_0]$ being a marginal distribution.
The probability $P_{\rm m}[\mathbf{D}|\mathbf{a}]$ is the probability  of recording the measurement outcomes' trajectory $\mathbf{D}$ given the fixed system's trajectory $\mathbf{a}$, which is signified by the subscript "m". We stress that due to feedback control, it is different from $P[\mathbf{D}|\mathbf{a}] = P[\mathbf{a}, \mathbf{D}]/P[\mathbf{a}]$, where $P[\mathbf{a}] = \int d\mathbf{D} P[\mathbf{a}, \mathbf{D}] $ is the marginal distribution~\cite{Sagawa2012, Potts2018}.
As demonstrated below, the probability $ P_{\rm m}[\mathbf{D}|\mathbf{a}]$ is given by
\begin{equation} \label{eq: D gaussians}
    P_{\rm m}[\mathbf{D}|\mathbf{a}] =  P_{\rm ini}[D_0|a_0] \left( \frac{2 \lambda \delta t}{\pi} \right)^{\frac{N+1}{2}} \prod_{n=0}^N \frac{1}{\gamma \delta t} e^{-2 \lambda \delta t \left(
    \frac{D_{n+1} - e^{-\gamma \delta t} D_{n}}{\gamma \delta t}- a_{n} \right)^2 } .
\end{equation}
The probability in Eq.~\eqref{eq:prob ad} is normalized, $\int d\mathbf{D} d\mathbf{a} P[\mathbf{a}, \mathbf{D}] = 1$, which can be shown by integrating in the order $D_{N+1} \to a_N \to D_N \to ... \to a_0 \to D_0$. The ensemble average of any trajectory-dependent quantity $O(\mathbf{a}, \mathbf{D})$ is obtained with 
\begin{equation}
    \erw{O} = \int d\mathbf{D} d\mathbf{a} O(\mathbf{a}, \mathbf{D}) P[\mathbf{a}, \mathbf{D}] .
\end{equation}

Equation~\eqref{eq: D gaussians} is derived as follows. In the formalism of the QFPME, the state of the detector at the time $t$ reads~\cite{Annby2022}
\begin{equation} \label{eq: filter}
    D(t) = \gamma \int_{-\infty}^t ds e^{-\gamma(t-s)} z(s),
\end{equation}
where $z(s)$ is the outcome of the continuous Gaussian measurement~\cite{Jacobs2006rev, Bednorz2012} at the time $s$. The integral represents the low-pass filter with the bandwidth $\gamma$. In the time-discrete notation, it reads 
\begin{equation} \label{eq: filter discrete}
    D_n = \gamma \delta t \sum_{k = 0}^n  e^{-\gamma(n-k)\delta t} z_k.
\end{equation}
In the Gaussian measurement for classical incoherent systems, the probability of the outcome $z$ given the state of the system $a$ is given by~\cite{Jacobs2006rev, Bednorz2012}
\begin{equation} \label{eq: gaussian}
    P[z|a] =  \sqrt{\frac{2 \lambda \delta t}{\pi} } e^{-2\lambda \delta t (z- a)^2 },
\end{equation}
which is a Gaussian distribution centered at $a$. Its variance is given by $1/(4 \lambda \delta t)$, which means increasing $\lambda$ results in a more precise measurement. The probability of obtaining a tuple of outcomes $\mathbf{z} = (z_0, z_1, ..., z_{N+1})$ given the trajectory of the system $\mathbf{a} = (a_0, a_1, ..., a_N)$ reads
\begin{equation}
    P_{\rm m} [\mathbf{z}| \mathbf{a}] = P_{\rm ini, z}[z_0|a_0] \left( \frac{2 \lambda \delta t}{\pi} \right)^{\frac{N+1}{2}} \prod_{n = 0}^N  e^{-2 \lambda \delta t (z_{n+1} - a_{n})^2 } ,
\end{equation}
where $P_{\rm ini, z}[z_0|a_0]$ is the probability of the initial $z_0$ before the start of the experiment. Using Eq.~\eqref{eq: filter discrete}, the probability of the trajectory of the detector $\mathbf{D}$ is obtained through
\begin{widetext}
    
\begin{equation} \label{eq: D gaussians app}
\begin{split}
    P_{\rm m} [\mathbf{D}| \mathbf{a}] &=  \left( \frac{2 \lambda \delta t}{\pi} \right)^{\frac{N+1}{2}} \int dz_{N+1} ... dz_0    \prod_{r = 0}^{N+1} \delta \left(D_r - \gamma \delta t \sum_{k=0}^r e^{-\gamma \delta t (r-k)}z_k\right) P_{\rm ini, z}[z_0|a_0] \prod_{n = 0}^N e^{-2 \lambda \delta t }(z_{n+1} - a_{n})^2 \\
    & = P_{\rm ini}[D_0|a_0] \left( \frac{2 \lambda \delta t}{\pi} \right)^{\frac{N+1}{2}} \prod_{n = 0}^N \frac{1}{\gamma \delta t} e^{-2 \lambda \delta t \left(
    \frac{D_{n+1} - e^{-\gamma \delta t} D_{n}}{\gamma \delta t}- a_{n} \right)^2 },
\end{split}
\end{equation}
\end{widetext}
where we recognize $P_{\rm ini}[D_0|a_0] := P_{\rm ini, z}[D_0/(\gamma \delta t)|a_0]/(\gamma \delta t)$ as the initial probability of the detector before the start of the experiment. Therefore, we recover $P_{\rm m} [\mathbf{D}| \mathbf{a}]$ in Eq.~\eqref{eq: D gaussians}.

\subsection{Derivation of the FTs in Eq.~\eqref{eq: FT detailed} and $\sigma_{\rm m}$ in Eq.~\eqref{eq: meas ent for}}
\label{app:derFTclas}

Suppose that in the forward experiment, the system and detector's measurement outcome follow particular trajectories $\mathbf{a}$ and $\mathbf{D}$, which happens with the probability $P[\mathbf{a},\mathbf{D}]$ given in Eq.~\eqref{eq:prob ad}.
We now introduce the backward experiment (signified by the subscript $"\mathrm{B}"$), where the measurement and feedback are implemented in the same way as in the forward, but we prepare the initial states of the system and detector according to a different probability $P_{\rm ini, B}[a, D]$.
We note that in the main text, we assumed that in both experiments the initial probability is the same. For this reason, we omitted the subscript $"\mathrm{B}"$.
Next, we introduce the reversed trajectories, $\bar{\mathbf{a}} = (\bar{a}_0, \bar{a}_1, ..., \bar{a}_N)$ and $\bar{\mathbf{D}} = (\bar{D}_0, \bar{D}_1, ..., \bar{D}_{N+1})$, which can be obtained as
$\bar{D}_n = D_{N + 1 - n}$ and $\bar{a}_n = a_{N-n}$, meaning $\bar{\mathbf{a}} = (a_N, ..., a_1, a_0)$ and $\bar{\mathbf{D}} = (D_{N+1}, ..., D_1, D_0)$.
The probability that in the backward experiment, the system and measurement outcome follow the reversed trajectories is given by
\begin{equation} \label{eq:prob ad rev}
    P_{\rm B}[\bar{\mathbf{a}}, \bar{\mathbf{D}}] = P_{\rm m, B}[\bar{\mathbf{D}}|\bar{\mathbf{a}}] P_{\rm B}[\bar{\mathbf{a}}|\mathbf{\Lambda}(\bar{\mathbf{D}})],
\end{equation}
which is exactly as in Eq.~\eqref{eq:prob ad} but with the probability $P_{\rm ini, B}$ instead of $P_{\rm ini}$ and with $\bar{a}_n$ and $\bar{D}_n$ replacing $a_n$ and $D_n$.
In particular, $P_{\rm m, B}[\bar{\mathbf{D}}|\bar{\mathbf{a}}]$ still follows Eq.\eqref{eq: D gaussians app}:
\begin{equation}
    P_{\rm m, B}[\bar{\mathbf{D}}|\bar{\mathbf{a}}] = P_{\rm ini, B}[\bar{D}_0|\bar{a}_0]   \left( \frac{2 \lambda \delta t}{\pi} \right)^{N/2} \prod_{n = 0}^N \frac{1}{\gamma \delta t} e^{-2 \lambda \delta t \left(
    \frac{\bar{D}_{n + 1} - e^{-\gamma \delta t} \bar{D}_{n}}{\gamma \delta t}- \bar{a}_{n} \right)^2 }.
\end{equation}
Using $\bar{a}_{n} = a_{N - n}$ and $\bar{D}_{n} = D_{N + 1 - n}$, we find
\begin{equation} \label{eq: D back gaussians app}
\begin{split}
    P_{\rm m, B}[\bar{\mathbf{D}}|\bar{\mathbf{a}}] &= P_{\rm ini, B}[\bar{D}_{N+1}|\bar{a}_N]   \left( \frac{2 \lambda \delta t}{\pi} \right)^{N/2} \prod_{n = 0}^N \frac{1}{\gamma \delta t} e^{-2 \lambda \delta t \left(
    \frac{D_{N-n} - e^{-\gamma \delta t} D_{N + 1 - n}}{\gamma \delta t}- a_{N-n} \right)^2 } \\
    &=
    P_{\rm ini, B}[\bar{D}_{N+1}|\bar{a}_N]   \left( \frac{2 \lambda \delta t}{\pi} \right)^{N/2} \prod_{n = 0}^N \frac{1}{\gamma \delta t} e^{-2 \lambda \delta t \left(
    \frac{D_{n} - e^{-\gamma \delta t} D_{n + 1}}{\gamma \delta t}- a_{n} \right)^2 }.
\end{split}
\end{equation}

The left-hand site of detailed FT in Eq.~\eqref{eq: FT detailed} is given by
\begin{equation} \label{eq: FT der}
    \frac{P_{\rm B}[\bar{\mathbf{a}}, \bar{\mathbf{D}}]}{P[\mathbf{a}, \mathbf{D}]} = 
    \frac{ P_{\rm m, B}[\bar{\mathbf{D}}|\bar{\mathbf{a}}] }{P_{\rm m}[\mathbf{D}| \mathbf{a} ] } \frac{ P_{\rm B}[\bar{\mathbf{a}}|\mathbf{\Lambda}(\bar{\mathbf{D}})]}{
    P[\mathbf{a}|\mathbf{\Lambda}(\mathbf{D})]
    } = e^{-\sigma_{\rm m}[\mathbf{a}, \mathbf{D}] }
    e^{-\sigma[\mathbf{a}, \mathbf{D}] } ,
\end{equation}
where we have recognized the measurement entropy $\sigma_{\rm m}$ defined in Eq.~\eqref{eq: meas ent def}.

To obtain the analytical expression for $\sigma_{\rm m}$ in Eq.~\eqref{eq: meas ent for}, we substitute Eq.~\eqref{eq: D gaussians app} for $P_{\rm m}[\mathbf{D}| \mathbf{a}]$ and Eq.~\eqref{eq: D back gaussians app} for $P_{\rm m, B}[\bar{\mathbf{D}}|\bar{\mathbf{a}}]$, resulting in
\begin{widetext}
\begin{equation} \label{eq: meas ent dis}
\begin{split}
     \sigma_{\rm m}[\mathbf{D}, \mathbf{a}] = -\ln \frac{P_{\rm m, B}[\bar{\mathbf{D}}|\bar{\mathbf{a}}]}{P_{\rm m} [\mathbf{D}| \mathbf{a}]} =& 2 \lambda \delta t \sum_{n = 0}^N \left(\left(
    \frac{D_{n} - e^{-\gamma \delta t} D_{n + 1}}{\gamma \delta t}- a_{n} \right)^2 - \left(
    \frac{D_{n+1} - e^{-\gamma \delta t} D_{n}}{\gamma \delta t}- a_{n} \right)^2 \right) - \ln \frac{P_{\rm ini, B}[D_{N+1}|a_N]}{P_{\rm ini}[D_0|a_0]} \\
    = &2 \lambda  \delta t \sum_{n=0}^N \left( \left(\frac{\gamma \delta t (D_{n+1} - D_{n}) - 2 (D_{n+1} - D_n)}{\gamma \delta t} \right) \left(\frac{\gamma \delta t (D_{n+1} + D_n)}{\gamma \delta t} -2 a_n \right) \right) - \ln \frac{P_{\rm ini, B}[D_{N+1}|a_N]}{P_{\rm ini}[D_0|a_0]} \\
    =&  \frac{4 \lambda}{\gamma} \sum_{n=0}^N \left( \left(2a_n - (D_{n+1} + D_n) \left(D_{n+1} - D_n \right) \right) \right) - \ln \frac{P_{\rm ini, B}[D_{N+1}|a_N]}{P_{\rm ini}[D_0|a_0]},
\end{split}
\end{equation}
\end{widetext}
where we inserted $e^{-\gamma \delta t} \simeq 1- \gamma \delta t$ to obtain the third equality and applied the limit $\delta t \to 0$ to obtain the fourth equality.
This is a time-discrete counterpart to Eq.~\eqref{eq: meas ent for}. For a continuous trajectory of the system, the above expression can be written in the following integral form:
\begin{equation}
    \sigma_{\rm m} = \frac{8 \lambda}{\gamma} \left( \int_0^\tau a_t dD_t - \int_0^\tau D_t \circ dD_t \right) -  \ln \frac{P_{\rm ini, B}[D_\tau|a_\tau]}{P_{\rm ini}[D_0|a_0]},
\end{equation}
where $dD_t$ given in Eq.~\eqref{eq: dD a}, and $ \circ dD_t$ denotes the Stratonovich integration:
\begin{equation}
    \circ dD_t = \gamma(a_t - D_t)dt + \frac{\gamma}{2\sqrt{\lambda}} \circ dW
\end{equation}
Converting into the It\^o integral leads to the expression for $\sigma_{\rm m}$ in Eq.~\eqref{eq: meas ent for}.

Next, to conclude the derivation of Eq.~\eqref{eq: FT der}, we show that
\begin{equation} \label{eq: crooks our}
    \frac{ P_{\rm B}[\bar{\mathbf{a}}|\mathbf{\Lambda}(\bar{\mathbf{D}})]}{P[\mathbf{a}|\mathbf{\Lambda}(\mathbf{D})]}  = e^{-\sigma[\mathbf{a}, \mathbf{D}]}.
\end{equation}
In our backward experiment, where measurement and feedback is applied in the same way as in the forward experiment,
\begin{equation} \label{eq:backrel}
    P_{\rm B}[\bar{\mathbf{a}}|\mathbf{\Lambda}(\bar{\mathbf{D}})] = P_{\rm B}[\bar{\mathbf{a}}|\bar{\mathbf{\Lambda}}(\mathbf{D})],
\end{equation}
which is a consequence of the relation
$\mathbf{\Lambda}(\bar{\mathbf{D}}) = \bar{\mathbf{\Lambda}}(\mathbf{D})$.
It means that applying the driving protocol with the reversed trajectory of measurement outcomes, $\mathbf{\Lambda}(\bar{\mathbf{D}})$, is equivalent to time reversing the driving protocol, where $\bar{\mathbf{\Lambda}} = (\bar{\Lambda}_1, ..., \bar{\Lambda}_N)$ and $\bar{\Lambda}_n = \Lambda_{N+1-n}$.
To show that $\mathbf{\Lambda}(\bar{\mathbf{D}}) = \bar{\mathbf{\Lambda}}(\mathbf{D})$, we recall that
in the formulation of QFPME-based feedback control, the n'th feedback step is realized with the protocol corresponding to the measurement outcome $D_n$, i.e., $\Lambda_n(\mathbf{D}) = \Lambda(D_n)$. Taking the reversed trajectory $\bar{\mathbf{D}}$ as the argument of the protocol results in $\Lambda_n(\bar{\mathbf{D}}) = \Lambda(\bar{D}_n) = \Lambda(D_{N+1-n}) = \Lambda_{N+1-n}(\mathbf{D}) $, where we used $\bar{D}_n = D_{N-1+n}$ in the second equality and $\bar{\Lambda}_n = \Lambda_{N-1+n}$ in the fourth equality.
The stochastic entropy production $\sigma$ is given by~\cite{Crooks_1999, Crooks_2000, Seifert_2005, Potts2018, Sagawa2012}
\begin{equation} \label{eq: crooks}         \frac{P_{\rm B}[\bar{\mathbf{a}}|\bar{\mathbf{\Lambda}}(\mathbf{D})]}{P[\mathbf{a}|\mathbf{\Lambda}(\mathbf{D})]} = e^{-\sigma[\mathbf{a}, \mathbf{D}]}.
\end{equation}
Inserting Eq.~\eqref{eq:backrel} leads to Eq.~\eqref{eq: crooks our}.
We note that in the frameworks of Refs.~\cite{Potts2018, Horowitz2010, Sagawa2012}, the relation $\mathbf{\Lambda}(\bar{\mathbf{D}}) = \bar{\mathbf{\Lambda}}(\mathbf{D})$ and, as a consequence,  Eqs.~\eqref{eq:backrel} and~\eqref{eq: crooks our} do not hold.

Finally, since both $P[\mathbf{D}|\mathbf{a}]$ and $P_{\rm B}[\bar{\mathbf{D}}|\bar{\mathbf{a}}]$ are normalized the integral FT follows from the detailed FT in Eq.~\eqref{eq: FT detailed}.

\subsection{The average measurement entropy in Eq.~\eqref{eq: meas ent avg}} \label{app: meas ent avg}

Here, we show how to obtain $\erw{\sigma_{\rm m}}$ in Eq.~\eqref{eq: meas ent avg} from $\sigma_{\rm m}$ in Eq.~\eqref{eq: meas ent for} using both time-continuous and time-discrete descriptions.

We insert $dD_t$ in Eq.~\eqref{eq: dD a} into $\sigma_{\rm m}$ in Eq.~\eqref{eq: meas ent for}, resulting in
\begin{equation}
\begin{split}
     \sigma_{\rm m}  =& \frac{8 \lambda}{\gamma} \int_0^\tau \gamma (a_t-D_t)^2 dt + \frac{8 \lambda}{\gamma} \int_0^\tau \frac{\gamma}{2 \sqrt{\lambda}} (a_t-D_t) dW  \\
     &-\gamma \tau -  \ln \frac{P_{\rm ini, B}[D_\tau|a_\tau]}{P_{\rm ini}[D_0|a_0]}.
\end{split}
\end{equation} 
Using the relation $\erw{  \int_0^\tau (a_t-D_t) dW} =0$, we obtain the rate of change of $\erw{\sigma_{\rm m}}$ in Eq.~\eqref{eq: meas ent avg} in the main text, for convenience reproduced here:
\begin{equation}
    \partial_t \langle \sigma_{\rm m} \rangle =  8 \lambda  \left( \left\langle (D_t - a_t)^2 \right\rangle - \frac{\gamma}{8\lambda} \right).
\end{equation}

To get an insight into this expressions, let us suppose that $a_t = a$ is constant. Then, the probability distribution of $D_t$ follows the standard Fokker-Planck equation and after a sufficient time period, is well approximated by
\begin{equation} \label{eq: fast det lim}
    P[D|a] \simeq \pi_a(D) :=  \sqrt{4 \lambda/( \pi \gamma )} e^{-\frac{4 \lambda}{\gamma} (D-a)^2 },
\end{equation}
which is a Gaussian distribution with the average $a$ and variance $\gamma/(8 \lambda)$. Therefore, $\partial_t \langle \sigma_{\rm m} \rangle$ vanishes when the system's state does not change, and the measurement entropy is only affected by updates of the systems's state.

The expression for the rate of change of $\erw{\sigma_{\rm m}}$ in Eq.~\eqref{eq: meas ent avg} can be also found by using the time-discrete notation. To this end, we rewrite $\sigma_{\rm m}$ in Eq.~\eqref{eq: meas ent dis} as
\begin{equation} \label{eq: meas ent alt}
\begin{split}
    &\sigma_{\rm m} = -  \ln \frac{P_{\rm ini, B}[D_{N+1}|a_N]}{P_{\rm ini}[D_0|a_0]} + \\ & \frac{4\lambda}{\gamma} \left( -  (D_{N+1}^2 - D_0^2)  + (2 D_{N+1} a_{N} - 2 D_0 a_0 ) +  \sum_{n=1}^{N} 2 D_n (a_{n-1} - a_{n} ) \right) ,
\end{split}  
\end{equation}
which consists of terms proportional to either $D_n$ or $D_n^2$. 
To evaluate $\langle \sigma_{\rm m} \rangle$, we integrate in the following order: $D_{N+1} \rightarrow a_{N} \rightarrow D_{N} \rightarrow ... \rightarrow a_0 \rightarrow D_0$. In the sum in Eq.~\eqref{eq: meas ent alt}, for each term $\bullet$ involving $D_n$ or $D_n^2$,  
the integral over each $D_{n \geq 1}$ is evaluated with respect to the probability
\begin{equation}
    \left( \frac{2 \lambda \delta t}{\pi} \right)^{1/2}  \frac{1}{\gamma \delta t} \int  dD_n \bullet e^{-2 \lambda \delta t \left(
    \frac{D_n - e^{-\gamma \delta t} D_{n-1}}{\gamma \delta t}- a_{n-1} \right)^2 } =: \mathcal{I}_n[\bullet] ,
\end{equation}
which can be seen from Eq.~\eqref{eq: D gaussians}.
For terms proportional to $D_n$ and to $D_n^2$, we have
\begin{equation}
\begin{split}
    \mathcal{I}_n[D_n] &= \gamma \delta t \left(
    \frac{e^{-\gamma \delta t} D_{n-1}}{\gamma \delta t} + a_{n-1} \right) = e^{-\gamma \delta t} D_{n-1} + \gamma \delta t  a_{n-1} \\
    &= (1 - \gamma \delta t) D_{n-1} + \gamma \delta t a_{n-1},
\end{split}
\end{equation}
and
\begin{equation}
\begin{split}
    \mathcal{I}_n[D_n^2] &= (\gamma \delta t)^2 \left(
    \frac{e^{-\gamma \delta t} D_{n-1}}{\gamma \delta t} + a_{n-1} \right)^2  +  \frac{(\gamma \delta t)^2}{4 \lambda \delta t} \\
     &= (D_{n-1}^2 - 2 \gamma \delta t  D_{n-1}^2 + 2 \gamma \delta t D_{n-1} a_{n-1}) + \frac{\gamma^2 \delta t}{4 \lambda},
\end{split}
\end{equation}
respectively, where we kept terms up to the first order in $\delta t$.
Applying these calculations to Eq.~\eqref{eq: meas ent alt} leads to
\begin{equation} \label{eq: recursive}
\begin{split}
    - \mathcal{I}_n[D_n^2] +  \mathcal{I}_n[2 D_n a_{n-1} ]  =& 
     - D_{n-1}^2 + 2  D_{n-1} a_{n-1}   - \frac{\gamma^2 \delta t}{4 \lambda} \\
     &+ 2 \gamma \delta t (D_{n-1} - a_{n-1})^2 .
\end{split}
\end{equation}
We now combine the first two elements in the equation above with $2 (a_{n-2} - a_{n-1} ) D_{n-1}$, resulting in
\begin{equation}
     2(a_{n-2} - a_{n-1} ) D_{n-1} - D_{n-1}^2 +2 a_{n-1} D_{n-1} = 
     - D_{n-1}^2 + 2 a_{n-2} D_{n-1},
\end{equation}
which leads to a recursive scheme when we repeatedly apply this reasoning to the sum in Eq.~\eqref{eq: meas ent alt} starting from $n = 1$. At the end, the first two terms in Eq.~\eqref{eq: recursive} for $n = N+1$ are $-D_0^2 + 2a_0 D_0$, which cancel out in Eq.~\eqref{eq: meas ent alt}. Therefore, the only remaining elements are the last two terms in Eq.~\eqref{eq: recursive}, and in the limit $\delta t \to 0$, we obtain
\begin{equation}
    \langle \sigma_{\rm m} \rangle = -\gamma \tau + 8 \lambda \delta t \sum_{n = 0}^N \langle (D_{n} - a_{n})^2 \rangle - \left\langle \ln \frac{P_{\rm ini, B}[D_{N+1}|a_N]}{P_{\rm ini}[D_0|a_0]} \right \rangle,
\end{equation}
which is equivalent to Eq.~\eqref{eq: meas ent avg}.

\subsection{Coarse-graining} \label{app:coarse-graining}

We have the marginal probability distributions $ P_{\rm B}[ \bar{\mathbf{D}}] = \int d\bar{ \mathbf {a}} P_{\rm B}[\bar{\mathbf{a}}, \bar{\mathbf{D}}]$ and  $ P[ \mathbf{D}] = \int d \mathbf {a} P[\mathbf{a}, \mathbf{D}]$.
Our starting point is Eq.~\eqref{eq: FT detailed}:
\begin{equation}
    P_{\rm B}[\bar{\mathbf{a}}, \bar{\mathbf{D}}]  = P[\mathbf{a}, \mathbf{D}] e^{-\sigma[\mathbf{a}, \mathbf{D}] - \sigma_{\rm m}[\mathbf{a}, \mathbf{D}]}.
\end{equation}
Inserting $P[\mathbf{a} | \mathbf{D}] = P[\mathbf{a}, \mathbf{D}] / P[\mathbf{D}]$ results in
\begin{equation}
    P_{\rm B}[\bar{\mathbf{a}}, \bar{\mathbf{D}}]  = P[\mathbf{D}] P[\mathbf{a}| \mathbf{D}] e^{-\sigma[\mathbf{a}, \mathbf{D}] - \sigma_{\rm m}[\mathbf{a}, \mathbf{D}]}.
\end{equation}
Upon integrating out the system' trajectory we obtain 
\begin{equation}
    P_{\rm B}[\bar{\mathbf{D}}]  = P[\mathbf{D}] \int d \mathbf{a}P[\mathbf{a}| \mathbf{D}] e^{-\sigma[\mathbf{a}, \mathbf{D}] - \sigma_{\rm m}[\mathbf{a}, \mathbf{D}]} = P[\mathbf{D}] e^{-\Sigma[\mathbf{D}]},
\end{equation}
where we used the definition of the coarse-grained $\Sigma[\mathbf{D}]$ in Eq.~\eqref{eq: ent cg}, which concludes the derivation of the detailed FT in Eq.~\eqref{eq: FT cg}. Further integrating of the detector's trajectory leads to the integral FT in Eq.~\eqref{eq: FT cg}. 

\section{Classical two-level system under bang-bang control: FTs and the second law}

For our general results, we denoted the measured observable as $\hat{A}=\sum_a a|a\rangle\langle a|$. Since in our example, $\hat{A}=|1\rangle\langle 1|- |0\rangle\langle 0|$, we will identify $|-1\rangle\equiv |0\rangle$ and similarly $x_{-1}\equiv x_{0}$ whenever the subscript denotes the state.

\subsection{Equation~\eqref{eq: meas ent fast} in the fast-detector limit} \label{app:fast_det}

Let us introduce a vector of probabilities $\vec{p} = [p_{-1}(D), p_1( D)]^T$, where $p_a(D)$ is the joint probability of $a$ and $D$ in the steady state. When the detector evolves much faster that the system ($\gamma \gg \kappa$), we can approximate~\cite{Annby2022}
\begin{equation}
    \vec{p} = \frac{1}{2} \left( [\pi_{-1}(D), \pi_{1}(D)]^T - \gamma^{-1} F^+ W(D) [\pi_{-1}(D), \pi_{1}(D)]^T \right),
\end{equation}
where the conditional probability of $D$ given $a$ reads $\pi_a(D) = \sqrt{4 \lambda/( \pi \gamma )} e^{-\frac{4 \lambda}{\gamma} (D-a)^2 }$, which was introduced in Eq.~\eqref{eq: fast det lim}, and the prefactor $1/2$ comes from the fact the system is equally likely to occupy both states. Here,
\begin{equation}
    W(D) = \theta(D) \begin{bmatrix}
-\kappa (1+n_{\rm B}) & \kappa n_{\rm B}\\
\kappa (1+n_{\rm B}) & -\kappa n_{\rm B}
\end{bmatrix} 
+ (1- \theta(D)) 
\begin{bmatrix}
-\kappa n_{\rm B} & \kappa (1+n_{\rm B})\\
\kappa n_{\rm B} & -\kappa (1+n_{\rm B}) 
\end{bmatrix},
\end{equation}
is a $D$-dependent transition matrix obtained from the Liouvillian in Eq.~\eqref{eq:LiouClas},
and $F^+ = \begin{bmatrix}
F^+_{-1} & 0\\
0 & F^+_{1} 
\end{bmatrix}$
is a diagonal matrix of Drazin inverses $F_a^+$ of the differential operator $F_a \bullet = \partial_D(D-a) \bullet + \frac{\gamma}{8 \lambda} \bullet$, which is a classical countertpart of the superoperator $\mc{J}$ introduced in Sec.~\ref{sec:ssClas}. 
We use the vector $\vec{p}$ to evaluate the expectation value in Eq.~\eqref{eq: meas ent avg}:
\begin{equation}
   \partial_t \langle \sigma_{\rm m} \rangle = -\gamma + 8 \lambda \langle (D - a)^2 \rangle =  -\gamma + 8 \lambda \sum_a \int_{-\infty}^\infty dD (D - a)^2 p_a(D) .
\end{equation}
The first term in $\vec{p}$ consisting of just $\pi_a(D)$ contributes a constant value,
\begin{equation}
    \frac{1}{2} \left(\int_{-\infty}^\infty dD (D+1)^2 \pi_0(D) + \int_{-\infty}^\infty dD (D-1)^2 \pi_1(D) \right) = \frac{\gamma}{8 \lambda},
\end{equation}
which cancels out with $-\gamma$.
The next term in $\vec{p}$ yields 
\begin{equation} \label{eq:fdl_temp1}
    \partial_t \langle \sigma_{\rm m} \rangle = 
        -  \frac{8\lambda}{\gamma} \frac{1}{2} \sum_a \left( \int_{-\infty}^\infty dD (D-a)^2 F^+_{a} v_{a}(D)   \right) ,
\end{equation}
where $[v_{-1}(D), v_1(D)]^T = W(D) [\pi_{-1}(D), \pi_1(D)]^T $.
To make progress, we use the generalized Hermite polynomials $\hex{n}{D-a}$, which were introduced in Eq.~\eqref{eq:He}. For a function $f(D)$ and $n \geq 1$, they satisfy the relation~\cite{Annby2022}
\begin{equation} \label{eq:fdl_temp2}
     \int_{-\infty}^\infty dD \hex{n}{D-a} F^+_a f(D) = -\frac{1}{n} \int dD He_n(D-a) f(D),
\end{equation}
and for $n = 0$, the left-hand side is $0$.
Since $\hex{0}{D-a} = 1$ and $\hex{2}{D-a} = (D-a)^2 + \gamma/(8 \lambda)$, we find
\begin{equation}
    (D-a)^2 = \hex{2}{D-a} + \frac{\gamma}{8 \lambda} \hex{0}{D-a} .
\end{equation}
Inserting this relation into Eq.~\eqref{eq:fdl_temp1} and applying Eq.~\eqref{eq:fdl_temp2} leads to
\begin{equation} 
\begin{split}
   \partial_t \langle \sigma_{\rm m} \rangle =& 
         \frac{ 4\lambda }{\gamma} \sum_a \left( \int_\infty^\infty dD \hex{2}{D-a}  v_{a}(D) \right) \\
         =&
          \frac{ 4 \lambda }{\gamma}  \int_{-\infty}^\infty dD  \left(D^2 +1 - \frac{\gamma}{8\lambda} \right) \left[ v_{-1}(D) + v_1(D) \right] \\
              \\
          & + \frac{ 4 \lambda }{\gamma}   \int_{-\infty}^\infty dD  (2D) \left[ v_{-1}(D) - v_1(D) \right]  \\
           = &
           \frac{ 8 \lambda }{\gamma} \left[\kappa (1+2n_{\rm B}) - \kappa \text{ Erf} \left(2 \sqrt{\lambda/\gamma} \right) -  \frac{\kappa e^{-4\lambda/\gamma}}{2 \sqrt{\pi}\sqrt{\lambda/\gamma} }  \right],
\end{split}
\end{equation}
where we recall that $\text{Erf}(\bullet)$
is the error function. We thus recover Eq.~\eqref{eq: meas ent fast}.

\subsection{FT for $m$ in Eq.~\eqref{eq:FTm}} \label{app:FTm}

The trajectories $\mathbf{a}$ and $\mathbf{D}$ uniquely determine $m[\mathbf{a}, \mathbf{D}]$, and, therefore, their joint probability is given by
\begin{equation}
    P[\mathbf{D}, \mathbf{a}, m] = \delta(m[\mathbf{D}, \mathbf{a}] - m) P[\mathbf{D}, \mathbf{a}].
\end{equation}
From the detailed FT in Eq.~\eqref{eq: FT detailed} and the relation $m[\bar{\mathbf{a}}, \bar{\mathbf{D}}] = -m[\mathbf{a}, \mathbf{D}]$, it follows that (here we drop the subscript "B")
\begin{equation}
P[\bar{\mathbf{D}}, \bar{\mathbf{a}}, -m] = e^{-\sigma[\mathbf{D}, \mathbf{a}] - \sigma_{\rm m}[\mathbf{D}, \mathbf{a}] } P[\mathbf{D}, \mathbf{a}, m].
\end{equation}
We define the joint probability of $m$ and $\mathbf{a}$, $P[\mathbf{a}, m] = \int d \mathbf{D} P[\mathbf{a}, \mathbf{D}, m]$, and find
\begin{equation} \label{eq:intcq}
P[\bar{\mathbf{a}}, -m] = P[\mathbf{a}, m] \int d \mathbf{D} e^{-\sigma[\mathbf{a}, \mathbf{D}] - \sigma_{\rm m}[\mathbf{a}, \mathbf{D}]} P[\mathbf{D}|\mathbf{a}, m],
\end{equation}
where the conditional probability $P[\mathbf{D}|\mathbf{a}, m] = P[\mathbf{D}, \mathbf{a}, m]/P[\mathbf{a}, m]$ can be calculated as
\begin{equation} \label{eq:intcon}
    P[\mathbf{D}|m, \mathbf{a}] = \frac{P[\mathbf{D}, m|\mathbf{a}]}{\int dD P[\mathbf{D}, m|\mathbf{a}]},
\end{equation}
with $P[\mathbf{D}, m|\mathbf{a}] = \delta(m[\mathbf{D}, \mathbf{a}] - m) P[\mathbf{D}|\mathbf{a}]$.
The stochastic entropy production in Eq.~\eqref{eq:intcq} is given by $e^{-\sigma[\mathbf{a}, \mathbf{D}]} = e^{m \omega/(k_{\rm B}T)}$, which does not explicitly depend on $\mathbf{D}$ and, for this reason, can be taken in front of the integral.

In the long-time limit, we can approximate
\begin{equation} \label{eq:prod_fast_det}
    e^{-\sigma_{\rm m}[\mathbf{D}, \mathbf{a}]} = \prod_{n=1}^N e^{-\frac{8\lambda}{\gamma} D_n (a_{n-1} - a_n)} ,
\end{equation}
which can be seen from Eq.~\eqref{eq: meas ent alt}. In the long time-limit, the net number of extracted quanta corresponds to the net number of energy quanta flowing from the reservoir into the system and can be decomposed into $m = m_+ - m_-$,  where $m_+$ and $m_-$ correspond to a total number of quanta flowing into and out of the system. We assume that the total number of jumps, $M = m_+ + m_-$, is small comparing to the number of steps $N$ and that all jumps are significantly separated in time. In each step $n$, when no jump occurs ($a_{n-1} - a_n = 0$), no heat exchange happens, and the corresponding element in the product in Eq.~\eqref{eq:prod_fast_det} is simply $1$ regardless of the value of $D_n$.
When a jump takes place, $a_{n-1} - a_n = 2 a_{n-1} = \pm 2$. If the signs of $a_{n-1}$ and $D_n$ are the same, the increment of entropy production is $-\omega/(k_{\rm B}T)$, and the jump contributes to $m_+$ (the system transitions from the ground to the excited state). But if they are different, the increment of entropy production is $\omega/(k_{\rm B}T)$, and the jump contributes to $m_-$ (the system transitions from the excited to the ground state). Since consecutive jumps happen in distant steps $n$, and the detector is much faster than the rate of jumps, for each step with a jump, we can approximate $P[D_n|a_{n-1}] = \sqrt{\frac{4 \lambda}{\pi \gamma} }   e^{-\frac{4 \lambda}{\gamma} (D_n-a_{n-1})^2}$ from Eq.~\eqref{eq: fast det lim}.
Therefore, we find 
\begin{equation} \label{eq:int1}
    \int d\mathbf{D} P[\mathbf{D}, m| \mathbf{a}] = \binom{M}{m_+} (1-\eta)^{m_+} \eta^{m_-},
\end{equation}
where, for instance, if the jump from $a_{n-1} = 1$ to $a_{n} = -1$ contributes to $m_-$, the corresponding term is $\int_{-\infty}^0 dD_n P[D_n|1] = \left[1 - \text{Erf} \left(2 \sqrt{\lambda/\gamma} \right) \right]/2 = \eta$, and the prefactor accounts for the number of combinations that the steps contributing to $m_+$ and $m_-$ can be arranged in.

Now, the product of the probability and the measurement entropy for the step with a jump from $a_{n-1}$ to $a_n = - a_{n-1}$ is given by $\sqrt{\frac{4 \lambda}{\pi \gamma} } e^{-\frac{4 \lambda}{\gamma} (D_n-a_{n-1})^2} e^{-\frac{8\lambda}{\gamma} D_n (a_{n-1} - a_n)} = \sqrt{\frac{4 \lambda}{\pi \gamma} } e^{-\frac{4 \lambda}{\gamma} (D_n+a_{n-1})^2}$, which equals $P[D_n|-a_{n-1}]$. Therefore, we find
\begin{equation} \label{eq:int2}
    \int d\mathbf{D} e^{-\sigma_{\rm m}[\mathbf{D}, \mathbf{a}]}P[\mathbf{D}, m| \mathbf{a}] = \binom{M}{m_+} \eta^{m_+} (1-\eta)^{m_-},
\end{equation}
since the roles of $\eta$ and $1-\eta$ are interchanged with respect to Eq.~\eqref{eq:int1}.
Combining Eqs.~\eqref{eq:intcq},~\eqref{eq:intcon},~\eqref{eq:int1}, and~\eqref{eq:int2} leads to
\begin{equation}
\begin{split}
    P[\bar{\mathbf{a}}, -m] =& P[\mathbf{a}, m] e^{m \omega/(k_{\rm B} T) } \frac{\binom{M}{m_+} \eta^{m_+} (1-\eta)^{m_-}}{\binom{M}{m_+} (1-\eta)^{m_+} \eta^{m_-}} \\
    =& P[\mathbf{a}, m] e^{m \omega/(k_{\rm B} T) } \left(\frac{\eta}{1-\eta} \right)^{m_+ - m_-} \\
    =&  P[\mathbf{a}, m] e^{m\left[\omega/(k_{\rm B}T) - \log{\left(\frac{1-\eta}{\eta} \right)} \right]} .
\end{split}
\end{equation}
Upon integrating out $\mathbf{a}$ from both sides, we obtain the FT for $m$ in Eq.~\eqref{eq:FTm}.

\section{Fluctuation theorems for
quantum systems} \label{app: quantum FT and SL}

\subsection{Forward trajectory}

\begin{widetext}
At the time $t = 0$, the system is prepared in the initial state
    $\op{\rho}_{\rm ini} = \sum_{v_{\rm i}} p_{v_{\rm i}}  |v_{\rm i} \rangle \langle v_{\rm i} |  $.
The probability of the trajectory of the measurement outcomes, $\mathbf{D} = (D_0, D_1, ..., D_{N+1})$, is given by~\cite{Annby2022}
\begin{equation}
    P[\mathbf{D}] =   P_{\rm ini}[D_0] \left(\frac{1}{\gamma \delta t} \right)^{N+1}  \text{Tr}\left\{ \mathcal{K}\left(\frac{D_{N+1} - e^{-\gamma 
    \delta t} D_{N} }{\gamma \delta t} \right) e^{\mathcal{L}(D_{N})\delta t} ... e^{\mathcal{L}(D_1)\delta t} \mathcal{K}\left(\frac{D_1 - e^{-\gamma 
    \delta t} D_{0} }{\gamma \delta t} \right) \op{\rho}_{\rm ini} \right\}  ,
\end{equation}
where $P_{\rm ini}$ is the initial distribution of the measurement outcome, and the measurement operator of the continuous Gaussian measurement is defined as $\mc{K}(z) \op{\rho} = K(z) \op{\rho} K(z)$, with the Hermitian operator $K(z) =  \left( \frac{2 \lambda \delta t}{\pi} \right)^{1/4} e^{-\lambda \delta t (z - \op{A})^2}$~\cite{Jacobs2006rev, Bednorz2012}. The superoperator $e^{\mathcal{L}(D_{n})\delta t}$ realizes a feedback step. Now, we insert the identity operator $\op{I} = \sum_{a_n} \proj{a_n} $ after the feedback superoperator corresponding to $D_n$ in two branches, which we here denote with $L$ and $R$, defining the Keldysh trajectory~\cite{Hofer2017, Nazarov2003, Hofer2016, Clerk2011}.
The quasi-probability of the trajectory reads
\begin{equation}
\begin{split}
    P[\mathbf{a}^L, \mathbf{a}^R, \mathbf{D}] =&   \left( \frac{2 \lambda \delta t}{\pi} \right)^{\frac{N+1}{2}} \left( \frac{1}{\gamma \delta t } \right)^{N+1} P_{\rm ini}[D_0]  \mathcal{A}( \mathbf{a}^L, \mathbf{a}^R, \mathbf{D} )  \\
    & \exp{ \left[ - \lambda \delta t \sum_{n=0}^N \left( \frac{D_{n+1} - e^{-\gamma \delta t} D_{n} }{\gamma \delta t} -a_{n}^L\right)^2 \right] } \exp{ \left[ - \lambda \delta t \sum_{n=0}^N \left( \frac{D_{n+1} - e^{-\gamma \delta t} D_{n} }{\gamma \delta t} -a_{n}^R\right)^2 \right] } ,
\end{split} 
\end{equation}
where
\begin{equation}
    \mathcal{A}(\mathbf{a}^L, \mathbf{a}^R, \mathbf{D}) = \langle a_{N}^L| e^{\mathcal{L}(D_{N}) \delta t } \left[ |a_{N-1}^L \rangle ... |a_1^L \rangle \langle a_1^L| e^{\mathcal{L}(D_1)  \delta t } \left[ |a_0^L \rangle \langle a_0^R| \right] |a_1^R \rangle \langle a_1^R | ... \langle a_{N-1}^R | \right] | a_{N}^R \rangle \times \delta_{a^L_{N}, a^R_{N}} \times \bra{a_0^L} \op{\rho}_{\rm ini} \ket{a_0^R} 
\end{equation}
and $\delta_{a^L_{N}, a^R_{N}} $ is the Kronecker delta.
We further introduce the quantum-jump unraveling~\cite{Manzano_2015, Manzano_2018, Manzano_2022, Horowitz_2013}. Each term $e^{\mathcal{L}(D_n) \delta t}$ can be expanded as
\begin{equation}
    e^{\mathcal{L}(D_n) \delta t} |a^L_{n-1} \rangle \langle a^R_{n-1} | =  
     e^{-i \delta t \op{H}_{\rm eff}} |a^L_{n-1} \rangle \langle a^R_{n-1} | e^{i \delta t \op{H}_{\rm eff}^\dagger}
    + \delta t \sum_k \op{L}_k  |a^L_{n-1} \rangle \langle a^R_{n-1} | \op{L}_k^\dagger  =: \mathcal{J}_0  |a^L_{n-1} \rangle \langle a^R_{n-1} | + \sum_k \mathcal{J}_k  |a^L_{n-1} \rangle \langle a^R_{n-1} | + ,
\end{equation}
where we used superoperators for convenience, with $\mathcal{J}_k$ corresponding to the quantum jump $k$ and $\mathcal{J}_0$ describing the non-Hermitian evolution with the effective Hamilonian $\op{H}_{\rm eff}= \op{H}_{\rm eff}(D_n) = \op{H}(D_n) -\frac{i}{2} \sum_k \op{L}^\dagger_k(D_n) \op{L}_k(D_n)$. The unraveling trajectory $\Gamma$ encompasses monitoring whether and if yes, which jump occurred in each time step, as well as the outcomes of the projective measurements at the initial time $0$ and the final time $\tau$, which is known as the two-point measurement scheme. In the two-point measurement scheme, we project onto $\proj{v_{\rm i}}$, the eigenstates of the initial state $\op{\rho}_{\rm ini}$, at the beginning. At the end we project onto $\proj{v_{\rm f}}$, the eigenstates of the final state of the system average over all measurement outcomes, 
$
    \op{\rho}_{\rm fin} = \sum_{v_{\rm f}} p_{v_{\rm f}}  |v_{\rm f} \rangle \langle v_{\rm f} |  
$.
Then, the quasi-probability distribution of the joint trajectory of the Keldysh path, unraveling, and detector is given by 
\begin{equation} \label{eq: prob quant}
\begin{split}
    P[\Gamma, \mathbf{a}^L, \mathbf{a}^R, \mathbf{D}] =&   \left( \frac{2 \lambda \delta t}{\pi} \right)^{\frac{N+1}{2}} \left( \frac{1}{\gamma \delta t } \right)^{N+1} P_{\rm ini}[D_0]  \mathcal{A}(\Gamma, \mathbf{a}^L, \mathbf{a}^R, \mathbf{D} )  \\
    & \exp{ \left[ - \lambda \delta t \sum_{n=0}^N \left( \frac{D_{n+1} - e^{-\gamma \delta t} D_{n} }{\gamma \delta t} -a_{n}^L\right)^2 \right] } \exp{ \left[ - \lambda \delta t \sum_{n=0}^N \left( \frac{D_{n+1} - e^{-\gamma \delta t} D_{n} }{\gamma \delta t} -a_{n}^R\right)^2 \right] } ,
\end{split} 
\end{equation}
where
\begin{align} \label{eq: A quant}
    \mathcal{A}(\Gamma, \mathbf{a}^L, \mathbf{a}^R, \mathbf{D} ) = p_{v_{\rm i}} \langle a_{N}^L| \mathcal{J}_{k_{N}}(D_{N}) \left[ |a_{N-1}^L \rangle ... |a_1^L \rangle \langle a_1^L| \mathcal{J}_{k_{1}}(D_{1}) \left[ |a_0^L \rangle \langle a_0^R| \right] |a_1^R \rangle \langle a_1^R | ... \langle a_{N-1}^R | \right] | a_{N}^R \rangle \times  \langle a_0^L |v_{\rm i} \rangle \langle v_{\rm i} | a^R_0 \rangle \times  \langle v_{\rm f} | a^L_{N} \rangle \langle a_{N}^R |v_{\rm f} \rangle,
\end{align}
with $k_n$ denoting the jump or non-Hermitian evolution governed by the measurement outcome $D_n$.
Next, we perform the Keldysh rotation, introducing the classical fields $a_n^{\rm c} = (a_n^L + a_n^R)/2$ and quantum fields $a_n^{\rm q} = (a_n^L - a_n^R)/2$. As a result, two last terms in Eq.~\eqref{eq: prob quant} can we expressed as
\begin{equation} \label{eq: Keldysh rotation}
\begin{split}
    &\exp{ \left[ - \lambda \delta t \sum_{n=0}^N \left( \frac{D_{n+1} - e^{-\gamma \delta t} D_{n} }{\gamma \delta t} -a_{n}^L\right)^2 \right] } \exp{ \left[ - \lambda \delta t \sum_{n=0}^N \left( \frac{D_{n+1} - e^{-\gamma \delta t} D_{n} }{\gamma \delta t} -a_{n}^R\right)^2 \right] } \\
    &= \exp{ \left[ - 2\lambda \delta t \sum_{n=0}^N \left( \frac{D_{n+1} - e^{-\gamma \delta t} D_{n} }{\gamma \delta t} -a_{n}^{\rm c}\right)^2 \right] } \exp{ \left[ - 2\lambda \delta t \sum_{n=0}^N \left(a_{n}^{\rm q} \right)^2 \right] } .
\end{split}
\end{equation}
We can notice that the classical fields $a_n^{\rm c}$ play the same role as the system's states $a_n$ in the probability in Eq.~\eqref{eq: D gaussians} for the classical systems. The quantum fields $a_n^{\rm q}$, on the other hand, introduce dephasing, which grows in strength with the distance $|a_n^L - a_n^R|$ between the two Keldysh branches.

\subsection{Backward trajectory}
In the backward experiment, signified by subscript "B", we perform measurement and feedback just like in the forward one, but the measurement operator is $\sum_a a \proj{\tilde{a}}$, where $\ket{\tilde{a}} = \Theta \ket{a}$, with $\Theta$ denoting the anti-Hermitian time-reversal operator, and we prepare the initial quantum state $\op{\rho}_{\rm ini, B} = \sum_{v_{\rm f}} p_{v_{\rm f}} \proj{\tilde{v}_{\rm f}} $. The time-reversed versions of the Keldysh and detector's trajectories are given by $\bar{a}^\ell_n = a^\ell_{N-n}$ (for $\ell = L, R$) and $\bar{D}_n = D_{N-n+1}$, respectively, where the Keldysh paths are defined with respect to the identity operators $\sum_a \proj{\tilde{a}}$. Under the Keldysh rotation, we have $\bar{a}_n^{\rm c} = a_{N-n}^{\rm c} $ and $\bar{a}_n^{\rm q} = a_{N-n}^{\rm q} $. As far as the time-reversed version of the quantum-jump unraveling, $\bar{\Gamma}$, we use the standard procedure~\cite{Manzano_2015, Manzano_2018, Manzano_2022, Horowitz_2013}: The outcomes of the two-point-measurement scheme are interchanged, with projectors onto $\proj{\tilde{v}_{\rm f}}$ at the initial time and onto $\proj{\tilde{v}_{\rm i}}$ at the final time.  Driving that implements feedback is performed under the time-reversal operator, i.e., the Hamiltonian and jump operators become $\Theta \op{H} \Theta^{-1}$  and $\Theta \op{L}_k \Theta^{-1}$, respectively.  For the effective non-Hermitian Hamiltonian, this results is $\Theta \op{H} \Theta^{-1} -\frac{i}{2} \sum_k \Theta \op{L}^\dagger_k \op{L}_k \Theta^{-1} = \Theta \op{H}_{\rm eff}^\dagger \Theta^{-1}$, where we used $\Theta i = -i \Theta$. 
The backward-trajectory jump superoperators are defined as (for $k \neq 0$)
\begin{equation} \label{eq: jump back k}
    \bar{\mc{J}}_{k_n} \op{\rho} = \Theta \op{L}_{k'_{N+1-n}} \Theta^{-1} \op{\rho} \Theta \op{L}_{k'_{N+1-n}}^\dagger \Theta^{-1} = e^{-\sigma_{k_{N+1-n}}} \Theta \op{L}^\dagger_{k_{N+1-n}} \Theta^{-1} \op{\rho} \Theta \op{L}_{k_{N+1-n}} \Theta^{-1},
\end{equation}
where for each $\op{L}_k$, we can uniquely identify $\op{L}_{k'} = e^{-\sigma_k/2} \op{L}_k^\dagger$, with $\sigma_k$ being the stochastic entropy increment due to dissipation, associated with the quantum jump $k$. For $k = 0$, we have 
\begin{equation} \label{eq: jump back k0}
    \bar{\mc{J}}_{0} \op{\rho} =  e^{-i \delta t \Theta \op{H}_{\rm eff}^\dagger \Theta^{-1}}  \op{\rho}  e^{+i \delta t \Theta \op{H}_{\rm eff} \Theta^{-1}} = \Theta e^{+i \delta t \op{H}_{\rm eff}^\dagger } \Theta^{-1}  \op{\rho}  \Theta e^{-i \delta t \op{H}_{\rm eff} } \Theta^{-1},
\end{equation}
The expression for the probability of the backward trajectory reads
\begin{equation} \label{eq:PB quant}
\begin{split}
     P_{\rm B}[\bar{\Gamma}, \bar{\mathbf{a}}^L, \bar{\mathbf{a}}^R, \bar{\mathbf{D}}] =&   \left( \frac{2 \lambda \delta t}{\pi} \right)^{\frac{N+1}{2}} \left( \frac{1}{\gamma \delta t } \right)^{N+1} P_{\rm ini, B}[\bar{D}_0]  \mathcal{A}_{\rm B}(\bar{\Gamma}, \bar{\mathbf{a}}^L, \bar{\mathbf{a}}^R, \bar{\mathbf{D}} )  \\
    & \exp{ \left[ - \lambda \delta t \sum_{n=0}^N \left( \frac{\bar{D}_{n+1} - e^{-\gamma \delta t} \bar{D}_{n} }{\gamma \delta t} -\bar{a}_{n}^L\right)^2 \right] } \exp{ \left[ - \lambda \delta t \sum_{n=0}^N \left( \frac{\bar{D}_{n+1} - e^{-\gamma \delta t} \bar{D}_{n} }{\gamma \delta t} -\bar{a}_{n}^R\right)^2 \right] } \\
    =&   \left( \frac{2 \lambda \delta t}{\pi} \right)^{\frac{N+1}{2}} \left( \frac{1}{\gamma \delta t } \right)^{N+1} P_{\rm ini, B}[D_{N+1}]  \mathcal{A}_{\rm B}(\bar{\Gamma}, \bar{\mathbf{a}}^L, \bar{\mathbf{a}}^R, \bar{\mathbf{D}} )  \\
    &  \exp{ \left[ - 2\lambda \delta t \sum_{n=0}^N \left( \frac{D_{n} - e^{-\gamma \delta t} D_{n+1} }{\gamma \delta t} -a_{n}^{\rm c}\right)^2 \right] } \exp{ \left[ - 2\lambda \delta t \sum_{n=0}^N \left(a_{n}^{\rm q} \right)^2 \right] }, \\
\end{split}
\end{equation}
where
\begin{equation} \label{eq:AB quant}
    \begin{split}
        \mathcal{A}_{\rm B}(\bar{\Gamma}, \bar{\mathbf{a}}^L, \bar{\mathbf{a}}^R, \bar{\mathbf{D}} ) &= p_{v_{\rm f}} \langle \bar{\tilde{a}}_{N}^L| \bar{\mathcal{J}}_{k_{N}}(\bar{D}_{N}) \left[ |\bar{\tilde{a}}_{N-1}^L \rangle ... |\bar{\tilde{a}}_1^L \rangle \langle \bar{\tilde{a}}_1^L| \bar{\mathcal{J}}_{k_{1}}(\bar{D}_{1}) \left[ |\bar{\tilde{a}}_0^L \rangle \langle \bar{\tilde{a}}_0^R| \right] |\bar{\tilde{a}}_1^R \rangle \langle \bar{\tilde{a}}_1^R | ... \langle \bar{\tilde{a}}_{N-1}^R | \right] | \bar{\tilde{a}}_{N}^R \rangle \times  \langle \bar{\tilde{a}}_0^L | \tilde{v}_{\rm f} \rangle \langle \tilde{v}_{\rm f} | \bar{\tilde{a}}^R_0 \rangle \times  \langle \tilde{v}_{\rm i} | \bar{\tilde{a}}^L_{N} \rangle \langle \bar{\tilde{a}}_{N}^R | \tilde{v}_{\rm i} \rangle \\
    & = 
    p_{v_{\rm f}} \langle \tilde{a}_{0}^L| \tilde{\mathcal{J}}_{k_{N}}(D_{1}) \left[ |\tilde{a}_{1}^L \rangle ... |\tilde{a}_{N-1}^L \rangle \langle \tilde{a}_{N-1}^L| \tilde{\mathcal{J}}_{k_{1}}(D_{N}) \left[ |\tilde{a}_N^L \rangle \langle \tilde{a}_N^R| \right] | \tilde{a}_{N-1}^R \rangle \langle \tilde{a}_{N-1}^R | ... \langle \tilde{a}_{1}^R | \right] | \tilde{a}_{0}^R \rangle \times  \langle \tilde{a}_N^L | \tilde{v}_{\rm f} \rangle \langle \tilde{v}_{\rm f} | \tilde{a}^R_N \rangle \times  \langle \tilde{v}_{\rm i} | \tilde{a}^L_{0} \rangle \langle \tilde{a}_{0}^R | \tilde{v}_{\rm i} \rangle
    .
    \end{split}
\end{equation}
Here, we have used $\bar{a}^\ell_n = a^\ell_{N-n}$ (for $\ell = L, R$) and $\bar{D}_n = D_{N-n+1}$ to obtain the second equalities in both Eq.~\eqref{eq:PB quant} and Eq.~\eqref{eq:AB quant}. Notice that, again the classical fields $a_n^{\rm c}$ play the same role as $a_n$ in the backward trajectory for a classical system [see Eq.~\eqref{eq: D back gaussians app}]. The quantum fields $a_n^{\rm q}$ are responsible for dephasing in exactly the same way as in the forward trajectory [cf. Eq.~\eqref{eq: Keldysh rotation}].

\subsection{Fluctuation theorem}
By combining the forward and backward trajectories, we find
\begin{equation} \label{eq:ftquantapp}
    \frac{ P_{\rm B}[\bar{\Gamma}, \bar{\mathbf{a}}^L, \bar{\mathbf{a}}^R, \bar{\mathbf{D}}]}{P[\Gamma, \mathbf{a}^L, \mathbf{a}^R, \mathbf{D}]} = \frac{\mathcal{A}_{\rm B}(\bar{\Gamma}, \bar{\mathbf{a}}^L, \bar{\mathbf{a}}^R, \bar{\mathbf{D}} )}{\mc{A}(\Gamma, \mathbf{a}^L, \mathbf{a}^R, \mathbf{D})} e^{\--\sigma_m[\mathbf{a}_{\rm c}, \mathbf{D} ]},
\end{equation}
where the measurement entropy is given by
\begin{equation}
   \sigma_m[\mathbf{a}_{\rm c}, \mathbf{D} ]  =  \frac{4 \lambda}{\gamma} \sum_{n=0}^N \left[ 2a^{\rm c}_n - (D_{n+1} + D_n) \left(D_{n+1} - D_n \right) \right] - \ln \frac{P_{\rm ini, B}[D_{N+1}]}{P_{\rm ini}[D_0]} ,
\end{equation}
which is analogous to the classical expression in Eq.~\eqref{eq: meas ent dis} but with $a_n^{\rm c} $ instead of $a_n$ . From Eqs.~\eqref{eq:AB quant} and~\eqref{eq: A quant}, we find
\begin{equation} \label{eq:ftA}
    \frac{\mathcal{A}_{\rm B}(\bar{\Gamma}, \bar{\mathbf{a}}^L, \bar{\mathbf{a}}^R, \bar{\mathbf{D}} )}{\mc{A}(\Gamma, \mathbf{a}^L, \mathbf{a}^R, \mathbf{D})} = e^{-\sigma[\Gamma, \mathbf{D}]},
\end{equation}
where we used the relations $\bra{\tilde{\bullet}} \tilde{\circ} \rangle = \bra{\circ} \bullet \rangle$ and $\bra{\tilde{\bullet}} \Theta \op{O}^\dagger \Theta^{-1} |\tilde{\circ} \rangle = \bra{\circ} \op{O} |\bullet \rangle$ for an operator $\op{O}$ as well as Eqs.~\eqref{eq: jump back k} and~\eqref{eq: jump back k0} for the backward jump superoperators. The stochastic entropy production is given by~\cite{Manzano_2015, Manzano_2018, Manzano_2022, Horowitz_2013}
\begin{equation} \label{eq:quantent}
    \sigma[\Gamma, \mathbf{D}] = -\log{p_{v_{\rm f}}} + \log{p_{v_{\rm i}}} + \sum_{n=1}^N \sigma_{k_n}(D_n).
\end{equation}
The detailed FT in Eq.~\eqref{eq:FT quant} in the main text can be obtained by inserting Eq.~\eqref{eq:ftA} into Eq.~\eqref{eq:ftquantapp} and integrating out the quantum fields $a_{\rm q}$.

\end{widetext}

\bibliography{refs}

\end{document}